\documentclass[12pt]{article}
\usepackage{amsmath,amssymb}
\usepackage{graphicx}
\newcommand{\eqdef}{\stackrel{\text{def}}{=}}

\newcommand{\n}{\nonumber\\}
\newcommand{\bm}{\boldsymbol}
\newcommand{\ignore}[1]{}
\numberwithin{equation}{section}
\newcommand{\Romannumeral}[1]{\uppercase\expandafter{\romannumeral#1}}

\newcommand{\V}{\text{\Romannumeral{5}}}

\newcounter{myremarkbangou}[section]

\newcounter{myremarkbangouA}[section]

\newcommand{\qbinom}[2]{\genfrac{[}{]}{0pt}{}{\,#1\,}{#2}}

\allowdisplaybreaks[4]

\begin{document}

\baselineskip=20pt

\newcommand{\preprint}{
\vspace*{-20mm}
   \begin{flushright}\normalsize \sf
    DPSU-22-2\\
  \end{flushright}}
\newcommand{\Title}[1]{{\baselineskip=26pt
  \begin{center} \Large \bf #1 \\ \ \\ \end{center}}}
\newcommand{\Author}{\begin{center}
  \large \bf Satoru Odake \end{center}}
\newcommand{\Address}{\begin{center}
     Faculty of Science, Shinshu University,
     Matsumoto 390-8621, Japan
   \end{center}}
\newcommand{\Accepted}[1]{\begin{center}
  {\large \sf #1}\\ \vspace{1mm}{\small \sf Accepted for Publication}
  \end{center}}

\preprint
\thispagestyle{empty}

\Title{New Finite Type Multi-Indexed Orthogonal Polynomials Obtained From
State-Adding Darboux Transformations}

\Author

\Address
\vspace{1cm}

\begin{abstract}
The Hamiltonians of finite type discrete quantum mechanics with real shifts
are real symmetric matrices of order $N+1$. We discuss the Darboux
transformations with higher degree ($>N$) polynomial solutions as seed
solutions. They are state-adding and the resulting Hamiltonians after $M$-steps
are of order $N+M+1$.
Based on twelve orthogonal polynomials (($q$-)Racah, (dual, $q$-)Hahn,
Krawtchouk and five types of $q$-Krawtchouk), new finite type multi-indexed
orthogonal polynomials are obtained, which satisfy second order difference
equations, and all the eigenvectors of the deformed Hamiltonian are described
by them.
We also present explicit forms of the Krein-Adler type multi-indexed
orthogonal polynomials and their difference equations, which are obtained from
the state-deleting Darboux transformations with lower degree ($\leq N$)
polynomial solutions as seed solutions.
\end{abstract}

\section{Introduction}
\label{sec:intro}

Exactly solvable quantum mechanical systems of one degree of freedom 
(Schr\"odinger equation: $\mathcal{H}\phi_n(x)=\mathcal{E}_n\phi_n(x)$) can be
deformed keeping solvability by the Darboux transformations.
By such deformations, new types of orthogonal polynomials, exceptional or
multi-indexed polynomials, are obtained \cite{gkm08}--\cite{idQMcH}
for the Askey-scheme of hypergeometric orthogonal polynomials \cite{ismail,kls}.
They satisfy second order differential or difference equations and form
a complete set of orthogonal basis in an appropriate Hilbert space in spite of
missing degrees, by which the restrictions of Bochner's theorem \cite{ismail}
are avoided.
We have studied orthogonal polynomials based on quantum mechanical formulations:
ordinary quantum mechanics (oQM) and two kinds of discrete quantum mechanics
(dQM), dQM with pure imaginary shifts (idQM) and dQM with real shifts (rdQM)
\cite{os24}. The Schr\"odinger equation for oQM is a differential equation and
that for dQM is a difference equation. The coordinate $x$ for oQM and idQM is
continuous and that for rdQM is discrete.

Depending on the choice of seed solution, the Darboux transformation can be
divided into three types: isospectral, state-deleting and state-adding.
When the wavefunction of the virtual state $\tilde{\phi}_{\text{v}}(x)$,
eigenstate $\phi_n(x)$ and pseudo virtual state
$\tilde{\phi}^{\text{pv}}_{\text{v}}(x)$ is used as a seed solution,
the Darboux transformation is isospectral, state-deleting and state-adding,
respectively.
The wavefunctions $\tilde{\phi}_{\text{v}}(x)$ and
$\tilde{\phi}^{\text{pv}}_{\text{v}}(x)$ are obtained from $\phi_n(x)$ by
twisting parameters.
For oQM and idQM, $\tilde{\phi}_{\text{v}}(x)$ and
$\tilde{\phi}^{\text{pv}}_{\text{v}}(x)$ are solutions of the Schr\"odinger
equation and not square integrable \cite{os25,os27,idQMcH,os29,os30}.
On the other hand, for rdQM, they satisfy the Schr\"odinger equation except
for the boundary \cite{os26,os35,casoidrdqm}.
The Darboux transformations with $\tilde{\phi}_{\text{v}}(x)$ give the case-(1)
multi-indexed polynomials and those with $\phi_n(x)$ or
$\tilde{\phi}^{\text{pv}}_{\text{v}}(x)$ give the case-(2) multi-indexed
polynomials.
Here, the case-(1) is the case that the set of missing degrees of the
multi-indexed polynomials is $\{0,1,\ldots,\ell-1\}$, and the case-(2) is
otherwise.
Another type of seed solution is discussed for the oQM systems with a finite
number of eigenstates $\phi_n(x)$ ($n=0,1,\ldots,n_{\text{max}}$) \cite{os28}.
We call them overshoot eigenfunctions
$\tilde{\phi}^{\text{os}}_n(x)$,
which have the same form as the eigenstates $\phi_n(x)$ but with
$n>n_{\text{max}}$ and are not square integrable.
The overshoot eigenfunctions correspond to the virtual or pseudo virtual
states.

The Hamiltonian of a finite type rdQM system is a real symmetric matrix of order
$N+1$. The coordinate $x$ takes a value in $\{0,1,\ldots,N\}$ and the number
of the eigenstates (eigenvectors) $\phi_n(x)$ ($n=0,1,\ldots,N$) is $N+1$.
By using the pseudo virtual state $\tilde{\phi}^{\text{pv}}_{\text{v}}(x)$
obtained from $\phi_n(x)$ by twisting parameters, the state-adding Darboux
transformation and the Casoratian identities are studied in \cite{casoidrdqm}.
Recently another type of seed solution that gives the state-adding Darboux
transformation was found by Miki, Tsujimoto and Vinet \cite{mtv22}.
They studied single-indexed exceptional Krawtchouk polynomials and one of them
(their type (\romannumeral2)) corresponds to the state-adding Darboux
transformation.
Motivated by their work, Sasaki and the present author studied the $M$-step
state-adding Darboux transformations, whose seed solutions have the same form
as the eigenstate $\phi_n(x)$ but with $n>N$ \cite{os40}. This situation
corresponds to the overshoot eigenfunctions $\tilde{\phi}^{\text{os}}_n(x)$ in
the oQM systems with a finite number of eigenstates.
The overshoot eigenfunction has an infinite norm, but the seed solution here
has a ``zero norm.''
(The meaning of ``zero norm'' is the following.
The Schr\"odinger equation (with $x=0,1,\ldots,N$) is a matrix eigenvalue
problem, and it can be interpreted as a difference equation with a continuous
$x$. The function $\phi_n(x)$ with $n>N$ satisfies this difference equation
and vanishes at $x=0,1,\ldots,N$.
So, $\phi_n(x)$ with $n>N$ satisfies the Schr\"odinger equation and it is a
zero vector, namely zero norm.)
After the $M$-step state-adding Darboux transformations with the seed solutions
$\phi_n(x)$ ($n\in\mathcal{D}=\{d_1,d_2,\ldots,d_M\}$, $d_j>N$), the order of
the Hamiltonian becomes $N+M+1$, and the coordinate $x$ takes a value in
$\{-M,-M+1,\ldots,N\}$. The deformed Hamiltonian has $N+M+1$ eigenvectors.
It is easy to find $N+1$ eigenvectors which correspond to the original
eigenvectors. They are expressed as
$\text{W}_{\text{C}}[\phi_{d_1},\ldots,\phi_{d_M},\phi_n](x)\times(\cdots)$
with $n\in\{0,1,\ldots,N\}$.
However, it is difficult to find extra $M$ eigenvectors.
In \cite{os40}, the special case $\mathcal{D}=\{N+1,N+2,\ldots,N+M\}$ is
studied in detail.

In this paper we study the deformations of finite type rdQM systems by the
$M$-step state-adding Darboux transformations with the seed solutions
$\phi_n(x)$ ($n>N$) and obtain all eigenvectors.
The original systems are described by twelve orthogonal polynomials: ($q$-)Racah,
(dual, $q$-)Hahn, Krawtchouk and five types of $q$-Krawtchouk.
In \cite{mtv22}, which corresponds to $M=1$ case, one extra eigenvector is
derived by two methods: (1) solving the difference equation, (2) shifting $N$
to $N+\varepsilon$ and taking $\varepsilon\to 0$ limit.
For general $M$ case, the first method is difficult and we adopt the second
method. Extra $M$ eigenvectors are obtained from
$\text{W}_{\text{C}}[\phi_{d_1},\ldots,\phi_{d_M},\phi_n](x)\times(\cdots)$
in the ``$n\to d_i$ limit'', which is achieved by shifting $N$ to
$N+\varepsilon$ and taking $\varepsilon\to 0$ limit.
The eigenvectors are described by new multi-indexed orthogonal polynomials
$\check{Q}_{\mathcal{D}',n}(x)$.

This paper is organized as follows.
In section \ref{sec:DT} the finite type rdQM systems are recapitulated and
the multi-step Darboux transformations with seed solutions $\phi_n(x)$ are
discussed.
In section \ref{sec:KAmiop} the results obtained in \S\,\ref{sec:DT} are
applied to the case of seed solutions $\phi_n(x)$ with $n\leq N$, which
corresponds to the state-deleting Darboux transformations.
The eigenvectors are described by the Krein-Adler type multi-indexed
orthogonal polynomials $\check{P}^{\text{KA}}_{\mathcal{D},n}(x)$
($n\in\{0,1,\ldots,N\}\backslash\mathcal{D}$).
This case was studied in \cite{os22}, but the explicit forms of their
difference equations, orthogonal relations etc. are new results.
Section \ref{sec:newmiop} is the main part of the paper.
The results obtained in \S\,\ref{sec:DT} are
applied to the case of seed solutions $\phi_n(x)$ with $n>N$, which
corresponds to the state-adding Darboux transformations.
The eigenvectors are described by new multi-indexed orthogonal polynomials
$\check{Q}_{\mathcal{D}',n}(x)$ ($n\in\{0,1,\ldots,N\}\cup\mathcal{D}$).
Section \ref{sec:summary} is for a summary and comments.
Data for the twelve orthogonal polynomials are presented in
Appendix\,\ref{app:data}. Data for various other quantities and several
formulas are also presented.
In Appendix\,\ref{app:n->di} we discuss the ``$n\to d_i$ limit.''

\section{Darboux Transformations}
\label{sec:DT}

In this section, after recapitulating the finite type rdQM systems
\cite{os12,os24}, we discuss the multi-step Darboux transformations with seed
solutions $\phi_n(x)$ \cite{os22}, especially their algebraic aspects.

\subsection{Original systems}
\label{sec:orgsys}

We consider the finite type rdQM systems as a starting point,
whose eigenvectors are described by the following twelve orthogonal polynomials
\cite{os12,os24}: Hahn (H), Krawtchouk (K), Racah (R), dual Hahn (dH),
dual quantum $q$-Krawtchouk (dq$q$K), $q$-Hahn ($q$H), $q$-Krawtchouk ($q$K),
quantum $q$-Krawtchouk (q$q$K), affine $q$-Krawtchouk (a$q$K), $q$-Racah ($q$R),
dual $q$-Hahn (d$q$H) and dual $q$-Krawtchouk (d$q$K).
The data of these polynomials are presented in Appendix\,\ref{app:poly}.

Let $N$ be a positive integer.
The Hamiltonian $\mathcal{H}$ of a finite type rdQM is a real symmetric
(tridiagonal in this case) matrix of order $N+1$,
\begin{align}
  \mathcal{H}=&(\mathcal{H}_{x,y})_{x,y=0,1,\ldots,N},\quad
  \mathcal{H}_{x,x}=B(x)+D(x),\quad\mathcal{H}_{x,y}=0\ \ (|x-y|>1),\n
  &\mathcal{H}_{x,x+1}=-\sqrt{B(x)D(x+1)},\quad
  \mathcal{H}_{x,x-1}=-\sqrt{B(x-1)D(x)}.
  \label{orgH}
\end{align}
Here the potential functions $B(x)$ and $D(x)$ are positive but vanish at the
boundary,
\begin{align}
  &B(x)>0\ \ (x=0,1,\ldots,N-1),\quad D(x)>0\ \ (x=1,2,\ldots,N),\n
  &B(N)=0,\quad D(0)=0.
  \label{B,D>0}
\end{align}
We write a matrix \eqref{orgH} as
\begin{align}
  \mathcal{H}&=-\sqrt{B(x)D(x+1)}\,e^{\partial}
  -\sqrt{B(x-1)D(x)}\,e^{-\partial}+B(x)+D(x)\n
  &=-\sqrt{B(x)}\,e^{\partial}\sqrt{D(x)}
  -\sqrt{D(x)}\,e^{-\partial}\sqrt{B(x)}+B(x)+D(x),
  \label{Hdef}
\end{align}
where $e^{\pm\partial}$ is a matrix whose $(x,y)$-element is $\delta_{x\pm1,y}$
($\Rightarrow$ $(e^{\partial})^{\dagger}=e^{-\partial}$),
and $A(x)$ means a diagonal matrix $A(x)=\text{diag}(A(0),A(1),\ldots,A(N))$.
Note that $e^{\pm\partial}e^{\mp\partial}\neq1$ due to the effect of boundaries,
\begin{equation}
  e^{\partial}e^{-\partial}=\text{diag}(1,1,\ldots,1,0),\quad
  e^{-\partial}e^{\partial}=\text{diag}(0,1,1,\ldots,1).
\end{equation}
We have $B(x)e^{\partial}e^{-\partial}=B(x)$,
$D(x)e^{-\partial}e^{\partial}=D(x)$, etc.
The Schr\"odinger equation of rdQM is a matrix eigenvalue problem,
\begin{equation}
  \mathcal{H}\phi_n(x)=\mathcal{E}_n\phi_n(x)\ \ (n=0,1,\ldots,N),
  \label{Hphin}
\end{equation}
where $\mathcal{H}\phi_n(x)\eqdef\sum_{y=0}^N\mathcal{H}_{x,y}\phi_n(y)$.
The Hamiltonian \eqref{Hdef} can be written in a factorized form,
\begin{equation}
  \mathcal{H}=\mathcal{A}^{\dagger}\mathcal{A},\quad
  \mathcal{A}\eqdef\sqrt{B(x)}-e^{\partial}\sqrt{D(x)},\quad
  \mathcal{A}^{\dagger}\eqdef\sqrt{B(x)}-\sqrt{D(x)}\,e^{-\partial}.
  \label{H=AdA}
\end{equation}
The tridiagonality \eqref{orgH} and factorization \eqref{H=AdA} imply
$0=\mathcal{E}_0<\mathcal{E}_1<\cdots<\mathcal{E}_N$.
The ground state eigenvector $\phi_0(x)$ is characterized by
$\mathcal{A}\phi_0(x)=0$,
\begin{equation}
  \sqrt{B(x)}\,\phi_0(x)=\sqrt{D(x+1)}\,\phi_0(x+1),
  \label{phi0eq}
\end{equation}
and given by
\begin{equation}
  \phi_0(x)\eqdef\sqrt{\prod_{y=0}^{x-1}\frac{B(y)}{D(y+1)}},
  \label{phi0def}
\end{equation}
which satisfies the normalization $\phi_0(0)=1$ by the convention
$\prod_{j=n}^{n-1}*\eqdef 1$.
For the twelve systems under consideration, the eigenvectors have the following
form,
\begin{equation}
  \phi_n(x)=\phi_0(x)\check{P}_n(x),\quad
  \check{P}_n(x)\eqdef P_n\bigl(\eta(x)\bigr),
  \label{phin=phi0cPn}
\end{equation}
where $P_n(\eta(x))$ is a polynomial of degree $n$ in the sinusoidal
coordinate $\eta(x)$ \cite{os7,os12}. We take the normalization as
\begin{equation}
  \check{P}_n(0)=P_n(0)=1.
  \label{cPn0=1}
\end{equation}
The similarity transformed Hamiltonian $\widetilde{\mathcal{H}}$ is defined by
\begin{align}
  &\widetilde{\mathcal{H}}\eqdef\phi_0(x)^{-1}\circ\mathcal{H}\circ\phi_0(x)
  =B(x)(1-e^{\partial})+D(x)(1-e^{-\partial})
  \label{Htdef}\\
  &\Bigl(\Rightarrow\widetilde{\mathcal{H}}_{x,x+1}=-B(x),
  \ \widetilde{\mathcal{H}}_{x,x-1}=-D(x),
  \ \widetilde{\mathcal{H}}_{x,x}=B(x)+D(x),
  \ \widetilde{\mathcal{H}}_{x,y}=0\ (|x-y|>1)\Bigr),
  \nonumber
\end{align}
and its eigenvalue problem is solved by the polynomial $\check{P}_n(x)$,
\begin{equation}
  \widetilde{\mathcal{H}}\check{P}_n(x)=\mathcal{E}_n\check{P}_n(x)
  \ \ (n=0,1,\ldots,N).
  \label{HtcPn}
\end{equation}
The orthogonality relations of $\check{P}_n(x)$ are
\begin{equation}
  \sum_{x=0}^N\phi_0(x)^2\check{P}_n(x)\check{P}_m(x)
  =\frac{\delta_{nm}}{d_n^2}\ \ (n,m=0,1,\ldots,N).
  \label{orthocPn}
\end{equation}

We have five families of the sinusoidal coordinates $\eta(x)$ \cite{os12},
\begin{equation}
  \begin{array}{rll}
  \text{(\romannumeral1)}:&\eta(x)=x
  &:\text{H,\,K}\\[3pt]
  \text{(\romannumeral2)}:&\eta(x)=x(x+d)
  &:\text{R},\,\text{dH}(d=a+b-1)\\[3pt]
  \text{(\romannumeral3)}:&\eta(x)=1-q^x
  &:\text{dq$q$K}\\[3pt]
  \text{(\romannumeral4)}:&\eta(x)=q^{-x}-1
  &:\text{$q$H,\,$q$K,\,q$q$K,\,a$q$K}\\[3pt]
  \text{(\romannumeral5)}:&\eta(x)=(q^{-x}-1)(1-dq^x)
  &:\text{$q$R},\,\text{d$q$H}(d=abq^{-1}),\,\text{d$q$K}(d=-p).
  \end{array}
  \label{etadef}
\end{equation}
We also have five families of the energy eigenvalues $\mathcal{E}_n$ \cite{os12},
\begin{equation}
  \begin{array}{rll}
  \text{(\romannumeral1)}':&\mathcal{E}_n=n
  &:\text{dH,\,K}\\[3pt]
  \text{(\romannumeral2)}':&\mathcal{E}_n=n(n+\tilde{d})
  &:\text{R},\,\text{H}(\tilde{d}=a+b-1)\\[3pt]
  \text{(\romannumeral3)}':&\mathcal{E}_n=1-q^n
  &:\text{q$q$K}\\[3pt]
  \text{(\romannumeral4)}':&\mathcal{E}_n=q^{-n}-1
  &:\text{d$q$H,\,d$q$K,\,dq$q$K,\,a$q$K}\\[3pt]
  \text{(\romannumeral5)}':&\mathcal{E}_n=(q^{-n}-1)(1-\tilde{d}q^n)
  &:\text{$q$R},\,\text{$q$H}(\tilde{d}=abq^{-1}),\,\text{$q$K}(\tilde{d}=-p).
  \end{array}
  \label{Endef}
\end{equation}
Note that $\eta(0)=\mathcal{E}_0=0$.
The constants $\rho$ and $\kappa$ are defined as follows,
\begin{equation}
  \rho\eqdef\left\{
  \begin{array}{ll}
  1&:\text{(\romannumeral1)},\,\text{(\romannumeral2)}\\
  q&:\text{(\romannumeral3)}\\
  q^{-1}&:\text{(\romannumeral4)},\,\text{(\romannumeral5)}
  \end{array}\right.,\quad
  \kappa\eqdef\left\{
  \begin{array}{ll}
  1&:\text{(\romannumeral1)}',\,\text{(\romannumeral2)}'\\
  q&:\text{(\romannumeral3)}'\\
  q^{-1}&:\text{(\romannumeral4)}',\,\text{(\romannumeral5)}'
  \end{array}\right..
  \label{defrhokappa}
\end{equation}
The rdQM systems have a set of parameters
$\bm{\lambda}=(\lambda_1,\lambda_2,\ldots)$ including the parameter $N$,
and various quantities depend on
$\bm{\lambda}$. Their dependence is expressed like, $f=f(\bm{\lambda})$,
$f(x)=f(x;\bm{\lambda})$.
The parameter $q$ is $0<q<1$ and $q^{\bm{\lambda}}$ stands for
$q^{(\lambda_1,\lambda_2,\ldots)}=(q^{\lambda_1},q^{\lambda_2},\ldots)$.
We omit writing $q$-dependence and sometimes omit writing
$\bm{\lambda}$-dependence, when it does not cause confusion.

\subsection{Darboux transformations}
\label{sec:Darb}

\subsubsection{difference equations}
\label{sec:sabuneq}

The matrix eigenvalue problem \eqref{HtcPn} is written in components as,
\begin{equation}
  B(x)\bigl(\check{P}_n(x)-\check{P}_n(x+1)\bigr)
  +D(x)\bigl(\check{P}_n(x)-\check{P}_n(x-1)\bigr)
  =\mathcal{E}_n\check{P}_n(x).
  \label{sabuneqcPn}
\end{equation}
Eq.\eqref{HtcPn} means that \eqref{sabuneqcPn} holds for $n=0,1,\ldots,N$ and
$x=0,1,\ldots,N$. However, we remark that this difference equation
\eqref{sabuneqcPn} holds for $x\in\mathbb{R}$. Moreover \eqref{sabuneqcPn}
holds for $n\in\mathbb{Z}_{\geq 0}$ (exactly speaking, we need replace
$\check{P}_n(x)$ with $\check{P}^{\text{monic}}_n(x)$, see
\S\,\ref{sec:newmiop}).
We also remark that the positive integer parameter $N$ can be extended to a real
value in \eqref{sabuneqcPn}.
Thus the difference equation \eqref{sabuneqcPn}
(with the replacement $\check{P}_n(x)\to\check{P}^{\text{monic}}_n(x)$) holds
for $x\in\mathbb{R}$, $n\in\mathbb{Z}_{\geq0}$ and $N\in\mathbb{R}$.
This is an important point for the Darboux transformations in this subsection
and constructions of new multi-indexed orthogonal polynomials in
\S\,\ref{sec:newmiop}.
The component form of \eqref{Hphin} is
\begin{equation}
  -\sqrt{B(x)D(x+1)}\,\phi_n(x+1)-\sqrt{B(x-1)D(x)}\,\phi_n(x-1)
  +\bigl(B(x)+D(x)\bigr)\phi_n(x)
  =\mathcal{E}_n\phi_n(x).
  \label{sabuneqphin}
\end{equation}
Since the explicit forms of $\phi_0(x)^2$ in Appendix\,\ref{app:poly} are
expressed in terms of $\alpha^x$,
$(\alpha)_x=\Gamma(\alpha+x)/\Gamma(\alpha)$ and
$(\alpha\,;q)_x=(\alpha\,;q)_{\infty}/(\alpha q^x\,;q)_{\infty}$,
we can consider $\phi_0(x)^2$ for $x\in\mathbb{R}$ (if needed, we shift $N$ to
a non-integer value).
Let us illustrate this situation using $q$R case as an example.
We rewrite $\phi_0(x)^2$ in Appendix\,\ref{app:qR} as follows:
\begin{align*}
  \phi_0(x)^2
  &=\frac{(a,b,c,d\,;q)_x}{(a^{-1}dq,b^{-1}dq,c^{-1}dq,q\,;q)_x\tilde{d}^x}
  \frac{1-dq^{2x}}{1-d}\n
  &=\frac{(a,b,c,d\,;q)_{\infty}}{(aq^x,bq^x,cq^x,dq^x\,;q)_{\infty}}
  \frac{(a^{-1}dq^{x+1},b^{-1}dq^{x+1},c^{-1}dq^{x+1},q^{x+1}\,;q)_{\infty}}
  {(a^{-1}dq,b^{-1}dq,c^{-1}dq,q\,;q)_{\infty}\tilde{d}^x}
  \frac{1-dq^{2x}}{1-d}.
\end{align*}
This is defined for generic values of $x\in\mathbb{R}$ and satisfies
$B(x)\phi_0(x)^2=D(x+1)\phi_0(x)^2$ ($x\in\mathbb{R}$).
However, the factor $(a\,;q)_{\infty}$ vanishes because of $a=q^{-N}$.
To avoid this, we shift $N$ slightly from an integer value.
For $\phi_0(x)^2$ at $x\in\mathbb{Z}$, after simplifying
$(a\,;q)_{\infty}/(aq^x\,;q)_{\infty}$,
we shift $N$ back to an integer value.
For $N\in\mathbb{Z}_{>0}$ and $x\in\mathbb{Z}$, $\phi_0(x;\bm{\lambda})^2$ is
non-vanishing only for $x=0,1,\ldots,N$.
If we ignore the positivity of the square root argument, the difference
equation \eqref{sabuneqphin} (with the replacement
$\phi_n(x)\to\phi^{\text{monic}}_n(x)=\phi_0(x)\check{P}^{\text{monic}}_n(x)$)
also holds for
$x\in\mathbb{R}$, $n\in\mathbb{Z}_{\geq0}$ and $N\in\mathbb{R}$.
Since these difference equations \eqref{sabuneqcPn} and \eqref{sabuneqphin} are
algebraic relations, they hold for any parameter range of $\bm{\lambda}$
(unless they are ill-defined).
The orthogonality relations \eqref{orthocPn} with positive weight restrict the
parameter range of $\bm{\lambda}$ (see the parameter range for the
positivity in Appendix\,\ref{app:poly}).
If we do not require the positivity of the weight factor, the relations
\eqref{orthocPn} themselves are valid for any parameter range of $\bm{\lambda}$
(unless they are ill-defined), because the relations \eqref{orthocPn} are
finite sums and algebraic relations.

To write the difference equations \eqref{sabuneqcPn} and \eqref{sabuneqphin}
for $x\in\mathbb{R}$ compactly, let us introduce the shift operators
$e^{\pm\hat{\partial}}$ acting on functions of $x\in\mathbb{R}$ as
$e^{\pm\hat{\partial}}f(x)\eqdef f(x\pm1)$ ($x\in\mathbb{R}$).
They are related as $(e^{\hat{\partial}})^{\dagger}=e^{-\hat{\partial}}$ and
satisfy $e^{\pm\hat{\partial}}e^{\mp\hat{\partial}}=1$.
By replacing $e^{\pm\partial}$ with $e^{\pm\hat{\partial}}$ in \eqref{Hdef} and
\eqref{Htdef}, we define the following operators acting on functions of
$x\in\mathbb{R}$,
\begin{align}
  \mathcal{H}^{\text{op}}&\eqdef-\sqrt{B(x)}\,e^{\hat{\partial}}\sqrt{D(x)}
  -\sqrt{D(x)}\,e^{-\hat{\partial}}\sqrt{B(x)}+B(x)+D(x),
  \label{Hdef2}\\
  \widetilde{\mathcal{H}}^{\text{op}}
  &\eqdef\phi_0(x)^{-1}\circ\mathcal{H}^{\text{op}}\circ\phi_0(x)
  =B(x)(1-e^{\hat{\partial}})+D(x)(1-e^{-\hat{\partial}}).
  \label{Htdef2}
\end{align}
Then the difference equations \eqref{sabuneqphin} and \eqref{sabuneqcPn} for
$x\in\mathbb{R}$ are expressed as
\begin{equation}
  \mathcal{H}^{\text{op}}\phi_n(x)=\mathcal{E}_n\phi_n(x),\quad
  \widetilde{\mathcal{H}}^{\text{op}}\check{P}_n(x)=\mathcal{E}_n\check{P}_n(x)
  \quad(n\in\mathbb{Z}_{\geq0}),
  \label{Hopphin}
\end{equation}
where we ignore the positivity of the square root argument, and $\phi_n(x)$ and
$\check{P}_n(x)$ should be replaced with $\phi^{\text{monic}}_n(x)$ and
$\check{P}^{\text{monic}}_n(x)$ for $n>N$, respectively.

\subsubsection{Darboux transformations}
\label{sec:Dar}

To describe the multi-step Darboux transformations, the Casorati determinant
(Casoratian) is needed. The Casoratian for $n$ functions $f_j(x)$ is defined as
\begin{equation}
  \text{W}_{\text{C}}[f_1,f_2,\ldots,f_n](x)
  \eqdef\det\Bigl(f_k(x+j-1)\Bigr)_{1\leq j,k\leq n},
\end{equation}
(for $n=0$, we set $\text{W}_{\text{C}}[\cdot](x)=1$) and the following
properties are used,
\begin{align}
  &\text{W}_{\text{C}}[gf_1,gf_2,\ldots,gf_n](x)
  =\prod_{k=0}^{n-1}g(x+k)\cdot\text{W}_{\text{C}}[f_1,f_2,\ldots,f_n](x),
  \label{WCformula1}\\
  &\text{W}_{\text{C}}\bigl[\text{W}_{\text{C}}[f_1,f_2,\ldots,f_n,g],
  \text{W}_{\text{C}}[f_1,f_2,\ldots,f_n,h]\,\bigr](x)\n
  &=\text{W}_{\text{C}}[f_1,f_2,\ldots,f_n](x+1)\,
  \text{W}_{\text{C}}[f_1,f_2,\ldots,f_n,g,h](x)
  \quad(n\geq 0).
  \label{WCformula2}
\end{align}
See \cite{wcid} for further properties of the Casoratian.

The Darboux transformations for rdQM with seed solutions $\phi_n(x)$ are
studied in \cite{os22}.
Since we are interested in their algebraic aspect here, we consider the
Darboux transformations for $\mathcal{H}^{\text{op}}$ rather than $\mathcal{H}$.
The algebraic calculations are exactly the same in both cases.
In this subsection, we ignore the positivity of the square root argument and
adopt the rule $\sqrt{A^2}=A$ (instead of $\sqrt{A^2}=|A|$ for $A\in\mathbb{R}$).
Although $\phi_n(x)$ should be replaced with
$\phi^{\text{monic}}_n(x)$ for $n>N$ case, we write it as
$\phi_n(x)$ in this subsection for simplicity of presentation.

We consider $M$-step Darboux transformations with the seed solutions
$\phi_{d_1}(x),\phi_{d_2}(x),\ldots,$ $\phi_{d_M}(x)$
satisfying \eqref{Hopphin}.
Let us denote a set of labels of seed solutions as $\mathcal{D}$
(exactly speaking an ordered set)
\begin{equation}
  \mathcal{D}=\{d_1,d_2,\ldots,d_M\}\ \ (d_j\in\mathbb{Z}_{\geq 0}
  \text{ : mutually distinct}).
  \label{Ddef}
\end{equation}
(Although this notation $d_j$ conflicts with the notation of the normalization
constant $d_n$ in \eqref{orthocPn}, we think this does not cause any confusion
because the latter appears as $\delta_{nm}/d_n^2$.)
For later use, let us define $\ell_{\mathcal{D}}$,
$\ell^{\text{KA}}_{\mathcal{D}}$ and $\mathcal{D}^{[i]}$ for
$i\geq-\min\mathcal{D}$,
\begin{align}
  &\ell_{\mathcal{D}}\eqdef\sum_{j=1}^Md_j-\frac12M(M-1),\quad
  \ell^{\text{KA}}_{\mathcal{D}}\eqdef\sum_{j=1}^Md_j-\frac12M(M+1),
  \label{ellDdef}\\
  &\mathcal{D}^{[i]}\eqdef\{d_1+i,d_2+i\ldots,d_M+i\}.
  \label{D[i]def}
\end{align}
The Hamiltonian \eqref{Hdef2} is factorized as
\begin{equation}
  \mathcal{H}^{\text{op}}
  =\mathcal{A}^{\text{op}\,\dagger}\mathcal{A}^{\text{op}},\quad
  \mathcal{A}^{\text{op}}\eqdef\sqrt{B(x)}-e^{\hat{\partial}}\sqrt{D(x)},\quad
  \mathcal{A}^{\text{op}\,\dagger}\eqdef
  \sqrt{B(x)}-\sqrt{D(x)}\,e^{-\hat{\partial}},
\end{equation}
and the difference equation and its solution are
\begin{equation}
  \mathcal{H}^{\text{op}}\phi_n(x)
  =\mathcal{E}_n\phi_n(x)\ \ (n\in\mathbb{Z}_{\geq0}).
  \label{Hopphin2}
\end{equation}
Let $s=0$ be the quantity at this starting point, and define the quantity at
the $s$-th step as follows \cite{os22}:
\begin{align}
  &\mathcal{H}^{\text{op}}_{d_1\ldots d_s}
  \eqdef\hat{\mathcal{A}}^{\text{op}}_{d_1\ldots d_s}
  \hat{\mathcal{A}}_{d_1\ldots d_s}^{\text{op}\,\dagger}+\mathcal{E}_{d_s},\\
  &\hat{\mathcal{A}}^{\text{op}}_{d_1\ldots d_s}
  \eqdef\sqrt{\hat{B}_{d_1\ldots d_s}(x)}
  -e^{\hat{\partial}}\sqrt{\hat{D}_{d_1\ldots d_s}(x)},
  \ \ \hat{\mathcal{A}}_{d_1\ldots d_s}^{\text{op}\,\dagger}
  \eqdef\sqrt{\hat{B}_{d_1\ldots d_s}(x)}
  -\sqrt{\hat{D}_{d_1\ldots d_s}(x)}\,e^{-\hat{\partial}},\\
  &\hat{B}_{d_1\ldots d_s}(x)\eqdef\left\{
  \begin{array}{ll}
  {\displaystyle\sqrt{\hat{B}_{d_1\ldots d_{s-1}}(x+1)
  \hat{D}_{d_1\ldots d_{s-1}}(x+1)}\,
  \frac{\phi_{d_1\ldots d_s}(x+1)}{\phi_{d_1\ldots d_s}(x)}}&(s\geq1)\\
  B(x-1)&(s=0)
  \end{array}\right.,\\
  &\hat{D}_{d_1\ldots d_s}(x)\eqdef\left\{
  \begin{array}{ll}
  {\displaystyle\sqrt{\hat{B}_{d_1\ldots d_{s-1}}(x)
  \hat{D}_{d_1\ldots d_{s-1}}(x)}\,
  \frac{\phi_{d_1\ldots d_s}(x-1)}{\phi_{d_1\ldots d_s}(x)}}&(s\geq1)\\
  D(x)&(s=0)
  \end{array}\right.,\\
  &\phi_{d_1\ldots d_s\,n}(x)\eqdef\hat{\mathcal{A}}^{\text{op}}_{d_1\ldots d_s}
  \phi_{d_1\ldots d_{s-1}\,n}(x).
\end{align}
Note that $\phi_{d_1\ldots d_s\,n}(x)$ is defined for $n\in\mathbb{Z}_{\geq 0}$,
but we have $\phi_{d_1\ldots d_s\,n}(x)=0$ for $n\in\{d_1,\ldots,d_s\}$.
Then we obtain
\begin{align}
  &\mathcal{H}^{\text{op}}_{d_1\ldots d_s}\phi_{d_1\ldots d_s\,n}(x)
  =\mathcal{E}_n\,\phi_{d_1\ldots d_s\,n}(x),\\
  &\phi_{d_1\ldots d_s\,n}(x)
  =(-1)^s\sqrt{\prod_{k=1}^s\hat{B}_{d_1\ldots d_k}(x)}\,
  \frac{\text{W}_{\text{C}}[\phi_{d_1},\ldots,\phi_{d_s},\phi_n](x)}
  {\text{W}_{\text{C}}[\phi_{d_1},\ldots,\phi_{d_s}](x+1)}\n
  &\phantom{\phi_{d_1\ldots d_s\,n}(x)}
  =(-1)^s\sqrt{\prod_{k=1}^s\hat{D}_{d_1\ldots d_k}(x+s+1-k)}\,
  \frac{\text{W}_{\text{C}}[\phi_{d_1},\ldots,\phi_{d_s},\phi_n](x)}
  {\text{W}_{\text{C}}[\phi_{d_1},\ldots,\phi_{d_s}](x)},\\
  &\phi_{d_1\ldots d_{s-1}\,n}(x)
  =\frac{\mathcal{A}_{d_1\ldots d_s}^{\text{op}\,\dagger}}
  {\mathcal{E}_n-\mathcal{E}_{d_s}}
  \phi_{d_1\ldots d_s\,n}(x)\ \ (s\geq 1),\\
  &\hat{B}_{d_1\ldots d_s}(x)=\hat{D}_{d_1\ldots d_s}(x+1)
  \Bigl(\frac{\phi_{d_1\ldots d_s}(x+1)}{\phi_{d_1\ldots d_s}(x)}\Bigr)^2,\\
  &\hat{B}_{d_1\ldots d_s}(x)\hat{D}_{d_1\ldots d_s}(x+1)
  =\hat{B}_{d_1\ldots d_{s-1}}(x+1)\hat{D}_{d_1\ldots d_{s-1}}(x+1)
  \ \ (s\geq1),\\
  &\hat{B}_{d_1\ldots d_s}(x)+\hat{D}_{d_1\ldots d_s}(x)+\mathcal{E}_{d_s}
  =\left\{\begin{array}{ll}
  \hat{B}_{d_1\ldots d_{s-1}}(x)+\hat{D}_{d_1\ldots d_{s-1}}(x+1)
  +\mathcal{E}_{d_{s-1}}&(s\geq2)\\[3pt]
  B(x)+D(x)&(s=1)
  \end{array}\right.,\\
  &\mathcal{H}^{\text{op}}_{d_1\ldots d_s}=
  \hat{\mathcal{A}}_{d_1\ldots d_{s+1}}^{\text{op}\,\dagger}
  \hat{\mathcal{A}}^{\text{op}}_{d_1\ldots d_{s+1}}+\mathcal{E}_{d_{s+1}},\\
  &\hat{B}_{d_1\ldots d_s}(x)=\sqrt{B(x+s-1)D(x+s)}\,
  \frac{w_{s-1}(x)}{w_{s-1}(x+1)}\frac{w_s(x+1)}{w_s(x)}\ \ (s\geq1),\\
  &\hat{D}_{d_1\ldots d_s}(x)=\sqrt{B(x-1)D(x)}\,
  \frac{w_{s-1}(x+1)}{w_{s-1}(x)}\frac{w_s(x-1)}{w_s(x)}\ \ (s\geq1),\\
  &\prod_{k=1}^s\hat{B}_{d_1\ldots d_k}(x)
  =\sqrt{\prod_{k=1}^sB(x+k-1)D(x+k)}\,
  \frac{w_s(x+1)}{w_s(x)},\\
  &\prod_{k=1}^s\hat{D}_{d_1\ldots d_k}(x+s+1-k)
  =\sqrt{\prod_{k=1}^sB(x+k-1)D(x+k)}\,
  \frac{w_s(x)}{w_s(x+1)},\\
  &\phi_{d_1\ldots d_s\,n}(x)
  =(-1)^s\Bigl(\prod_{k=1}^sB(x+k-1)D(x+k)\Bigr)^{\frac14}\,
  \frac{\text{W}_{\text{C}}[\phi_{d_1},\ldots,\phi_{d_s},\phi_n](x)}
  {\sqrt{w_s(x)w_s(x+1)}},
\end{align}
etc. Here $w_s(x)$ is
\begin{equation}
  w_s(x)=\text{W}_{\text{C}}[\phi_{d_1},\ldots,\phi_{d_s}](x).
\end{equation}

For $s=M$, we write $\mathcal{H}^{\text{op}}_{d_1\ldots d_M}
=\mathcal{H}^{\text{op}}_{\mathcal{D}}$,
$\hat{\mathcal{A}}^{\text{op}}_{d_1\ldots d_M}
=\hat{\mathcal{A}}^{\text{op}}_{\mathcal{D}}$,
$\hat{B}_{d_1\ldots d_M}(x)=\hat{B}_{\mathcal{D}}(x)$,
$\hat{D}_{d_1\ldots d_M}(x)=\hat{D}_{\mathcal{D}}(x)$,
$\phi_{d_1\ldots d_M\,n}(x)=\phi_{\mathcal{D}\,n}(x)$,
then we have
\begin{equation}
  \mathcal{H}^{\text{op}}_{\mathcal{D}}
  =\hat{\mathcal{A}}^{\text{op}}_{\mathcal{D}}
  \hat{\mathcal{A}}_{\mathcal{D}}^{\text{op}\,\dagger}
  +\mathcal{E}_{d_M},\quad
  \mathcal{H}^{\text{op}}_{\mathcal{D}}\phi_{\mathcal{D}\,n}(x)
  =\mathcal{E}_n\phi_{\mathcal{D}\,n}(x).
\end{equation}
By expressing this Hamiltonian in the standard form, we have
\begin{align}
  &\mathcal{H}^{\text{op}}_{\mathcal{D}}
  =\mathcal{A}_{\mathcal{D}}^{\text{op}\,\dagger}
  \mathcal{A}^{\text{op}}_{\mathcal{D}}+\mathcal{E}_{\mu},\quad
  \mu\eqdef\min(\mathbb{Z}_{\geq0}\backslash\mathcal{D}),
  \label{HopD}\\
  &\mathcal{A}^{\text{op}}_{\mathcal{D}}\eqdef\sqrt{B_{\mathcal{D}}(x)}
  -e^{\hat{\partial}}\sqrt{D_{\mathcal{D}}(x)},\quad
  \mathcal{A}_{\mathcal{D}}^{\text{op}\,\dagger}\eqdef\sqrt{B_{\mathcal{D}}(x)}
  -\sqrt{D_{\mathcal{D}}(x)}\,e^{-\hat{\partial}},\\
  &B_{\mathcal{D}}(x)\eqdef
  \sqrt{\hat{B}_{\mathcal{D}}(x+1)\hat{D}_{\mathcal{D}}(x+1)}\,
  \frac{\phi_{\mathcal{D}\,\mu}(x+1)}{\phi_{\mathcal{D}\,\mu}(x)},
  \label{BD}\\
  &D_{\mathcal{D}}(x)\eqdef
  \sqrt{\hat{B}_{\mathcal{D}}(x)\hat{D}_{\mathcal{D}}(x)}\,
  \frac{\phi_{\mathcal{D}\,\mu}(x-1)}{\phi_{\mathcal{D}\,\mu}(x)},
  \label{DD}\\
  &\phi_{\mathcal{D}\,n}(x)
  =(-1)^M\Bigl(\prod_{k=1}^MB(x+k-1)D(x+k)\Bigr)^{\frac14}\,
  \frac{\text{W}_{\text{C}}[\phi_{d_1},\ldots,\phi_{d_M},\phi_n](x)}
  {\sqrt{w_M(x)w_M(x+1)}},
  \label{phiDn}
\end{align}
and
\begin{align}
  &B_{\mathcal{D}}(x)+D_{\mathcal{D}}(x)+\mathcal{E}_{\mu}
  =\hat{B}_{\mathcal{D}}(x)+\hat{D}_{\mathcal{D}}(x+1)+\mathcal{E}_{d_M},\\
  &B_{\mathcal{D}}(x)=\sqrt{B(x+M)D(x+M+1)}\,
  \frac{w_M(x)}{w_M(x+1)}
  \frac{\text{W}_{\text{C}}[\phi_{d_1},\ldots,\phi_{d_M},\phi_{\mu}](x+1)}
  {\text{W}_{\text{C}}[\phi_{d_1},\ldots,\phi_{d_M},\phi_{\mu}](x)},\\
  &D_{\mathcal{D}}(x)=\sqrt{B(x-1)D(x)}\,
  \frac{w_M(x+1)}{w_M(x)}
  \frac{\text{W}_{\text{C}}[\phi_{d_1},\ldots,\phi_{d_M},\phi_{\mu}](x-1)}
  {\text{W}_{\text{C}}[\phi_{d_1},\ldots,\phi_{d_M},\phi_{\mu}](x)}.
\end{align}

We remark that the results so far have used \eqref{Hopphin2}, but not
\eqref{phin=phi0cPn}. The information \eqref{phin=phi0cPn} will be used in
\S\,\ref{sec:KAmiop} and \S\,\ref{sec:newmiop}.
In the following, we will denote $\phi_{\mathcal{D}\,n}(x)$ in \eqref{phiDn}
as $\phi_{\mathcal{D}\,n}^{\text{gen}}(x)$.

\subsubsection{orthogonality relations}
\label{sec:orthorel}

In \S\,\ref{sec:Dar} we consider a continuous variable $x\in\mathbb{R}$, and
the results are valid for a discrete variable $x\in\mathbb{Z}$.
Let us consider the case $x\in\mathbb{Z}$. This means the embedding of the
finite system in \S\,\ref{sec:orgsys} into the infinite system.
The matrices $e^{\pm\partial}$ satisfy $e^{\pm\partial}e^{\mp\partial}=1$.
The inner product of vectors $f$ and $g$ is defined by
$(f,g)=\sum_{x\in\mathbb{Z}}f(x)g(x)$.
Then the formal calculation gives
\begin{align}
  (\phi_{d_1\ldots d_s\,n},\phi_{d_1\ldots d_s\,m})
  &=(\hat{\mathcal{A}}_{d_1\ldots d_s}\phi_{d_1\ldots d_{s-1}\,n},
  \hat{\mathcal{A}}_{d_1\ldots d_s}\phi_{d_1\ldots d_{s-1}\,m})\n
  &=(\hat{\mathcal{A}}_{d_1\ldots d_s}^{\dagger}
  \hat{\mathcal{A}}_{d_1\ldots d_s}
  \phi_{d_1\ldots d_{s-1}\,n},\phi_{d_1\ldots d_{s-1}\,m})\n
  &=\bigl((\mathcal{H}_{d_1\ldots d_{s-1}}-\mathcal{E}_{d_s})
  \phi_{d_1\ldots d_{s-1}\,n},\phi_{d_1\ldots d_{s-1}\,m}\bigr)\n
  &=(\mathcal{E}_n-\mathcal{E}_{d_s})
  (\phi_{d_1\ldots d_{s-1}\,n},\phi_{d_1\ldots d_{s-1}\,m}\bigr)
  \ \ (s\geq 1),
\end{align}
and this gives
\begin{equation}
  (\phi_{d_1\ldots d_s\,n},\phi_{d_1\ldots d_s\,m})
  =(\phi_n,\phi_m)\prod_{j=1}^s(\mathcal{E}_n-\mathcal{E}_{d_j}).
\end{equation}
If everything goes well after $M$-steps, we obtain
\begin{equation}
  \sum_{x\in\mathbb{Z}}\phi^{\text{gen}}_{\mathcal{D}\,n}(x;\bm{\lambda})
  \phi^{\text{gen}}_{\mathcal{D}\,m}(x;\bm{\lambda})
  =\frac{\delta_{nm}}{d_n(\bm{\lambda})^2}\prod_{j=1}^M
  \bigl(\mathcal{E}_n(\bm{\lambda})-\mathcal{E}_{d_j}(\bm{\lambda})\bigr).
  \label{sumphigen}
\end{equation}
As discussed in \S\,\ref{sec:sabuneq}, $\phi_0(x;\bm{\lambda})^2$ for
$N\in\mathbb{Z}_{>0}$ and $x\in\mathbb{Z}$ is non-vanishing only for
$x=0,1,\ldots,N$.
The function $\phi^{\text{gen}}_{\mathcal{D}\,n}(x;\bm{\lambda})$ contains
the function ``$\phi_0(x)$'' as a factor.
For the original system ($M=0$ case), the ``$\phi_0(x)$ factor'' is
$\phi_0(x;\bm{\lambda})$, and the sum in \eqref{sumphigen} is reduced to
$\sum_{x=0}^N$.
In \S\,\ref{sec:KAmiop} and \S\,\ref{sec:newmiop} we will see the following
situations.
For the systems in \S\,\ref{sec:KAmiop}, the ``$\phi_0(x)$ factor'' is
$\phi_0(x;\bm{\lambda}+M\bm{\delta})$, whose parameter $N$ is $N-M$, and
the sum in \eqref{sumphigen} is reduced to $\sum_{x=0}^{N-M}$.
For the systems in \S\,\ref{sec:newmiop}, the ``$\phi_0(x)$ factor'' is
$\phi_0(x+M;\bm{\lambda}-M\bm{\bar{\delta}})$, whose parameter $N$ is $N+M$,
and the sum in \eqref{sumphigen} is reduced to $\sum_{x=-M}^N$.

\section{Krein-Adler Type Multi-Indexed Orthogonal Polynomials From
State-Deleting Darboux Transformations}
\label{sec:KAmiop}

The Darboux transformations with seed solutions $\phi_n(x)$ ($n\leq N$) are
studied in \cite{os22} and we call the resulting multi-indexed polynomials
Krein-Adler type (KA-type) multi-indexed polynomials.
Explicit forms of their difference equations, orthogonal relations etc. are
not given in \cite{os22}, and we present them here.

The multi-index set we will consider is
\begin{equation}
  \mathcal{D}=\{d_1,d_2,\ldots,d_M\}\ \ (0\leq d_j\leq N
  \text{ : mutually distinct}),
  \label{DdefKA}
\end{equation}
and $M$ should be $M\leq N$.
By using the information \eqref{phin=phi0cPn}, the eigenfunctions
$\phi_{\mathcal{D}\,n}^{\text{gen}}(x)$ are expressed in terms of $\phi_0$
and multi-indexed polynomials \cite{os22}.

The denominator polynomial $\Xi^{\text{KA}}_{\mathcal{D}}(\eta)$ and the
multi-indexed polynomial $P^{\text{KA}}_{\mathcal{D},n}(\eta)$ are defined by
(see Appendix\,\ref{app:poly} for $\bm{\delta}$ and Appendix\,\ref{app:varphiM}
for $\varphi_M(x)$)
\begin{align}
  \check{\Xi}^{\text{KA}}_{\mathcal{D}}(x;\bm{\lambda})
  &\eqdef\Xi^{\text{KA}}_{\mathcal{D}}
  \bigl(\eta(x;\bm{\lambda}+(M-1)\bm{\delta});\bm{\lambda}\bigr)\n
  &\eqdef\mathcal{C}^{\text{KA}}_{\mathcal{D}}(\bm{\lambda})^{-1}
  \varphi_M(x;\bm{\lambda})^{-1}
  \text{W}_{\text{C}}[\check{P}_{d_1},\ldots,\check{P}_{d_M}](x;\bm{\lambda}),
  \label{cXiDKAdef}\\
  \check{P}^{\text{KA}}_{\mathcal{D},n}(x;\bm{\lambda})
  &\eqdef P^{\text{KA}}_{\mathcal{D},n}\bigl(\eta(x;\bm{\lambda}+M\bm{\delta});
  \bm{\lambda}\bigr)\quad(n=0,1,\ldots,N)\n
  &\eqdef\mathcal{C}^{\text{KA}}_{\mathcal{D},n}(\bm{\lambda})^{-1}
  \varphi_{M+1}(x;\bm{\lambda})^{-1}
  \text{W}_{\text{C}}[\check{P}_{d_1},\ldots,\check{P}_{d_M},\check{P}_n]
  (x;\bm{\lambda}).
  \label{cPDnKAdef}
\end{align}
We remark that $\check{P}^{\text{KA}}_{\mathcal{D},n}(x;\bm{\lambda})=0$
for $n\in\mathcal{D}$, and $\check{P}^{\text{KA}}_{\mathcal{D},n}(x)$ can be
defined for $n\in\mathbb{Z}_{\geq0}$ by replacing $\check{P}_n$ with
$\check{P}^{\text{monic}}_n$ for $n>N$, see \S\,\ref{sec:newmiop}.
They are
\begin{align}
  \check{\Xi}^{\text{KA}}_{\mathcal{D}}(x;\bm{\lambda})&:
  \text{a polynomial of degree
  $\ell^{\text{KA}}_{\mathcal{D}}+M=\ell_{\mathcal{D}}$ in
  $\eta(x;\bm{\lambda}+(M-1)\bm{\delta})$},\n
  \check{P}^{\text{KA}}_{\mathcal{D},n}(x;\bm{\lambda})&:
  \text{a polynomial of degree $\ell^{\text{KA}}_{\mathcal{D}}+n$ in
  $\eta(x;\bm{\lambda}+M\bm{\delta})$},
  \label{cPDnKAdeg}
\end{align}
where $\ell^{\text{KA}}_{\mathcal{D}}$ and $\ell_{\mathcal{D}}$ are given by
\eqref{ellDdef}.
The constants $\mathcal{C}^{\text{KA}}_{\mathcal{D}}(\bm{\lambda})$ and
$\mathcal{C}^{\text{KA}}_{\mathcal{D},n}(\bm{\lambda})$ are
determined by the following normalization conditions,
\begin{equation}
  \check{\Xi}^{\text{KA}}_{\mathcal{D}}(0;\bm{\lambda})
  =\check{P}^{\text{KA}}_{\mathcal{D},n}(0;\bm{\lambda})=1.
  \label{cPKADn(0)=1}
\end{equation}
In contrast to the case-(1) multi-indexed polynomials in \cite{os26},
the denominator polynomial and the multi-indexed polynomial are essentially
the same,
\begin{equation}
  \check{P}^{\text{KA}}_{\mathcal{D},n}(x;\bm{\lambda})
  =\check{\Xi}^{\text{KA}}_{\mathcal{D}'}(x;\bm{\lambda}),\quad
  \mathcal{D}'=\{d_1,d_2,\ldots,d_M,n\}.
\end{equation}
Thus we have
\begin{equation}
  \mathcal{C}^{\text{KA}}_{\mathcal{D},n}(\bm{\lambda})
  =\mathcal{C}^{\text{KA}}_{\mathcal{D}'}(\bm{\lambda}),\quad
  \mathcal{D}'=\{d_1,d_2,\ldots,d_M,n\}.
  \label{cCXiP}
\end{equation}
The constant $\mathcal{C}^{\text{KA}}_{\mathcal{D}}(\bm{\lambda})$ is given by
((A.16) of \cite{os22})
\begin{equation}
  \mathcal{C}^{\text{KA}}_{\mathcal{D}}(\bm{\lambda})
  =(-1)^{\binom{M}{2}}\kappa^{-\binom{M}{3}}
  \prod_{1\leq j<k\leq M}
  \frac{\mathcal{E}_{d_k}(\bm{\lambda})-\mathcal{E}_{d_j}(\bm{\lambda})}
  {B(0;\bm{\lambda}+(j-1)\bm{\delta})},
\end{equation}
and \eqref{cCXiP} gives the constant
$\mathcal{C}^{\text{KA}}_{\mathcal{D},n}(\bm{\lambda})$,
\begin{equation}
  \mathcal{C}^{\text{KA}}_{\mathcal{D},n}(\bm{\lambda})
  =\mathcal{C}^{\text{KA}}_{\mathcal{D}}(\bm{\lambda})
  (-1)^M\kappa^{-\binom{M}{2}}\prod_{j=1}^M
  \frac{\mathcal{E}_n(\bm{\lambda})-\mathcal{E}_{d_j}(\bm{\lambda})}
  {B(0;\bm{\lambda}+(j-1)\bm{\delta})}.
  \label{CKADn=}
\end{equation}

Let us calculate $\phi_{\mathcal{D}\,n}^{\text{gen}}(x)$ \eqref{phiDn}.
{}From the property \eqref{WCformula1} and the definitions
\eqref{cXiDKAdef}--\eqref{cPDnKAdef}, we have
\begin{align}
  &w_M(x)=\text{W}_{\text{C}}[\phi_{d_1},\ldots,\phi_{d_M}](x)
  =\prod_{k=1}^M\phi_0(x+k-1;\bm{\lambda})\cdot
  \mathcal{C}^{\text{KA}}_{\mathcal{D}}(\bm{\lambda})
  \varphi_M(x;\bm{\lambda})
  \check{\Xi}^{\text{KA}}_{\mathcal{D}}(x;\bm{\lambda}),\\
  &\text{W}_{\text{C}}[\phi_{d_1},\ldots,\phi_{d_M},\phi_n](x)
  =\prod_{k=1}^{M+1}\phi_0(x+k-1;\bm{\lambda})\cdot
  \mathcal{C}^{\text{KA}}_{\mathcal{D},n}(\bm{\lambda})
  \varphi_{M+1}(x;\bm{\lambda})
  \check{P}^{\text{KA}}_{\mathcal{D},n}(x;\bm{\lambda}).
\end{align}
By using these and \eqref{phi0eq}, \eqref{varphiM+1M} and \eqref{B(x+j),D(x)},
we obtain
\begin{align}
  \phi_{\mathcal{D}\,n}^{\text{gen}}(x;\bm{\lambda})
  &=\kappa^{-\frac12\binom{M}{2}}\prod_{j=1}^M
  \frac{\mathcal{E}_n(\bm{\lambda})-\mathcal{E}_{d_j}(\bm{\lambda})}
  {\sqrt{B(0;\bm{\lambda}+(j-1)\bm{\delta})}}
  \times\phi^{\text{KA}}_{\mathcal{D}\,n}(x;\bm{\lambda}),
  \label{phiDngenKA}\\
  \phi^{\text{KA}}_{\mathcal{D}\,n}(x;\bm{\lambda})
  &\eqdef\psi^{\text{KA}}_{\mathcal{D}}(x;\bm{\lambda})
  \check{P}^{\text{KA}}_{\mathcal{D},n}(x;\bm{\lambda}),
  \label{phiDnKA}\\
  \psi^{\text{KA}}_{\mathcal{D}}(x;\bm{\lambda})&\eqdef
  \frac{\phi_0(x;\bm{\lambda}+M\bm{\delta})}
  {\sqrt{\check{\Xi}^{\text{KA}}_{\mathcal{D}}(x;\bm{\lambda})
  \check{\Xi}^{\text{KA}}_{\mathcal{D}}(x+1;\bm{\lambda})}}.
  \label{psiDKA}
\end{align}
Next let us calculate the potential functions \eqref{BD}--\eqref{DD}.
By using \eqref{varphiM+1M} and \eqref{B(x+j),D(x)}, we obtain
\begin{align}
  B_{\mathcal{D}}(x;\bm{\lambda})&=\kappa^MB(x;\bm{\lambda}+M\bm{\delta})
  \frac{\check{\Xi}^{\text{KA}}_{\mathcal{D}}(x;\bm{\lambda})}
  {\check{\Xi}^{\text{KA}}_{\mathcal{D}}(x+1;\bm{\lambda})}
  \frac{\check{P}^{\text{KA}}_{\mathcal{D},\mu}(x+1;\bm{\lambda})}
  {\check{P}^{\text{KA}}_{\mathcal{D},\mu}(x;\bm{\lambda})},
  \label{BDKA}\\
  D_{\mathcal{D}}(x;\bm{\lambda})&=\kappa^MD(x;\bm{\lambda}+M\bm{\delta})
  \frac{\check{\Xi}^{\text{KA}}_{\mathcal{D}}(x+1;\bm{\lambda})}
  {\check{\Xi}^{\text{KA}}_{\mathcal{D}}(x;\bm{\lambda})}
  \frac{\check{P}^{\text{KA}}_{\mathcal{D},\mu}(x-1;\bm{\lambda})}
  {\check{P}^{\text{KA}}_{\mathcal{D},\mu}(x;\bm{\lambda})}.
  \label{DDKA}
\end{align}
The deformed Hamiltonian \eqref{HopD} is similarity transformed to
\begin{align}
  \widetilde{\mathcal{H}}^{\text{KA\,op}}_{\mathcal{D}}(\bm{\lambda})&\eqdef
  \psi^{\text{KA}}_{\mathcal{D}}(x;\bm{\lambda})^{-1}\circ
  \mathcal{H}^{\text{op}}_{\mathcal{D}}(\bm{\lambda})\circ
  \psi^{\text{KA}}_{\mathcal{D}}(x;\bm{\lambda})\n
  &=\kappa^MB(x;\bm{\lambda}+M\bm{\delta})
  \frac{\check{\Xi}^{\text{KA}}_{\mathcal{D}}(x;\bm{\lambda})}
  {\check{\Xi}^{\text{KA}}_{\mathcal{D}}(x+1;\bm{\lambda})}\Bigl(
  \frac{\check{P}^{\text{KA}}_{\mathcal{D},\mu}(x+1;\bm{\lambda})}
  {\check{P}^{\text{KA}}_{\mathcal{D},\mu}(x;\bm{\lambda})}
  -e^{\hat{\partial}}\Bigr)\n
  &\quad+\kappa^MD(x;\bm{\lambda}+M\bm{\delta})
  \frac{\check{\Xi}^{\text{KA}}_{\mathcal{D}}(x+1;\bm{\lambda})}
  {\check{\Xi}^{\text{KA}}_{\mathcal{D}}(x;\bm{\lambda})}\Bigl(
  \frac{\check{P}^{\text{KA}}_{\mathcal{D},\mu}(x-1;\bm{\lambda})}
  {\check{P}^{\text{KA}}_{\mathcal{D},\mu}(x;\bm{\lambda})}
  -e^{-\hat{\partial}}\Bigr)
  +\mathcal{E}_{\mu}(\bm{\lambda}).
  \label{HtDopdef}
\end{align}
{}From the result in \S\,\ref{sec:Dar}, the multi-indexed polynomials
$\check{P}^{\text{KA}}_{\mathcal{D},n}(x;\bm{\lambda})$ satisfy the difference
equation,
\begin{equation}
  \widetilde{\mathcal{H}}^{\text{KA\,op}}_{\mathcal{D}}(\bm{\lambda})
  \check{P}^{\text{KA}}_{\mathcal{D},n}(x;\bm{\lambda})
  =\mathcal{E}_n(\bm{\lambda})
  \check{P}^{\text{KA}}_{\mathcal{D},n}(x;\bm{\lambda}),
  \label{HtDopcPDnKA}
\end{equation}
for $x\in\mathbb{R}$.
We remark that this difference equation holds for any $\bm{\lambda}$.
In the expressions \eqref{psiDKA}--\eqref{DDKA}, they contain $\phi_0$, $B$
and $D$ with the parameter $\bm{\lambda}+M\bm{\delta}$, which means that the
parameter $N$ is $N-M$.
Therefore the matrix
$\widetilde{\mathcal{H}}^{\text{KA}}_{\mathcal{D}}(\bm{\lambda})$ is of order
$N-M+1$,
\begin{align}
  \widetilde{\mathcal{H}}^{\text{KA}}_{\mathcal{D}}(\bm{\lambda})&=
  \bigl(\widetilde{\mathcal{H}}^{\text{KA}}_{\mathcal{D}\,x,y}
  (\bm{\lambda})\bigr)_{x,y=0,1,\ldots,N-M},\n
  \widetilde{\mathcal{H}}^{\text{KA}}_{\mathcal{D}}(\bm{\lambda})&\eqdef
  \widetilde{\mathcal{H}}^{\text{KA\,op}}_{\mathcal{D}}(\bm{\lambda})
  \bigm|_{\hat{\partial}\to\partial}.
  \label{HtDKAdef}
\end{align}
Since $B(x;\bm{\lambda}+M\bm{\delta})$ and $D(x;\bm{\lambda}+M\bm{\delta})$ 
satisfy the boundary conditions $B(N-M;\bm{\lambda}+M\bm{\delta})=0$ at $x=N-M$
and $D(0;\bm{\lambda}+M\bm{\delta})=0$ at $x=0$, the difference equation
\eqref{HtDopcPDnKA} implies the following.
The matrix eigenvalue problem for $\widetilde{\mathcal{H}}_{\mathcal{D}}$ is
solved by the multi-indexed polynomials
$\check{P}^{\text{KA}}_{\mathcal{D},n}(x)$,
\begin{equation}
  \widetilde{\mathcal{H}}^{\text{KA}}_{\mathcal{D}}(\bm{\lambda})
  \check{P}^{\text{KA}}_{\mathcal{D},n}(x;\bm{\lambda})
  =\mathcal{E}_n(\bm{\lambda})
  \check{P}^{\text{KA}}_{\mathcal{D},n}(x;\bm{\lambda})
  \ \ (n\in\{0,1,\ldots,N\}\backslash\mathcal{D}).
\end{equation}

The orthogonality relations for $\check{P}^{\text{KA}}_{\mathcal{D},n}(x)$
can be read from \eqref{sumphigen}.
{}From \eqref{phiDngenKA}--\eqref{psiDKA}, 
$\phi_{\mathcal{D}\,n}^{\text{gen}}(x;\bm{\lambda})$ contains
$\phi_0(x;\bm{\lambda}+M\bm{\delta})$, which does not vanish
at $x=0,1,\ldots,N-M$ for $x\in\mathbb{Z}$.
So the sum in \eqref{sumphigen} is reduced to $\sum_{x=0}^{N-M}$.
We obtain the orthogonality relations,
\begin{align}
  &\quad\sum_{x=0}^{N-M}\frac{\phi_0(x;\bm{\lambda}+M\bm{\delta})^2}
  {\check{\Xi}^{\text{KA}}_{\mathcal{D}}(x;\bm{\lambda})
  \check{\Xi}^{\text{KA}}_{\mathcal{D}}(x+1;\bm{\lambda})}
  \check{P}^{\text{KA}}_{\mathcal{D},n}(x;\bm{\lambda})
  \check{P}^{\text{KA}}_{\mathcal{D},m}(x;\bm{\lambda})\n
  &=\frac{\delta_{nm}}{d^{\text{KA}}_{\mathcal{D},n}(\bm{\lambda})^2}
  \ \ (n,m\in\{0,1,\ldots,N\}\backslash\mathcal{D}),
  \label{orthoKA}
\end{align}
where the normalization constants
$d^{\text{KA}}_{\mathcal{D},n}(\bm{\lambda})^2$ are obtained from
\eqref{sumphigen} and \eqref{phiDngenKA}--\eqref{psiDKA} as follows,
\begin{equation}
  d^{\text{KA}}_{\mathcal{D},n}(\bm{\lambda})^2
  =d_n(\bm{\lambda})^2\kappa^{-\binom{M}{2}}
  \prod_{j=1}^M\frac{\mathcal{E}_n(\bm{\lambda})-\mathcal{E}_{d_j}(\bm{\lambda})}
  {B(0;\bm{\lambda}+(j-1)\bm{\delta})}.
\end{equation}
In order to call the relations \eqref{orthoKA} truly orthogonality relations,
the positivity of the weight factor is necessary.
The positivity of $\phi_0(x;\bm{\lambda}+M\bm{\delta})^2$ can be easily
achieved by choosing parameters $\bm{\lambda}$ so that
$B(x;\bm{\lambda}+M\bm{\delta})$ and $D(x;\bm{\lambda}+M\bm{\delta})$ are
positive (except for the boundaries). The positivity of
$\check{\Xi}^{\text{KA}}_{\mathcal{D}}(x;\bm{\lambda})
\check{\Xi}^{\text{KA}}_{\mathcal{D}}(x+1;\bm{\lambda})$
(for $x=0,1,\ldots,N-M$) can be achieved by imposing the following
Krein-Adler condition on $\mathcal{D}$ \cite{os22},
\begin{equation}
  \prod_{j=1}^M(m-d_j)\geq 0\ \ (\forall m\in\mathbb{Z}_{\geq0}).
  \label{KA}
\end{equation}
We remark that this condition \eqref{KA} is rewritten as
\begin{equation}
  e_j-e_{j-1}\equiv 1\ \ (\text{mod }2)\ \ (j=2,3,\ldots),
  \label{KA2}
\end{equation}
where $\mathbb{Z}_{\geq0}\backslash\mathcal{D}=\{e_1,e_2,\ldots\}$
with $e_1<e_2<\cdots$.
In this situation, the Hamiltonian of the deformed system
\begin{align}
  \mathcal{H}^{\text{KA}}_{\mathcal{D}}(\bm{\lambda})&=
  \bigl(\mathcal{H}^{\text{KA}}_{\mathcal{D}\,x,y}
  (\bm{\lambda})\bigr)_{x,y=0,1,\ldots,N-M},\n
  \mathcal{H}^{\text{KA}}_{\mathcal{D}}(\bm{\lambda})&\eqdef
  \psi^{\text{KA}}_{\mathcal{D}}(x;\bm{\lambda})\circ
  \widetilde{\mathcal{H}}^{\text{KA}}_{\mathcal{D}}(\bm{\lambda})\circ
  \psi^{\text{KA}}_{\mathcal{D}}(x;\bm{\lambda})^{-1}\n
  &=-\sqrt{B_{\mathcal{D}}(x;\bm{\lambda})D_{\mathcal{D}}(x+1;\bm{\lambda})}
  \,e^{\partial}
  -\sqrt{B_{\mathcal{D}}(x-1;\bm{\lambda})D_{\mathcal{D}}(x;\bm{\lambda})}
  \,e^{-\partial}\n
  &\quad+B_{\mathcal{D}}(x;\bm{\lambda})+D_{\mathcal{D}}(x;\bm{\lambda})
  +\mathcal{E}_{\mu}(\bm{\lambda})
\end{align}
is a real symmetric matrix, and the matrix eigenvalue problem for
$\mathcal{H}^{\text{KA}}_{\mathcal{D}}$ is solved by
$\phi^{\text{KA}}_{\mathcal{D}\,n}(x)$ \eqref{phiDnKA},
\begin{equation}
  \mathcal{H}^{\text{KA}}_{\mathcal{D}}(\bm{\lambda})
  \phi^{\text{KA}}_{\mathcal{D}\,n}(x;\bm{\lambda})
  =\mathcal{E}_n(\bm{\lambda})
  \phi^{\text{KA}}_{\mathcal{D}\,n}(x;\bm{\lambda})
  \ \ (n\in\{0,1,\ldots,N\}\backslash\mathcal{D}).
\end{equation}
However, we remark that the relations \eqref{orthoKA} hold for any
$\bm{\lambda}$ and $\mathcal{D}$, because the expressions are algebraic and
the sum in \eqref{orthoKA} is a finite sum.

We have shown the state-deleting property of the Darboux transformations by
using the embedding of the finite system into the infinite system.
On the other hand, in \cite{os22}, the state-deleting property was shown without
using such a trick.
It was shown that each step of the Darboux transformation reduces the size of
the matrix (Hamiltonian) by one.

The original systems in \S\,\ref{sec:orgsys} have shape invariance.
As its consequence, the forward and backward shift relations hold.
Here we consider the forward shift relation,
\begin{equation}
  \mathcal{F}^{\text{op}}(\bm{\lambda})=B(0;\bm{\lambda})\varphi(x)^{-1}
  (1-e^{\hat{\partial}}),\quad
  \mathcal{F}^{\text{op}}(\bm{\lambda})\check{P}_n(x;\bm{\lambda})
  =\mathcal{E}_n(\bm{\lambda})\check{P}_{n-1}(x;\bm{\lambda}+\bm{\delta}).
  \label{FopcP=}
\end{equation}
By using this and properties of determinant and the notation \eqref{D[i]def},
we can show the following for $d_j\geq1$,
\begin{equation}
  \check{P}^{\text{KA}}_{\mathcal{D},0}(x;\bm{\lambda})
  =\check{\Xi}^{\text{KA}}_{\mathcal{D}^{[-1]}}(x;\bm{\lambda}+\bm{\delta}).
  \label{cPKAD0}
\end{equation}
A slightly different but similar relation exists for the case-(1) multi-indexed
polynomials in \cite{os26}.
Let us consider the case $d_j\geq1$.
Since this condition means $\mu=0$, \eqref{HtDopdef} is expressed as
\begin{align}
  \widetilde{\mathcal{H}}^{\text{KA\,op}}_{\mathcal{D}}(\bm{\lambda})
  &=\kappa^MB(x;\bm{\lambda}+M\bm{\delta})
  \frac{\check{\Xi}^{\text{KA}}_{\mathcal{D}}(x;\bm{\lambda})}
  {\check{\Xi}^{\text{KA}}_{\mathcal{D}}(x+1;\bm{\lambda})}\Bigl(
  \frac{\check{\Xi}^{\text{KA}}_{\mathcal{D}^{[-1]}}
  (x+1;\bm{\lambda}+\bm{\delta})}
  {\check{\Xi}^{\text{KA}}_{\mathcal{D}^{[-1]}}(x;\bm{\lambda}+\bm{\delta})}
  -e^{\hat{\partial}}\Bigr)\n
  &\quad+\kappa^MD(x;\bm{\lambda}+M\bm{\delta})
  \frac{\check{\Xi}^{\text{KA}}_{\mathcal{D}}(x+1;\bm{\lambda})}
  {\check{\Xi}^{\text{KA}}_{\mathcal{D}}(x;\bm{\lambda})}\Bigl(
  \frac{\check{\Xi}^{\text{KA}}_{\mathcal{D}^{[-1]}}
  (x-1;\bm{\lambda}+\bm{\delta})}
  {\check{\Xi}^{\text{KA}}_{\mathcal{D}^{[-1]}}(x;\bm{\lambda}+\bm{\delta})}
  -e^{-\hat{\partial}}\Bigr).
  \label{HtDop0}
\end{align}
By replacing $\hat{\partial}$ with $\partial$, an expression of
$\widetilde{\mathcal{H}}^{\text{KA}}_{\mathcal{D}}(\bm{\lambda})$ is obtained.
This resembles the similarity transformed Hamiltonian for the case-(1)
multi-indexed polynomials in \cite{os26}.

\section{New Multi-Indexed Orthogonal Polynomials From State-Adding
Darboux Transformations}
\label{sec:newmiop}

In this section we consider the Darboux transformations with seed solutions
$\phi_n(x)$ ($n>N$) \cite{os40} and present new multi-indexed orthogonal
polynomials $\check{Q}_{\mathcal{D}',n}(x)$.

The multi-index set we will consider is
\begin{align}
  \mathcal{D}&=\{d_1,d_2,\ldots,d_M\}\ \ (d_j>N
  \text{ : mutually distinct}),\\
  \label{Ddefnew}
  &\quad d_j=N+1+m_j,\ \ m_j\in\mathbb{Z}_{\geq0},
\end{align}
and we assume $M\leq N$.
In this section we use the following notations,
\begin{align}
  \mathcal{D}'&=\mathcal{D}^{[-N-1]}=\{m_1,m_2,\ldots,m_M\},
  \label{D'def}\\
  \bm{\lambda}'&=\bm{\lambda}+(N+1)(\bm{\delta}+\bm{\bar{\delta}}),
  \label{lambda'def}
\end{align}
where $\bm{\bar{\delta}}$ is given in Appendix\,\ref{app:poly}.
We remark that $N$ is one element of $\bm{\lambda}$.

The twelve polynomials $\check{P}_n(x)$ in Appendix\,\ref{app:poly} with the
normalization \eqref{cPn0=1} are ill-defined for $n>N$, because there is a
factor $1/(-N)_k$ or $1/(q^{-N}\,;q)_k$ in the sum $\sum_{k=0}^n$ of the
(basic) hypergeometric series expansion.
To avoid this, let us consider the monic polynomial,
\begin{equation}
  \check{P}^{\text{monic}}_n(x;\bm{\lambda})
  =c_n(\bm{\lambda})^{-1}\check{P}_n(x;\bm{\lambda}),
\end{equation}
where $c_n(\bm{\lambda})$ is the coefficient of the highest degree term,
\begin{equation}
  \check{P}_n(x;\bm{\lambda})
  =c_n(\bm{\lambda})\eta(x;\bm{\lambda})^n+(\text{lower degree terms}).
  \label{cndef}
\end{equation}
The monic polynomial $\check{P}^{\text{monic}}_n(x;\bm{\lambda})$ is
well-defined for $n\in\mathbb{Z}_{\geq0}$.
Explicit forms of $c_n(\bm{\lambda})$ are given in Appendix\,\ref{app:poly} and
the universal expression of $c_n(\bm{\lambda})$ is ((A.14) in \cite{os22})
\begin{equation}
  c_n(\bm{\lambda})=(-1)^n\kappa^{-\binom{n}{2}}
  \prod_{j=1}^n\frac{\mathcal{E}_n(\bm{\lambda})
  -\mathcal{E}_{j-1}(\bm{\lambda})}
  {\eta(j;\bm{\lambda})B(0;\bm{\lambda}+(j-1)\bm{\delta})}.
  \label{cnuniv}
\end{equation}
For the monic polynomial, the orthogonality relations \eqref{orthocPn} become
\begin{equation}
  \sum_{x=0}^N\phi_0(x;\bm{\lambda})^2
  \check{P}^{\text{monic}}_n(x;\bm{\lambda})
  \check{P}^{\text{monic}}_m(x;\bm{\lambda})
  =\frac{\delta_{nm}}{d^{\text{monic}}_n(\bm{\lambda})^2},\quad
  d^{\text{monic}}_n(\bm{\lambda})
  =c_n(\bm{\lambda})d_n(\bm{\lambda}).
  \label{orthocPnmonic}
\end{equation}

The polynomials 
$\check{\Xi}^{\text{KA}}_{\mathcal{D}}(x;\bm{\lambda})$ \eqref{cXiDKAdef} and
$\check{P}^{\text{KA}}_{\mathcal{D},n}(x;\bm{\lambda})$ \eqref{cPDnKAdef} with
the normalization conditions \eqref{cPKADn(0)=1} are defined for $d_j,n\leq N$.
By replacing $\check{P}_{d_j}$ and $\check{P}_n$ with
$\check{P}^{\text{monic}}_{d_j}$ and $\check{P}^{\text{monic}}_n$, their
definitions are extended to $d_j,n\in\mathbb{Z}_{\geq0}$.
The monic version of these polynomials are given by
\begin{align}
  \check{\Xi}^{\text{KA\,monic}}_{\mathcal{D}}(x;\bm{\lambda})
  &=c^{\eta}_{\mathcal{D}}(\bm{\lambda})^{-1}
  \varphi_M(x;\bm{\lambda})^{-1}
  \text{W}_{\text{C}}[\check{P}^{\text{monic}}_{d_1},\ldots,
  \check{P}^{\text{monic}}_{d_M}](x;\bm{\lambda}),
  \label{cXiDKAmonic}\\
  \check{P}^{\text{KA\,monic}}_{\mathcal{D},n}(x;\bm{\lambda})
  &=c^{\eta}_{\mathcal{D},n}(\bm{\lambda})^{-1}
  \varphi_{M+1}(x;\bm{\lambda})^{-1}
  \text{W}_{\text{C}}[\check{P}^{\text{monic}}_{d_1},\ldots,
  \check{P}^{\text{monic}}_{d_M},
  \check{P}^{\text{monic}}_n](x;\bm{\lambda}),
  \label{cPDnKAmonic}
\end{align}
where the constants $c^{\eta}_{\mathcal{D}}(\bm{\lambda})$ and
$c^{\eta}_{\mathcal{D},n}(\bm{\lambda})$ are given by \eqref{cetaD} and
\eqref{cetaDn} respectively.
These monic polynomials and their non-monic versions are related as
\begin{align}
  c^{\eta}_{\mathcal{D}}(\bm{\lambda})
  \check{\Xi}^{\text{KA\,monic}}_{\mathcal{D}}(x;\bm{\lambda})
  &=\mathcal{C}^{\text{KA}}_{\mathcal{D}}(\bm{\lambda})
  \frac{\check{\Xi}^{\text{KA}}_{\mathcal{D}}(x;\bm{\lambda})}
  {\prod_{j=1}^Mc_{d_j}(\bm{\lambda})},
  \label{cXiDKAmonicnonmonic}\\
  c^{\eta}_{\mathcal{D},n}(\bm{\lambda})
  \check{P}^{\text{KA\,monic}}_{\mathcal{D},n}(x;\bm{\lambda})
  &=\mathcal{C}^{\text{KA}}_{\mathcal{D},n}(\bm{\lambda})
  \frac{\check{P}^{\text{KA}}_{\mathcal{D},n}(x;\bm{\lambda})}
  {\prod_{j=1}^Mc_{d_j}(\bm{\lambda})\cdot c_n(\bm{\lambda})}.
  \label{cPDnKAmonicnonmonic}
\end{align}
This $\check{P}^{\text{KA\,monic}}_{\mathcal{D},n}(x;\bm{\lambda})$
satisfies the difference equation \eqref{HtDopcPDnKA} by replacing
$\check{\Xi}^{\text{KA}}_{\mathcal{D}}$ and
$\check{P}^{\text{KA}}_{\mathcal{D},n}$ with
$\check{\Xi}^{\text{KA\,monic}}_{\mathcal{D}}$ and
$\check{P}^{\text{KA\,monic}}_{\mathcal{D},n}$ in \eqref{HtDopdef}.

For $n>N$, by setting $n=N+1+m$ ($m\in\mathbb{Z}_{\geq 0}$),
the monic polynomial $\check{P}_n(x;\bm{\lambda})$ has a factorization property
(Theorem\,2.1 in \cite{os40}, with $\check{Q}_m(x;N,\bm{\lambda})\propto
\check{P}^{\text{monic}}_m(x-N-1;-N-2,\bm{\lambda}')$
(see (2.54) in \cite{os40}).
This proportional constant is determined by \eqref{eta(x-N-1)} with $M=1$),
\begin{equation}
  \check{P}^{\text{monic}}_n(x;\bm{\lambda})
  =\Lambda(x;\bm{\lambda})\rho^{(N+1)m}
  \check{P}^{\text{monic}}_m(x-N-1;\bm{\lambda}'),
  \label{fac_cPn}
\end{equation}
where $\Lambda(x;\bm{\lambda})$ is defined by
\begin{equation}
  \Lambda(x;\bm{\lambda})\eqdef\prod_{k=0}^N\bigl(
  \eta(x;\bm{\lambda})-\eta(k;\bm{\lambda})\bigr),
  \label{Lambdadef}
\end{equation}
and its properties are given in Appendix\,\ref{app:Lam}.
{}From \eqref{cXiDKAmonic} and \eqref{fac_cPn}, we have
\begin{align}
  &\quad\check{\Xi}^{\text{KA\,monic}}_{\mathcal{D}}(x;\bm{\lambda})\n
  &=c^{\eta}_{\mathcal{D}}(\bm{\lambda})^{-1}
  \varphi_M(x;\bm{\lambda})^{-1}
  \prod_{j=1}^M\Lambda(x+j-1;\bm{\lambda})\cdot
  \prod_{k=1}^M\rho^{(N+1)m_k}\n
  &\quad\times
  \text{W}_{\text{C}}[\check{P}^{\text{monic}}_{m_1},\ldots,
  \check{P}^{\text{monic}}_{m_M}](x-N-1;\bm{\lambda}'),\n
  &=\frac{c^{\eta}_{\mathcal{D}'}(\bm{\lambda}')}
  {c^{\eta}_{\mathcal{D}}(\bm{\lambda})}
  \frac{\varphi_M(0;\bm{\lambda}')}{\varphi_M(N+1;\bm{\lambda})}
  \prod_{j=1}^M\Lambda(x+j-1;\bm{\lambda})\cdot
  \prod_{k=1}^M\rho^{(N+1)m_k}\cdot
  \check{\Xi}^{\text{KA\,monic}}_{\mathcal{D}'}(x-N-1;\bm{\lambda}'),\!
  \label{cXiDKAmonic2}
\end{align}
where \eqref{WCformula1} and \eqref{varphiM/varphiM} are used.
So $\check{\Xi}^{\text{KA\,monic}}_{\mathcal{D}}(x;\bm{\lambda})$ is
divisible by $\prod_{j=1}^M\Lambda(x+j-1;\bm{\lambda})$, which is a polynomial
in $\eta(x;\bm{\lambda}+(M-1)\bm{\delta})$, see \eqref{Lam_poly}
(with $M\to M-1$).
The relation \eqref{cPKAD0} implies that
$\check{P}^{\text{KA\,monic}}_{\mathcal{D},0}(x;\bm{\lambda})$ is
divisible by $\prod_{j=1}^M\Lambda(x+j-1;\bm{\lambda}+\bm{\delta})$,
which is a polynomial in $\eta(x;\bm{\lambda}+M\bm{\delta})$ \eqref{Lam_poly2}.
It is expected that
$\check{P}^{\text{KA\,monic}}_{\mathcal{D},n}(x;\bm{\lambda})$ is also
divisible by $\prod_{j=1}^M\Lambda(x+j-1;\bm{\lambda}+\bm{\delta})$, and
this is indeed the case.

Based on $\check{P}^{\text{KA\,monic}}_{\mathcal{D},n}(x;\bm{\lambda})$,
we define new multi-indexed polynomials
$\check{Q}^{\text{monic}}_{\mathcal{D}',n}(x;\bm{\lambda})$ for
$n\in\mathbb{Z}_{\geq0}$.
First, let us define $\check{Q}^{\text{monic}}_{\mathcal{D}',n}(x;\bm{\lambda})$
for $n\in\mathbb{Z}_{\geq0}\backslash\mathcal{D}$ as follows
(see \eqref{Lam_poly2}),
\begin{align}
  \check{Q}^{\text{monic}}_{\mathcal{D}',n}(x;\bm{\lambda})
  &\eqdef\rho^{\binom{M}{2}N}
  \prod_{j=1}^M\Lambda(x+j-1;\bm{\lambda}+\bm{\delta})^{-1}\cdot
  \check{P}^{\text{KA\,monic}}_{\mathcal{D},n}(x;\bm{\lambda})
  \label{cQmonicdef}\\
  &=c^{\eta}_{\mathcal{D},n}(\bm{\lambda})^{-1}\rho^{\binom{M}{2}N}
  \prod_{j=1}^M\Lambda(x+j-1;\bm{\lambda}+\bm{\delta})^{-1}\n
  &\quad\times\varphi_{M+1}(x;\bm{\lambda})^{-1}
  \text{W}_{\text{C}}[\check{P}^{\text{monic}}_{d_1},\ldots,
  \check{P}^{\text{monic}}_{d_M},\check{P}^{\text{monic}}_n](x;\bm{\lambda}).
  \label{cQmonic}
\end{align}
For $n>N$ and $n\not\in\mathcal{D}$ (we set $n=N+1+m$), by using
\eqref{fac_cPn} and \eqref{WCformula1}, we have
\begin{align}
  &\quad\check{Q}^{\text{monic}}_{\mathcal{D}',n}(x;\bm{\lambda})\n
  &=c^{\eta}_{\mathcal{D},n}(\bm{\lambda})^{-1}\rho^{\binom{M}{2}N}
  \prod_{k=1}^M\rho^{(N+1)m_k}\cdot\rho^{(N+1)m}
  \frac{\prod_{j=1}^{M+1}\Lambda(x+j-1;\bm{\lambda})}
  {\prod_{j=1}^M\Lambda(x+j-1;\bm{\lambda}+\bm{\delta})}
  \frac{\varphi_{M+1}(x-N-1;\bm{\lambda}')}{\varphi_{M+1}(x;\bm{\lambda})}\n
  &\quad\times
  \varphi_{M+1}(x-N-1;\bm{\lambda}')^{-1}
  \text{W}_{\text{C}}[\check{P}^{\text{monic}}_{m_1},\ldots,
  \check{P}^{\text{monic}}_{m_M},\check{P}^{\text{monic}}_m]
  (x-N-1;\bm{\lambda}'),
  \label{cQmonic_n=N+1+m}
\end{align}
which vanishes for $x\in\{-M,-M+1,\ldots,N\}$ by \eqref{prodLam/Lam} and
\eqref{varphiM/varphiM}.
Next, let us define $\check{Q}^{\text{monic}}_{\mathcal{D}',n}(x;\bm{\lambda})$
for $n\in\mathcal{D}$, based on the expression \eqref{cQmonic}.
For $n=d_i\in\mathcal{D}$, the Casoratian $\text{W}_{\text{C}}[\cdots]$ in
\eqref{cQmonic} vanishes, but
$c^{\eta}_{\mathcal{D},n}(\bm{\lambda})$ \eqref{cetaDn} also vanishes.
So, by taking certain appropriate ``$n\to d_i$ limit'', we may obtain a finite
quantity.
Following the prescription explained in Appendix\,\ref{app:n->di}, we define
$\check{Q}^{\text{monic}}_{\mathcal{D}',n}(x;\bm{\lambda})$ with
$n=d_i\in\mathcal{D}$ as follows,
\begin{align}
  \check{Q}^{\text{monic}}_{\mathcal{D}',n}(x;\bm{\lambda})
  &\eqdef\rho^{-n}
  c^{\prime\,\eta}_{\mathcal{D},n}(\bm{\lambda})^{-1}\rho^{\binom{M}{2}N}
  \prod_{j=1}^M\Lambda(x+j-1;\bm{\lambda}+\bm{\delta})^{-1}
  \label{cQmonic2}\\
  &\quad\times\varphi_{M+1}(x;\bm{\lambda})^{-1}
  \left|
  \begin{array}{cccc}
  \check{P}^{\text{monic}}_{d_1}(x_1;\bm{\lambda})&\cdots
  &\check{P}^{\text{monic}}_{d_M}(x_1;\bm{\lambda})
  &R_1(x)\\
  \check{P}^{\text{monic}}_{d_1}(x_2;\bm{\lambda})&\cdots
  &\check{P}^{\text{monic}}_{d_M}(x_2;\bm{\lambda})
  &R_2(x)\\
  \vdots&\cdots&\vdots&\vdots\\
  \check{P}^{\text{monic}}_{d_1}(x_{M+1};\bm{\lambda})&\cdots
  &\check{P}^{\text{monic}}_{d_M}(x_{M+1};\bm{\lambda})
  &R_{M+1}(x)
  \end{array}\right|,
  \nonumber
\end{align}
where $x_j=x+j-1$ and $R_j(x)$ are given by
\begin{align}
  &\quad R_j(x)=R_{M,j,n}(x;\bm{\lambda})
  \label{Rjdef}\\
  &\eqdef\rho^{(N+1)m_i}Y_{M,j}(x;\bm{\lambda})
  \check{P}^{\text{monic}}_{m_i}(x_j-N-1;\bm{\lambda}')
  +\check{\tilde{P}}^{(1)}_n(x_j;\bm{\lambda})
  +\rho^{(N+1)m_i}\Lambda(x_j;\bm{\lambda})
  \check{\tilde{P}}^{(2)}_{m_i}(x_j;\bm{\lambda}),
  \nonumber
\end{align}
and $Y_{M,j}(x;\bm{\lambda})$, $\check{\tilde{P}}^{(1)}_n(x;\bm{\lambda})$,
$\check{\tilde{P}}^{(2)}_m(x;\bm{\lambda})$ and
$c^{\prime\,\eta}_{\mathcal{D},n}(\bm{\lambda})$ are defined by \eqref{YMj},
\eqref{ctP1def}, \eqref{ctP2def} and \eqref{cetaDn'}, respectively.
We remark that the contribution to
$\check{Q}^{\text{monic}}_{\mathcal{D}',n}(x)$ \eqref{cQmonic2} coming from
the third term of $R_j(x)$ \eqref{Rjdef} (the term containing
$\check{\tilde{P}}^{(2)}_{m_i}(x)$ : contribution (c) in
Appendix\,\ref{app:(c)cPm}) has the following form,
\begin{equation*}
  \Lambda(x+M;\bm{\lambda}-M\bm{\bar{\delta}})\times
  \bigl(\text{a polynomial in $\eta(x;\bm{\lambda}+M\bm{\delta})$}\bigr),
\end{equation*}
which vanishes for $x\in\{-M,-M+1,\ldots,N\}$.
So, when dealing with $\check{Q}^{\text{monic}}_{\mathcal{D}',d_i}(x)$ 
for $x\in\{-M,-M+1,\ldots,N\}$, we can ignore the third term of $R_j(x)$.
These $\check{Q}^{\text{monic}}_{\mathcal{D}',n}(x;\bm{\lambda})$
($n\in\mathbb{Z}_{\geq0}$) are
\begin{align}
  &\check{Q}^{\text{monic}}_{\mathcal{D}',n}(x;\bm{\lambda}):
  \text{a monic polynomial of degree $\ell_{\mathcal{D}'}+n$ in
  $\eta(x;\bm{\lambda}+M\bm{\delta})$},\n
  &\check{Q}^{\text{monic}}_{\mathcal{D}',n}(x;\bm{\lambda})
  \eqdef Q^{\text{monic}}_{\mathcal{D}',n}\bigl(
  \eta(x;\bm{\lambda}+M\bm{\delta});\bm{\lambda}\bigr),
  \label{cQD'nmonic}
\end{align}
where $\ell_{\mathcal{D}'}$ is given by \eqref{ellDdef}.
We define the non-monic version $\check{Q}_{\mathcal{D}',n}(x;\bm{\lambda})$
as follows,
\begin{equation}
  \check{Q}_{\mathcal{D}',n}(x;\bm{\lambda})
  \eqdef\frac{\check{Q}^{\text{monic}}_{\mathcal{D}',n}(x;\bm{\lambda})}
  {\check{Q}^{\text{monic}}_{\mathcal{D}',n}(-M;\bm{\lambda})}
  \ \ \bigl(\Rightarrow\check{Q}_{\mathcal{D}',n}(-M;\bm{\lambda})=1\bigr).
\end{equation}

Let us calculate $\phi_{\mathcal{D}\,n}^{\text{gen}}(x)$ \eqref{phiDn} for
$n\in\mathbb{Z}_{\geq0}\backslash\mathcal{D}$.
By using \eqref{phiDngenKA},
\eqref{cXiDKAmonicnonmonic}--\eqref{cPDnKAmonicnonmonic},
\eqref{cXiDKAmonic2}, \eqref{cQmonicdef}, \eqref{phi0Lam} and \eqref{CKADn=},
we obtain
\begin{align}
  \phi_{\mathcal{D}\,n}^{\text{gen}}(x;\bm{\lambda})
  &=(-1)^Mc_n(\bm{\lambda})\kappa^{\frac12\binom{M}{2}}
  \rho^{-(2N+1)\binom{M}{2}}
  \prod_{j=1}^M\sqrt{B(0;\bm{\lambda}+(j-1)\bm{\delta})}\n
  &\quad\times
  \frac{c^{\eta}_{\mathcal{D},n}(\bm{\lambda})}
  {c^{\eta}_{\mathcal{D}'}(\bm{\lambda}')}
  \frac{\Lambda_M(0;\bm{\lambda})^{\frac12}}
  {\phi_0(M;\bm{\lambda}-M\bm{\bar{\delta}})}
  \frac{\varphi_M(N+1;\bm{\lambda})}{\varphi_M(0;\bm{\lambda}')}
  \times\phi^{Q\,\text{monic}}_{\mathcal{D}'\,n}(x;\bm{\lambda}),
  \label{phiDngenQ}\\
  \phi^{Q\,\text{monic}}_{\mathcal{D}'\,n}(x;\bm{\lambda})&\eqdef
  \psi^{Q\,\text{monic}}_{\mathcal{D}'}(x;\bm{\lambda})
  \check{Q}^{\text{monic}}_{\mathcal{D}',n}(x;\bm{\lambda}),
  \label{phiDnQ}\\
  \psi^{Q\,\text{monic}}_{\mathcal{D}'}(x;\bm{\lambda})&\eqdef
  \frac{\phi_0(x+M;\bm{\lambda}-M\bm{\bar{\delta}})}
  {\sqrt{\check{\Xi}^{Q\,\text{monic}}_{\mathcal{D}'}(x;\bm{\lambda})
  \check{\Xi}^{Q\,\text{monic}}_{\mathcal{D}'}(x+1;\bm{\lambda})}}.
  \label{psiDQ}
\end{align}
Here $\check{\Xi}^{Q\,\text{monic}}_{\mathcal{D}'}(x;\bm{\lambda})$ is given by
\begin{align}
  &\check{\Xi}^{Q\,\text{monic}}_{\mathcal{D}'}(x;\bm{\lambda})
  \eqdef\rho^{(N+1)\ell_{\mathcal{D}'}}
  \check{\Xi}^{\text{KA\,monic}}_{\mathcal{D}'}(x-N-1;\bm{\lambda}')\n
  &\qquad:\text{a monic polynomial of degree $\ell_{\mathcal{D}'}$ in
  $\eta(x;\bm{\lambda}+(M-1)\bm{\delta})$},
\end{align}
which is shown by \eqref{cPDnKAdeg} and
\begin{equation}
  \eta(x-N-1;\bm{\lambda}'+(M-1)\bm{\delta})
  =\rho^{-N-1}\bigl(\eta(x;\bm{\lambda}+(M-1)\bm{\delta})
  -\eta(N+1;\bm{\lambda}+(M-1)\bm{\delta})\bigr).
  \label{eta(x-N-1)}
\end{equation}
Since the parameter $N$ for $\bm{\lambda}+\bm{\delta}$ is $N-1$, we have
\begin{equation}
  \check{\Xi}^{Q\,\text{monic}}_{\mathcal{D}'}(x;\bm{\lambda}+\bm{\delta})
  =\rho^{N\ell_{\mathcal{D}'}}
  \check{\Xi}^{\text{KA\,monic}}_{\mathcal{D}'}
  \bigl(x-N;\bm{\lambda}+\bm{\delta}+N(\bm{\delta}+\bm{\bar{\delta}})\bigr)
  =\rho^{N\ell_{\mathcal{D}'}}
  \check{\Xi}^{\text{KA\,monic}}_{\mathcal{D}'}
  (x-N;\bm{\lambda}'-\bm{\bar{\delta}}).
\end{equation}
We define the non-monic version $\check{\Xi}^Q_{\mathcal{D}'}(x;\bm{\lambda})$
as follows,
\begin{equation}
  \check{\Xi}^Q_{\mathcal{D}'}(x;\bm{\lambda})
  \eqdef\frac{\check{\Xi}^{Q\,\text{monic}}_{\mathcal{D}'}(x;\bm{\lambda})}
  {\check{\Xi}^{Q\,\text{monic}}_{\mathcal{D}'}(-M;\bm{\lambda})}
  \ \ \bigl(\Rightarrow\check{\Xi}^Q_{\mathcal{D}'}(-M;\bm{\lambda})=1\bigr).
\end{equation}
Next let us calculate the potential functions \eqref{BD}--\eqref{DD}.
Recall \eqref{BDKA}--\eqref{DDKA}, $\mu=0$ now and \eqref{cPKAD0}.
{}From \eqref{cXiDKAmonic2} with the replacements
$\mathcal{D}\to\mathcal{D}^{[-1]}$ and
$\bm{\lambda}\to\bm{\lambda}+\bm{\delta}$, we have
\begin{align}
  &\quad\check{\Xi}^{\text{KA}\,\text{monic}}_{\mathcal{D}^{[-1]}}
  (x;\bm{\lambda}+\bm{\delta})\n
  &=(\text{const})\times\prod_{j=1}^M\Lambda(x+j-1;\bm{\lambda}+\bm{\delta})
  \cdot\check{\Xi}^{\text{KA}\,\text{monic}}_{(\mathcal{D}^{[-1]})^{[-N]}}
  \bigl(x-N;\bm{\lambda}+\bm{\delta}+N(\bm{\delta}+\bm{\bar{\delta}})\bigr)\n
  &=(\text{const})\times\prod_{j=1}^M\Lambda(x+j-1;\bm{\lambda}+\bm{\delta})
  \cdot\check{\Xi}^{\text{KA}\,\text{monic}}_{\mathcal{D}^{[-N-1]}}
  (x-N;\bm{\lambda}'-\bm{\bar{\delta}})\n
  &=(\text{const})\times\prod_{j=1}^M\Lambda(x+j-1;\bm{\lambda}+\bm{\delta})
  \cdot\rho^{-N\ell_{\mathcal{D}'}}\check{\Xi}^{Q\,\text{monic}}_{\mathcal{D}'}
  (x;\bm{\lambda}+\bm{\delta}).
  \label{cXiD[-1]KAmonic}
\end{align}
By using \eqref{cXiDKAmonicnonmonic}--\eqref{cPDnKAmonicnonmonic},
\eqref{cPKAD0}, \eqref{cXiDKAmonic2}, \eqref{cXiD[-1]KAmonic} and
\eqref{B(x+M),D(x+M)}, \eqref{BDKA}--\eqref{DDKA} become
\begin{align}
  B_{\mathcal{D}}(x;\bm{\lambda})&=B(x+M;\bm{\lambda}-M\bm{\bar{\delta}})
  \frac{\check{\Xi}^{Q\,\text{monic}}_{\mathcal{D}'}(x;\bm{\lambda})}
  {\check{\Xi}^{Q\,\text{monic}}_{\mathcal{D}'}(x+1;\bm{\lambda})}
  \frac{\check{\Xi}^{Q\,\text{monic}}_{\mathcal{D}'}
  (x+1;\bm{\lambda}+\bm{\delta})}
  {\check{\Xi}^{Q\,\text{monic}}_{\mathcal{D}'}
  (x;\bm{\lambda}+\bm{\delta})},
  \label{BDnew}\\
  D_{\mathcal{D}}(x;\bm{\lambda})&=D(x+M;\bm{\lambda}-M\bm{\bar{\delta}})
  \frac{\check{\Xi}^{Q\,\text{monic}}_{\mathcal{D}'}(x+1;\bm{\lambda})}
  {\check{\Xi}^{Q\,\text{monic}}_{\mathcal{D}'}(x;\bm{\lambda})}
  \frac{\check{\Xi}^{Q\,\text{monic}}_{\mathcal{D}'}
  (x-1;\bm{\lambda}+\bm{\delta})}
  {\check{\Xi}^{Q\,\text{monic}}_{\mathcal{D}'}
  (x;\bm{\lambda}+\bm{\delta})}.
  \label{DDnew}
\end{align}
These potential functions resemble those of the case-(1) multi-indexed
polynomials in \cite{os26}.
The deformed Hamiltonian \eqref{HopD} is similarity transformed to
\begin{align}
  &\quad\widetilde{\mathcal{H}}^{Q\,\text{op}}_{\mathcal{D}'}(\bm{\lambda})
  \eqdef\psi^{Q\,\text{monic}}_{\mathcal{D}'}(x;\bm{\lambda})^{-1}\circ
  \mathcal{H}^{\text{op}}_{\mathcal{D}}(\bm{\lambda})\circ
  \psi^{Q\,\text{monic}}_{\mathcal{D}'}(x;\bm{\lambda})\n
  &=B(x+M;\bm{\lambda}-M\bm{\bar{\delta}})
  \frac{\check{\Xi}^{Q\,\text{monic}}_{\mathcal{D}'}(x;\bm{\lambda})}
  {\check{\Xi}^{Q\,\text{monic}}_{\mathcal{D}'}(x+1;\bm{\lambda})}\Bigl(
  \frac{\check{\Xi}^{Q\,\text{monic}}_{\mathcal{D}'}
  (x+1;\bm{\lambda}+\bm{\delta})}
  {\check{\Xi}^{Q\,\text{monic}}_{\mathcal{D}'}
  (x;\bm{\lambda}+\bm{\delta})}
  -e^{\hat{\partial}}\Bigr)\n
  &\quad+D(x+M;\bm{\lambda}-M\bm{\bar{\delta}})
  \frac{\check{\Xi}^{Q,\,\text{monic}}_{\mathcal{D}'}(x+1;\bm{\lambda})}
  {\check{\Xi}^{Q\,\text{monic}}_{\mathcal{D}'}(x;\bm{\lambda})}\Bigl(
  \frac{\check{\Xi}^{Q\,\text{monic}}_{\mathcal{D}'}
  (x-1;\bm{\lambda}+\bm{\delta})}
  {\check{\Xi}^{Q\,\text{monic}}_{\mathcal{D}'}
  (x;\bm{\lambda}+\bm{\delta})}
  -e^{-\hat{\partial}}\Bigr).
  \label{HtDQop}
\end{align}
{}From the result in \S\,\ref{sec:Dar}, the multi-indexed polynomials
$\check{Q}^{\text{monic}}_{\mathcal{D}',n}(x;\bm{\lambda})$ satisfy the
difference equation,
\begin{equation}
  \widetilde{\mathcal{H}}^{Q\,\text{op}}_{\mathcal{D}'}(\bm{\lambda})
  \check{Q}^{\text{monic}}_{\mathcal{D}',n}(x;\bm{\lambda})
  =\mathcal{E}_n(\bm{\lambda})
  \check{Q}^{\text{monic}}_{\mathcal{D}',n}(x;\bm{\lambda}),
  \label{HtDQopcQDn}
\end{equation}
for $x\in\mathbb{R}$ and $n\in\mathbb{Z}_{\geq0}$.
The results \eqref{HtDQopcQDn} for $n\in\mathcal{D}$ are obtained from those
for $n\in\mathbb{Z}_{\geq0}\backslash\mathcal{D}$ by taking the
``$n\to d_i$ limit.''
Exactly speaking, the results \eqref{HtDQopcQDn} for $n\in\mathcal{D}$ may not
be proven, because there is ambiguity in the ``$n\to d_i$ limit'' and the
prescription given in Appendix\,\ref{app:n->di} is one way of taking that limit.
However, we can verify \eqref{HtDQopcQDn} for $n\in\mathbb{Z}_{\geq0}$ by
direct calculation (using Mathematica) for small $N$, $M$, $d_j$ and $n$.
We remark that this difference equation holds for any $\bm{\lambda}$.
In the expressions \eqref{psiDQ} and \eqref{BDnew}--\eqref{DDnew}, they contain
$\phi_0$, $B$ and $D$ with the coordinate $x+M$ and the parameter
$\bm{\lambda}-M\bm{\bar{\delta}}$, which means that the parameter $N$ is $N+M$.
Therefore the matrix $\widetilde{\mathcal{H}}^Q_{\mathcal{D}'}(\bm{\lambda})$ is
of order $N+M+1$,
\begin{align}
  \widetilde{\mathcal{H}}^Q_{\mathcal{D}'}(\bm{\lambda})&=
  \bigl(\widetilde{\mathcal{H}}^Q_{\mathcal{D}'\,x,y}
  (\bm{\lambda})\bigr)_{x,y=-M,-M+1,\ldots,N},\n
  \widetilde{\mathcal{H}}^Q_{\mathcal{D}'}(\bm{\lambda})&\eqdef
  \widetilde{\mathcal{H}}^{Q\,\text{op}}_{\mathcal{D}'}(\bm{\lambda})
  \bigm|_{\hat{\partial}\to\partial}.
  \label{HtDQdef}
\end{align}
Since $B(x+M;\bm{\lambda}-M\bm{\bar{\delta}})$ and
$D(x+M;\bm{\lambda}-M\bm{\bar{\delta}})$ satisfy the boundary conditions
$B(N+M;\bm{\lambda}-M\bm{\bar{\delta}})=0$ at $x=N$ and
$D(0;\bm{\lambda}-M\bm{\bar{\delta}})=0$ at $x=-M$, the difference equation
\eqref{HtDQopcQDn} implies the following.
The matrix eigenvalue problem for $\widetilde{\mathcal{H}}^Q_{\mathcal{D}'}$ is
solved by the multi-indexed polynomials
$\check{Q}^{\text{monic}}_{\mathcal{D}',n}(x)$,
\begin{equation}
  \widetilde{\mathcal{H}}^Q_{\mathcal{D}'}(\bm{\lambda})
  \check{Q}^{\text{monic}}_{\mathcal{D}',n}(x;\bm{\lambda})
  =\mathcal{E}_n(\bm{\lambda})
  \check{Q}^{\text{monic}}_{\mathcal{D}',n}(x;\bm{\lambda})
  \ \ (n\in\{0,1,\ldots,N\}\cup\mathcal{D}).
\end{equation}
For $n>N$ and $n\not\in\mathcal{D}$, this equation also holds as $0=0$,
see \eqref{cQmonic_n=N+1+m} and its comment.

The orthogonality relations for $\check{Q}^{\text{monic}}_{\mathcal{D}',n}(x)$
can be read from \eqref{sumphigen}.
{}From \eqref{phiDngenQ}--\eqref{psiDQ},
$\phi_{\mathcal{D}\,n}^{\text{gen}}(x;\bm{\lambda})$ contains
$\phi_0(x+M;\bm{\lambda}-M\bm{\bar{\delta}})$, which does not vanish
at $x=-M,-M+1,\ldots,N$ for $x\in\mathbb{Z}$.
So the sum in \eqref{sumphigen} is reduced to $\sum_{x=-M}^N$.
We obtain the orthogonality relations,
\begin{align}
  &\quad\sum_{x=-M}^N\frac{\phi_0(x+M;\bm{\lambda}-M\bm{\bar{\delta}})^2}
  {\check{\Xi}^{Q\,\text{monic}}_{\mathcal{D}'}(x;\bm{\lambda})
  \check{\Xi}^{Q\,\text{monic}}_{\mathcal{D}'}(x+1;\bm{\lambda})}
  \check{Q}^{\text{monic}}_{\mathcal{D}',n}(x;\bm{\lambda})
  \check{Q}^{\text{monic}}_{\mathcal{D}',m}(x;\bm{\lambda})\n
  &=\frac{\delta_{nm}}{d^{Q\,\text{monic}}_{\mathcal{D}',n}(\bm{\lambda})^2}
  \ \ (n,m\in\{0,1,\ldots,N\}\cup\mathcal{D}).
  \label{orthoQ}
\end{align}
Here the normalization constants
$d^{Q\,\text{monic}}_{\mathcal{D}',n}(\bm{\lambda})^2$ for
$n\in\{0,1,\ldots,N\}$ are obtained from \eqref{sumphigen} and
\eqref{phiDngenQ}--\eqref{psiDQ} as follows,
\begin{align}
  d^{Q\,\text{monic}}_{\mathcal{D}',n}(\bm{\lambda})^2
  &=d^{\text{monic}}_n(\bm{\lambda})^2\,
  \kappa^{\binom{M}{2}}\rho^{-2(2N+1)\binom{M}{2}}
  \prod_{j=1}^M\frac{B(0;\bm{\lambda}+(j-1)\bm{\delta})}
  {\mathcal{E}_n(\bm{\lambda})-\mathcal{E}_{d_j}(\bm{\lambda})}\n
  &\quad\times
  \frac{c^{\eta}_{\mathcal{D},n}(\bm{\lambda})^2}
  {c^{\eta}_{\mathcal{D}'}(\bm{\lambda}')^2}
  \frac{\Lambda_M(0;\bm{\lambda})}
  {\phi_0(M;\bm{\lambda}-M\bm{\bar{\delta}})^2}
  \frac{\varphi_M(N+1;\bm{\lambda})^2}{\varphi_M(0;\bm{\lambda}')^2}.
  \label{dQmonicD'n}
\end{align}
For $n=d_i\in\mathcal{D}$, we take the ``$n\to d_i$ limit'' of
\eqref{dQmonicD'n}.
By using \eqref{cetaDn=-A}, \eqref{En-Edi=-A} and \eqref{dnmonic=-A},
we obtain $d^{Q\,\text{monic}}_{\mathcal{D}',n}(\bm{\lambda})^2$ with
$n=d_i\in\mathcal{D}$,
\begin{align}
  d^{Q\,\text{monic}}_{\mathcal{D}',n}(\bm{\lambda})^2
  &=d^{\,\prime\,\text{monic}}_n(\bm{\lambda})^2\,
  \kappa^{\binom{M}{2}}\rho^{-2(2N+1)\binom{M}{2}}
  \frac{\prod_{j=1}^MB(0;\bm{\lambda}+(j-1)\bm{\delta})}
  {\prod_{\substack{j=1\\ j\neq i}}^M
  (\mathcal{E}_n(\bm{\lambda})-\mathcal{E}_{d_j}(\bm{\lambda}))}\n
  &\quad\times
  \frac{c^{\prime\,\eta}_{\mathcal{D},n}(\bm{\lambda})^2}
  {c^{\eta}_{\mathcal{D}'}(\bm{\lambda}')^2}
  \frac{\Lambda_M(0;\bm{\lambda})}
  {\phi_0(M;\bm{\lambda}-M\bm{\bar{\delta}})^2}
  \frac{\varphi_M(N+1;\bm{\lambda})^2}{\varphi_M(0;\bm{\lambda}')^2}\n
  &\quad\times\rho^{2n}\kappa^{-n}\times\left\{
  \begin{array}{ll}
  1&:\text{(\romannumeral1)}',\,\text{(\romannumeral3)}',\,
  \text{(\romannumeral4)}'\\
  (2n+\tilde{d}\,)^{-1}&:\text{(\romannumeral2)}'\\[2pt]
  (1-\tilde{d}q^{2n})^{-1}&:\text{(\romannumeral5)}'
  \end{array}\right..
  \label{dQminicD'di}
\end{align}
In order to call the relations \eqref{orthoQ} truly orthogonality relations,
the positivity of the weight factor is necessary.
We will discuss this problem in the next subsection.
If the positivity is satisfied, the Hamiltonian of the deformed system
\begin{align}
  \mathcal{H}^Q_{\mathcal{D}'}(\bm{\lambda})&=
  \bigl(\mathcal{H}^Q_{\mathcal{D}'\,x,y}
  (\bm{\lambda})\bigr)_{x,y=-M,-M+1,\ldots,N},\n
  \mathcal{H}^Q_{\mathcal{D}'}(\bm{\lambda})&\eqdef
  \psi^{Q\,\text{monic}}_{\mathcal{D}'}(x;\bm{\lambda})\circ
  \widetilde{\mathcal{H}}^Q_{\mathcal{D}'}(\bm{\lambda})\circ
  \psi^{Q\,\text{monic}}_{\mathcal{D}'}(x;\bm{\lambda})^{-1}\n
  &=-\sqrt{B_{\mathcal{D}}(x;\bm{\lambda})D_{\mathcal{D}}(x+1;\bm{\lambda})}
  \,e^{\partial}
  -\sqrt{B_{\mathcal{D}}(x-1;\bm{\lambda})D_{\mathcal{D}}(x;\bm{\lambda})}
  \,e^{-\partial}\n
  &\quad+B_{\mathcal{D}}(x;\bm{\lambda})+D_{\mathcal{D}}(x;\bm{\lambda})
\end{align}
is a real symmetric matrix, and the matrix eigenvalue problem for
$\mathcal{H}^Q_{\mathcal{D}'}$ is solved by
$\phi^{Q\,\text{monic}}_{\mathcal{D}'\,n}(x)$ \eqref{phiDnQ},
\begin{equation}
  \mathcal{H}^Q_{\mathcal{D}'}(\bm{\lambda})
  \phi^{Q\,\text{monic}}_{\mathcal{D}'\,n}(x;\bm{\lambda})
  =\mathcal{E}_n(\bm{\lambda})
  \phi^{Q\,\text{monic}}_{\mathcal{D}'\,n}(x;\bm{\lambda})
  \ \ (n\in\{0,1,\ldots,N\}\cup\mathcal{D}).
\end{equation}
However, we remark that the relations \eqref{orthoQ} hold for any $\bm{\lambda}$
and $\mathcal{D}$, because the expressions are algebraic and the sum in
\eqref{orthoQ} is a finite sum.
We can verify \eqref{orthoQ} by direct calculation (using Mathematica) for
small $N$, $M$, $d_j$ and $n$.

We comment on the special case $\mathcal{D}=\{N+1,N+2,\ldots,N+M\}$ studied in
\cite{os40}. In this case we have $\mathcal{D}'=\{0,1,\ldots,M-1\}$ and the
denominator polynomial $\check{\Xi}^{\text{KA}}_{\mathcal{D}'}(x)$ becomes
a constant, $\check{\Xi}^{\text{KA}}_{\mathcal{D}'}(x)=1$ \cite{os22},
namely $\check{\Xi}^{Q\,\text{monic}}_{\mathcal{D}'}(x)=1$.
So $\widetilde{\mathcal{H}}^{Q\,\text{op}}_{\mathcal{D}'}(\bm{\lambda})$
\eqref{HtDQop} becomes
\begin{equation}
  \widetilde{\mathcal{H}}^{Q\,\text{op}}_{\mathcal{D}'}(\bm{\lambda})
  =B(x+M;\bm{\lambda}-M\bm{\bar{\delta}})(1-e^{\hat{\partial}})
  +D(x+M;\bm{\lambda}-M\bm{\bar{\delta}})(1-e^{-\hat{\partial}}).
\end{equation}
This and \eqref{HtDQopcQDn} imply
\begin{equation}
  \check{Q}^{\text{monic}}_{\mathcal{D}',n}(x;\bm{\lambda})
  =\rho^{-nM}\check{P}^{\text{monic}}_n(x+M;\bm{\lambda}-M\bm{\bar{\delta}})
  \ \ (n=0,1,\ldots,N+M),
  \label{cQD'n=cPn}
\end{equation}
which is shown by \eqref{cQD'nmonic} and
\begin{equation}
  \eta(x+M;\bm{\lambda}-M\bm{\bar{\delta}})
  =\rho^M\bigl(\eta(x;\bm{\lambda}+M\bm{\delta})
  -\eta(-M;\bm{\lambda}+M\bm{\delta})\bigr).
\end{equation}
We remark that verification of this relation \eqref{cQD'n=cPn} based on
\eqref{cQmonic} and \eqref{cQmonic2} is non-trivial.
In \cite{os40}, where
$\check{Q}^{\text{monic}}_{\mathcal{D}',n}(x;\bm{\lambda})$ is not given,
the fact that the eigenpolynomials are given by
$\check{P}^{\text{monic}}_n(x+M;\bm{\lambda}-M\bm{\bar{\delta}})$
is shown based on direct calculations of Casoratians.

\subsection{Positivity of the weight}
\label{sec:positivity}

Let us consider the condition for the positivity of the weight factor in
\eqref{orthoQ}.
The positivity of $\phi_0(x+M;\bm{\lambda}-M\bm{\bar{\delta}})^2$ can be easily
achieved by choosing parameters $\bm{\lambda}$ so that
$B(x+M;\bm{\lambda}-M\bm{\bar{\delta}})$ and
$D(x+M;\bm{\lambda}-M\bm{\bar{\delta}})$ are positive (except for the
boundaries). From the positivity conditions for $B(x;\bm{\lambda})$ and
$D(x;\bm{\lambda})$ given in Appendix\,\ref{app:poly}, we have
\begin{align}
  \text{H}:&\ \ a,b>0\n[-5pt]
  \text{K}:&\ \ 0<p<1\n[-5pt]
  \text{R}:&\ \ b>N+d,\ \ d>M,\ \ 0<c<1+d-M,\ \ (d\neq M+1)\n[-5pt]
  \text{dH}:&\ \ a>0,\ \ b>M,\ \ (a+b\neq M+1,M+2)\n[-5pt]
  \text{dq$q$K}:&\ \ p>q^{-N}\n[-5pt]
  \text{$q$H}:&\ \ 0<a,b<1\n[-5pt]
  \text{$q$K}:&\ \ p>0
  \label{rangeM}\\[-5pt]
  \text{q$q$K}:&\ \ p>q^{-N-M}\n[-5pt]
  \text{a$q$K}:&\ \ 0<p<q^{-1}\n[-5pt]
  \text{$q$R}:&\ \ 0<b<dq^N,\ \ d<q^M,\ \ dq^{1-M}<c<1,
  \ \ (d\neq q^{M+1})\n[-5pt]
  \text{d$q$H}:&\ \ 0<a<1,\ \ 0<bq^{-M}<1,\ \ (ab\neq q^{M+1},q^{M+2})\n[-5pt]
  \text{d$q$K}:&\ \ p>0.
  \nonumber
\end{align}
We divide these into two classes,
\begin{equation}
  \text{(a)}:\text{H, K, dq$q$K, $q$H, $q$K, a$q$K, d$q$K},\quad
  \text{(b)}:\text{R, dH, q$q$K, $q$R, d$q$H}.
  \label{classab}
\end{equation}
The parameter ranges \eqref{rangeM} are independent of $M$ for class (a) and
dependent on $M$ for class (b).
The positivity of
$\check{\Xi}^{Q\,\text{monic}}_{\mathcal{D}'}(x;\bm{\lambda})
\check{\Xi}^{Q\,\text{monic}}_{\mathcal{D}'}(x+1;\bm{\lambda})$
(for $x=-M,-M+1,\ldots,N$) is highly non-trivial.
We try this problem by numerical calculations.
In the rest of this subsection we assume $d_1<d_2<\cdots<d_M$ and
the parameters $\bm{\lambda}$ satisfying \eqref{rangeM}.

Based on numerical calculations, we have observed the followings.
For class (a), the condition for the positivity is given by
\begin{align}
  &d_j-d_{j-1}\equiv 1\ \ (\text{mod }2)
  \ \ (j=1,2,\ldots,M\ ;\ d_0=N),\n
  &\Bigl(\Leftrightarrow
  m_{2k-1}:\text{even}\ \bigl(k=1,2,\ldots,[\tfrac{M+1}{2}]\bigr),
  \ m_{2k}:\text{odd}\ \bigl(k=1,2,\ldots,[\tfrac{M}{2}]\bigr)\Bigr).
  \label{djsaodd}
\end{align}
For class (b), the situation is more complicated.
For $\mathcal{D}$ satisfying \eqref{djsaodd}, the positivity is satisfied by
further restricting the range of parameters \eqref{rangeM}.
We have not yet found a definite range, but the following ranges seem to work
well: q$q$K: large $p$, dH: large $b$, d$q$H: small $b$, R: large $d$,
$q$R: small $d$.
Even for $\mathcal{D}$ not satisfying \eqref{djsaodd}, the positivity
may be satisfied.
For example, in the case of $M=1$, if the parameters are well chosen, there
are no restrictions on $d_1$. For general $M$, we have not yet found the
conditions for $\mathcal{D}$.

\section{Summary and Comments}
\label{sec:summary}

The Hamiltonian of a finite type rdQM is a real symmetric matrix of order $N+1$
(the coordinate $x\in\{0,1,\ldots,N\}$) and the Schr\"odinger equation is a
matrix eigenvalue problem, whose eigenvectors are $\phi_n(x)$ with
$n\in\{0,1,\ldots,N\}$. This eigenvector $\phi_n(x)$ is extended to the
function $\phi_n(x)$ ($x\in\mathbb{R}$, $n\in\mathbb{Z}_{\geq 0}$) (for $n>N$,
$\phi_n(x)$ is replaced with $\phi^{\text{monic}}_n(x)$), which satisfy the
difference equation.
Based on such rdQM systems described by the twelve orthogonal polynomials,
we have considered their deformation by the multi-step Darboux transformations
with seed solutions $\phi_n(x)$.
The seed solution $\phi_n(x)$ with $n>N$ corresponds to the overshoot
eigenfunction $\tilde{\phi}^{\text{os}}_n(x)$ in oQM systems with a finite
number of eigenstates \cite{os28}, and it is a ``zero norm'' eigenvector because
it vanishes at $x\in\{0,1,\ldots,N\}$ \cite{os40}.
For seed solutions $\phi_n(x)$ with $n\leq N$
($n\in\mathcal{D}=\{d_1,d_2,\ldots,d_M\}$), the Darboux transformations are
state-deleting and the deformed Hamiltonian has order $N-M+1$
($x\in\{0,1,\ldots,N-M\}$) and the eigenvectors are described by the
Krein-Alder type multi-indexed orthogonal polynomials
$\check{P}^{\text{KA}}_{\mathcal{D},n}(x)$
($n\in\{0,1,\ldots,N\}\backslash\mathcal{D}$) \cite{os22}.
For seed solutions $\phi_n(x)$ with $n>N$ ($n\in\mathcal{D}$), the Darboux
transformations are state-adding and the deformed Hamiltonian has order $N+M+1$
($x\in\{-M,-M+1,\ldots,N\}$) and the eigenvectors are described by new
multi-indexed orthogonal polynomials $\check{Q}_{\mathcal{D}',n}(x)$
($n\in\{0,1,\ldots,N\}\cup\mathcal{D}$).
Explicit forms of the difference equation and the orthogonality relations for
$\check{P}^{\text{KA}}_{\mathcal{D},n}(x)$ are new results and given in
\S\,\ref{sec:KAmiop}.
New multi-indexed orthogonal polynomials $\check{Q}_{\mathcal{D}',n}(x)$ are
main results of this paper. Their definitions, the difference equation and the
orthogonality relations are given in \S\,\ref{sec:newmiop}.
The energy eigenvalues of $M$ added states are $\mathcal{E}_n$
($n\in\mathcal{D}$). This is in contrast to the case of state-adding Darboux
transformations with the pseudo virtual states as seed solutions
\cite{casoidrdqm}, in which the energy eigenvalues of $M$ added states are
$\mathcal{E}_{-n-1}$ ($n\in\mathcal{D}$).
The positivity condition of the weight factor is discussed in
\S\,\ref{sec:positivity}. Partial results are obtained, but they are still
unsatisfactory. It is an important problem to clarify the positivity condition.
In \S\,\ref{sec:newmiop} the condition $M\leq N$ is assumed, and it is an
interesting problem to consider the case $M>N$.

We have shown the state-adding property of the Darboux transformations with
seed solutions $\phi_n(x)$ ($n>N$) by using the embedding of the finite system
into the infinite system.
In Appendix B of \cite{casoidrdqm}, where the state-adding Darboux
transformations with the pseudo virtual states
$\tilde{\phi}^{\text{pv}}_{\text{v}}(x)$ as seed solutions are discussed,
the state-adding property was shown without using such a trick.
It was shown that each step of the Darboux transformation increases the size of
the matrix (Hamiltonian) by one.
We think that such a proof is possible in the present case as well.
We hope that we will be able to report on this subject in detail elsewhere.

The multi-indexed orthogonal polynomials do not satisfy the three term
recurrence relations, which characterize the ordinary orthogonal polynomials
\cite{ismail}. They satisfy the recurrence relations with more terms
(\cite{d14,d17,rrmiop5,mtv22} for rdQM).
It is an interesting problem to study the recurrence relations for the new
multi-indexed orthogonal polynomials $\check{Q}_{\mathcal{D}',n}(x)$.

\section*{Acknowledgements}

I thank Ryu Sasaki for discussions in the early stage of this work.
This work is supported by JSPS KAKENHI Grant Number JP19K03667.

\bigskip
\appendix
\section{Various Data And Formulas}
\label{app:data}

In this appendix we present various data and formulas on orthogonal polynomials,
$\eta(x)$, $\Lambda(x)$, $\varphi_M(x)$, $B(x)$, $D(x)$ and $\phi_0(x)$.
The Pochhammer symbol (shifted factorial), the hypergeometric series and their
$q$-versions are defined by \cite{ismail,kls},
\begin{align}
  &(a)_n\eqdef\prod_{j=0}^{n-1}(a+j),\quad
  (a_1,\ldots,a_r)_n\eqdef\prod_{k=1}^r(a_k)_n,\\
  &(a\,;q)_n\eqdef\prod_{j=0}^{n-1}(1-aq^j),\quad
  (a_1,\ldots,a_r\,;q)_n\eqdef\prod_{k=1}^r(a_k\,;q)_n,\\
  &{}_rF_s\Bigl(\genfrac{}{}{0pt}{}{a_1,\ldots,a_r}{b_1,\ldots,b_s}
  \Bigm|x\Bigr)
  \eqdef\sum_{k=0}^{\infty}\frac{(a_1,\ldots,a_r)_k}{(b_1,\ldots,b_s)_k}
  \frac{x^k}{k!},\\
  &{}_r\phi_s\Bigl(\genfrac{}{}{0pt}{}{a_1,\ldots,a_r}{b_1,\ldots,b_s}
  \Bigm|q\,;z\Bigr)
  \eqdef\sum_{k=0}^{\infty}\frac{(a_1,\ldots,a_r\,;q)_k}{(b_1,\ldots,b_s\,;q)_k}
  (-1)^{(1+s-r)k}q^{(1+s-r)\binom{k}{2}}\frac{z^k}{(q\,;q)_k},
\end{align}
with the conventions $\sum_{j=n}^{n-1}*=0$ and $\prod_{j=n}^{n-1}*=1$.
The binomial coefficient and its $q$-version are
\begin{align}
  \binom{N}{n}&=\frac{N!}{n!\,(N-n)!}=\frac{(-1)^n(-N)_n}{(1)_n},\\
  \qbinom{N}{n}&=\frac{(q\,;q)_N}{(q\,;q)_n\,(q\,;q)_{N-n}}
  =\frac{(-1)^n(q^{-N}\,;q)_n}{(q\,;q)_n}q^{Nn-\binom{n}{2}}.
\end{align}

\subsection{Orthogonal polynomials}
\label{app:poly}

We give the data for the twelve orthogonal polynomials in the order of
\eqref{etadef}.
The parameterization of some polynomials are different from the conventional
ones, see \cite{os12}.
We consider $\epsilon=\epsilon'=1$ cases in \cite{os12}.
The universal expression of $c_n$ \eqref{cndef} is given by \eqref{cnuniv}.
The polynomials $\check{\tilde{P}}^{(1)}_n(x;\bm{\lambda})$ and
$\check{\tilde{P}}^{(2)}_m(x;\bm{\lambda})$ are defined by \eqref{ctP1def}
and \eqref{ctP2def}, respectively.

\subsubsection{Hahn (H)}
\label{app:H}

Parameter range for the positivity \eqref{B,D>0}: $a,b>0$.
\begin{align*}
  &\bm{\lambda}=(a,b,N),\quad\bm{\delta}=(1,1,-1),\quad
  \bm{\bar{\delta}}=(0,0,-1),\quad\kappa=1,\quad\rho=1,\\
  &\mathcal{E}_n(\bm{\lambda})=n(n+a+b-1),\quad
  \eta(x;\bm{\lambda})=x,\quad
  \varphi(x;\bm{\lambda})=1,\\
  &\check{P}_n(x;\bm{\lambda})={}_3F_2\Bigl(
  \genfrac{}{}{0pt}{}{-n,\,n+a+b-1,\,-x}{a,\,-N}\Bigm|1\Bigr),\quad
  c_n(\bm{\lambda})=\frac{(n+a+b-1)_n}{(a,-N)_n},\\
  &\check{P}^{\text{monic}}_n(x;\bm{\lambda})
  =\sum_{k=0}^n\frac{(a+k,-N+k)_{n-k}}{(n+a+b-1+k)_{n-k}}
  \frac{(-n,-x)_k}{k!},\\
  &B(x;\bm{\lambda})=(x+a)(N-x),\quad
  D(x;\bm{\lambda})=x(b+N-x),\\
  &\phi_0(x;\bm{\lambda})^2
  =\binom{N}{x}\frac{(a)_x}{(b+N-x)_x},\\
  &d_n(\bm{\lambda})^2
  =\binom{N}{n}\frac{(a)_n\,(2n+a+b-1)(a+b)_N}{(b)_n\,(n+a+b-1)_{N+1}}
  \times\frac{(b)_N}{(a+b)_N},\\
  &\check{\tilde{P}}^{(1)}_n(x;\bm{\lambda})
  =\sum_{k=0}^{n-1}\frac{(a+k)_{n-k}}{(n+a+b-1+k)_{n-k}}
  \frac{(-n,-x)_k}{k!}\\
  &\phantom{\check{\tilde{P}}^{(1)}_n(x;\bm{\lambda})=}\quad
  \times\sum_{l=0}^{n-k-1}(-N+k)_l(-N+k+l+1)_{n-k-1-l},\\
  &\check{\tilde{P}}^{(2)}_m(x;\bm{\lambda})
  =\sum_{k=0}^m\frac{(a+N+1+k,N+2+k)_{m-k}}{(a+b+2N+1+m+k)_{m-k}}
  \frac{(-m,-x+N+1)_k}{k!}\\
  &\phantom{\check{\tilde{P}}^{(2)}_m(x;\bm{\lambda})=}\quad
  \times\biggl(\sum_{l=0}^{m-k-1}\Bigl(\frac{1}{a+N+1+k+l}
  +\frac{1}{N+2+k+l}\\
  &\phantom{\check{\tilde{P}}^{(2)}_m(x;\bm{\lambda})=}\quad
  \qquad\qquad\quad
  -\frac{2}{a+b+2N+1+m+k+l}\Bigr)
  +\sum_{l=0}^{k-1}\frac{1}{-x+N+1+l}\biggr).
\end{align*}

\subsubsection{Krawtchouk (K)}
\label{app:K}

Parameter range for the positivity \eqref{B,D>0}: $0<p<1$.
\begin{align*}
  &\bm{\lambda}=(p,N),\quad\bm{\delta}=(0,-1),\quad
  \bm{\bar{\delta}}=(0,-1),\quad\kappa=1,\quad\rho=1\\
  &\mathcal{E}_n(\bm{\lambda})=n,\quad
  \eta(x;\bm{\lambda})=x,\quad
  \varphi(x;\bm{\lambda})=1,\\
  &\check{P}_n(x;\bm{\lambda})={}_2F_1\Bigl(
  \genfrac{}{}{0pt}{}{-n,\,-x}{-N}\Bigm|p^{-1}\Bigr),\quad
  c_n(\bm{\lambda})=\frac{1}{(-N)_n\,p^n},\\
  &\check{P}^{\text{monic}}_n(x;\bm{\lambda})
  =\sum_{k=0}^n(-N+k)_{n-k}
  \frac{(-n,-x)_k}{k!}p^{n-k},\\
  &B(x;\bm{\lambda})=p(N-x),\quad
  D(x;\bm{\lambda})=(1-p)x,\\
  &\phi_0(x;\bm{\lambda})=
  \binom{N}{x}\Bigl(\frac{p}{1-p}\Bigr)^x,\quad
  d_n(\bm{\lambda})^2
  =\binom{N}{n}\Bigl(\frac{p}{1-p}\Bigr)^n\times(1-p)^N,\\
  &\check{\tilde{P}}^{(1)}_n(x;\bm{\lambda})
  =\sum_{k=0}^{n-1}\frac{(-n,-x)_k}{k!}p^{n-k}
  \sum_{l=0}^{n-k-1}(-N+k)_l(-N+k+l+1)_{n-k-1-l},\\
  &\check{\tilde{P}}^{(2)}_m(x;\bm{\lambda})
  =\sum_{k=0}^m(N+2+k)_{m-k}\frac{(-m,-x+N+1)_k}{k!}p^{m-k}\\
  &\phantom{\check{\tilde{P}}^{(2)}_m(x;\bm{\lambda})=}\quad
  \times\biggl(\sum_{l=0}^{m-k-1}\frac{1}{N+2+k+l}
  +\sum_{l=0}^{k-1}\frac{1}{-x+N+1+l}\biggr).
\end{align*}

\subsubsection{Racah (R)}
\label{app:R}

We take $a=-N$ and define $\tilde{d}=a+b+c-d-1$.\\
Parameter range for the positivity \eqref{B,D>0}:
$0<d<a+b$, $0<c<1+d$, ($d\neq 1$).
\begin{align*}
  &\bm{\lambda}=(a,b,c,d),\quad\bm{\delta}=(1,1,1,1),\quad
  \bm{\bar{\delta}}=(1,0,0,1),\quad\kappa=1,\quad\rho=1,\\
  &\mathcal{E}_n(\bm{\lambda})=n(n+\tilde{d}),\quad
  \eta(x;\bm{\lambda})=x(x+d),\quad
  \varphi(x;\bm{\lambda})=\frac{2x+1+d}{1+d},\\
  &\check{P}_n(x;\bm{\lambda})={}_4F_3\Bigl(
  \genfrac{}{}{0pt}{}{-n,\,n+\tilde{d},\,-x,\,x+d}
  {a,\,b,\,c}\Bigm|1\Bigr),\quad
  c_n(\bm{\lambda})=\frac{(\tilde{d}+n)_n}{(a,b,c)_n},\\
  &\check{P}^{\text{monic}}_n(x;\bm{\lambda})
  =\sum_{k=0}^n\frac{(a+k,b+k,c+k)_{n-k}}{(\tilde{d}+n+k)_{n-k}}
  \frac{(-n,-x,x+d)_k}{k!},\\
  &B(x;\bm{\lambda})=-\frac{(x+a)(x+b)(x+c)(x+d)}{(2x+d)(2x+1+d)},\\
  &D(x;\bm{\lambda})=-\frac{(x+d-a)(x+d-b)(x+d-c)x}{(2x-1+d)(2x+d)},\\
  &\phi_0(x;\bm{\lambda})^2=\frac{(a,b,c,d)_x}{(1+d-a,1+d-b,1+d-c,1)_x}\,
  \frac{2x+d}{d},\\
  &d_n(\bm{\lambda})^2=\frac{(a,b,c,\tilde{d})_n}
  {(1+\tilde{d}-a,1+\tilde{d}-b,1+\tilde{d}-c,1)_n}\,
  \frac{2n+\tilde{d}}{\tilde{d}}\\
  &\phantom{d_n(\bm{\lambda})^2=}
  \times\frac{(-1)^N(1+d-a,1+d-b,1+d-c)_N}{(\tilde{d}+1)_N(d+1)_{2N}},\\
  &\check{\tilde{P}}^{(1)}_n(x;\bm{\lambda})
  =\sum_{k=0}^{n-1}\frac{(b+k,c+k)_{n-k}}{(\tilde{d}+n+k)_{n-k}}
  \frac{(-n,-x,x+d)_k}{k!}\\
  &\phantom{\check{\tilde{P}}^{(1)}_n(x;\bm{\lambda})=}\quad
  \times\sum_{l=0}^{n-k-1}\Bigl((-N+k)_l(-N+k+l+1)_{n-k-1-l}
  -\frac{(-N+k)_{n-k}}{\tilde{d}+n+k+l}\Bigr),\\
  &\check{\tilde{P}}^{(2)}_m(x;\bm{\lambda})
  =\sum_{k=0}^m\frac{(N+2+k,b+N+1+k,c+N+1+k)_{m-k}}
  {(\tilde{d}+2N+2+m+k)_{m-k}}\\
  &\phantom{\check{\tilde{P}}^{(2)}_m(x;\bm{\lambda})=}\quad
  \times\frac{(-m,-x+N+1,x+N+1+d)_k}{k!}\\
  &\phantom{\check{\tilde{P}}^{(2)}_m(x;\bm{\lambda})=}
  \times\biggl(\sum_{l=0}^{m-k-1}\Bigl(\frac{1}{N+2+k+l}
  +\frac{1}{b+N+1+k+l}\\
  &\phantom{\check{\tilde{P}}^{(2)}_m(x;\bm{\lambda})=}\qquad\qquad\quad
  +\frac{1}{c+N+1+k+l}-\frac{1}{\tilde{d}+2N+2+m+k+l}\Bigr)\\
  &\phantom{\check{\tilde{P}}^{(2)}_m(x;\bm{\lambda})=}\qquad
  +\sum_{l=0}^{k-1}\Bigl(\frac{1}{-x+N+1+l}+\frac{1}{x+N+1+d+l}\Bigr)\biggr).
\end{align*}

\subsubsection{dual Hahn (dH)}
\label{app:dH}

Parameter range for the positivity \eqref{B,D>0}: $a,b>0$, ($a+b\neq 1,2$).
\begin{align*}
  &\bm{\lambda}=(a,b,N),\quad\bm{\delta}=(1,0,-1),\quad
  \bm{\bar{\delta}}=(0,1,-1),\quad\kappa=1,\quad\rho=1,\\
  &\mathcal{E}_n(\bm{\lambda})=n,\quad
  \eta(x;\bm{\lambda})=x(x+a+b-1),\quad
  \varphi(x;\bm{\lambda})=\frac{2x+a+b}{a+b},\\
  &\check{P}_n(x;\bm{\lambda})={}_3F_2\Bigl(
  \genfrac{}{}{0pt}{}{-n,\,x+a+b-1,\,-x}{a,\,-N}\Bigm|1\Bigr),\quad
  c_n(\bm{\lambda})=\frac{1}{(a,-N)_n},\\
  &\check{P}^{\text{monic}}_n(x;\bm{\lambda})
  =\sum_{k=0}^n(a+k,-N+k)_{n-k}
  \frac{(-n,x+a+b-1,-x)_k}{k!},\\
  &B(x;\bm{\lambda})=\frac{(x+a)(x+a+b-1)(N-x)}
  {(2x-1+a+b)(2x+a+b)},\\
  &D(x;\bm{\lambda})=\frac{x(x+b-1)(x+a+b+N-1)}
  {(2x-2+a+b)(2x-1+a+b)},\\
  &\phi_0(x;\bm{\lambda})^2
  =\binom{N}{x}\frac{(a)_x\,(2x+a+b-1)(a+b)_N}{(b)_x\,(x+a+b-1)_{N+1}\,},\\
  &d_n(\bm{\lambda})^2
  =\binom{N}{n}\frac{(a)_n}{(b+N-n)_n}\times\frac{(b)_{N}}{(a+b)_N},\\
  &\check{\tilde{P}}^{(1)}_n(x;\bm{\lambda})
  =\sum_{k=0}^{n-1}(a+k)_{n-k}
  \frac{(-n,x+a+b-1,-x)_k}{k!}\\
  &\phantom{\check{\tilde{P}}^{(1))}_n(x;\bm{\lambda})=}\quad
  \times\sum_{l=0}^{n-k-1}(-N+k)_l(-N+k+l+1)_{n-k-1-l},\\
  &\check{\tilde{P}}^{(2)}_m(x;\bm{\lambda})
  =\sum_{k=0}^m(a+N+1+k,N+2+k)_{m-k}\frac{(-m,x+a+b+N,-x+N+1)_k}{k!}\\
  &\phantom{\check{\tilde{P}}^{(2)}_m(x;\bm{\lambda})=}\quad
  \times\biggl(\sum_{l=0}^{m-k-1}\Bigl(\frac{1}{a+N+1+k+l}
  +\frac{1}{N+2+k+l}\Bigr)\\
  &\phantom{\check{\tilde{P}}^{(2)}_m(x;\bm{\lambda})=}\quad\qquad
  +\sum_{l=0}^{k-1}\Bigl(\frac{1}{x+a+b+N+l}+\frac{1}{-x+N+1+l}\Bigr)\biggr).
\end{align*}

\subsubsection{dual quantum $\bm{q}$-Krawtchouk (dq$\bm{q}$K)}
\label{app:dqqH}

Parameter range for the positivity \eqref{B,D>0}: $p>q^{-N}$.
\begin{align*}
  &q^{\bm{\lambda}}=(p,q^N),\quad\bm{\delta}=(0,-1),\quad
  \bm{\bar{\delta}}=(1,-1),\quad\kappa=q^{-1},\quad\rho=q,\\
  &\mathcal{E}_n(\bm{\lambda})=q^{-n}-1,\quad
  \eta(x;\bm{\lambda})=1-q^x,\quad
  \varphi(x;\bm{\lambda})=q^x,\\
  &\check{P}_n(x;\bm{\lambda})={}_2\phi_1\Bigl(
  \genfrac{}{}{0pt}{}{q^{-n},\,q^{-x}}
  {q^{-N}}\Bigm|q\,;pq^{x+1}\Bigr),\quad
  c_n(\bm{\lambda})=\frac{p^nq^{-\frac12n(n-1)}}{(q^{-N}\,;q)_n},\\
  &\check{P}^{\text{monic}}_n(x;\bm{\lambda})
  =\sum_{k=0}^n(q^{-N+k}\,;q)_{n-k}
  \frac{(q^{-n},q^{-x}\,;q)_k}{(q\,;q)_k}p^{k-n}q^{kx+k+\frac12n(n-1)},\\
  &B(x;\bm{\lambda})=p^{-1}q^{-x-N-1}(1-q^{N-x}),\quad
  D(x;\bm{\lambda})=(q^{-x}-1)(1-p^{-1}q^{-x}),\\
  &\phi_0(x;\bm{\lambda})^2
  =\qbinom{N}{x}\frac{p^{-x}q^{-Nx}}{(p^{-1}q^{-x}\,;q)_x},\quad
  d_n(\bm{\lambda})^2
  =\qbinom{N}{n}\frac{p^{-n}q^{n(n-1-N)}}{(p^{-1}q^{-N}\,;q)_n}\,
  \times(p^{-1}q^{-N}\,;q)_N,\\
  &\check{\tilde{P}}^{(1)}_n(x;\bm{\lambda})
  =\sum_{k=0}^{n-1}
  \frac{(q^{-n},q^{-x}\,;q)_k}{(q\,;q)_k}p^{k-n}q^{kx+k+\frac12n(n-1)}\\
  &\phantom{\check{\tilde{P}}^{(1)}_n(x;\bm{\lambda})=}\quad
  \times\sum_{l=0}^{n-k-1}q^{-N+k+l}(q^{-N+k}\,;q)_l
  (q^{-N+k+l+1}\,;q)_{n-k-1-l},\\
  &\check{\tilde{P}}^{(2)}_m(x;\bm{\lambda})
  =\sum_{k=0}^m(q^{N+2+k}\,;q)_{m-k}
  \frac{(q^{-m},q^{-x+N+1}\,;q)_k}{(q\,;q)_k}
  p^{k-m}q^{k(x+1)-(N+1)m+\frac12m(m-1)}\\
  &\phantom{\check{\tilde{P}}^{(2)}_m(x;\bm{\lambda})=}\quad
  \times\biggl(\sum_{l=0}^{m-k-1}\frac{1}{1-q^{N+2+k+l}}
  +\sum_{l=0}^{k-1}\frac{1}{1-q^{-x+N+1+l}}\biggr).
\end{align*}

\subsubsection{$\bm{q}$-Hahn ($\bm{q}$H)}
\label{app:qH}

Parameter range for the positivity \eqref{B,D>0}: $0<a,b<1$.
\begin{align*}
  &q^{\bm{\lambda}}=(a,b,q^N),\quad\bm{\delta}=(1,1,-1),\quad
  \bm{\bar{\delta}}=(0,0,-1),\quad\kappa=q^{-1},\quad\rho=q^{-1},\\
  &\mathcal{E}_n(\bm{\lambda})=(q^{-n}-1)(1-abq^{n-1}),\quad
  \eta(x;\bm{\lambda})=q^{-x}-1,\quad
  \varphi(x;\bm{\lambda})=q^{-x},\\
  &\check{P}_n(x;\bm{\lambda})={}_3\phi_2\Bigl(
  \genfrac{}{}{0pt}{}{q^{-n},\,abq^{n-1},\,q^{-x}}
  {a,\,q^{-N}}\Bigm|q\,;q\Bigr),\quad
  c_n(\bm{\lambda})=\frac{(abq^{n-1}\,;q)_n}{(a,q^{-N}\,;q)_n},\\
  &\check{P}^{\text{monic}}_n(x;\bm{\lambda})
  =\sum_{k=0}^n\frac{(aq^k,q^{-N+k}\,;q)_{n-k}}{(abq^{n-1+k}\,;q)_{n-k}}
  \frac{(q^{-n},q^{-x}\,;q)_k}{(q\,;q)_k}q^k,\\
  &B(x;\bm{\lambda})=(1-aq^x)(q^{x-N}-1),\quad
  D(x;\bm{\lambda})=aq^{-1}(1-q^x)(q^{x-N}-b),\\
  &\phi_0(x;\bm{\lambda})^2
  =\qbinom{N}{x}\frac{(a\,;q)_x}{(bq^{N-x};q)_x\,a^x},\\
  &d_n(\bm{\lambda})^2
  =\qbinom{N}{n}\frac{(a,abq^{-1};q)_n}{(abq^N,b\,;q)_n\,a^n}\,
  \frac{1-abq^{2n-1}}{1-abq^{-1}}
  \times\frac{(b\,;q)_N\,a^N}{(ab\,;q)_N},\\
  &\check{\tilde{P}}^{(1)}_n(x;\bm{\lambda})
  =\sum_{k=0}^{n-1}\frac{(aq^k\,;q)_{n-k}}{(abq^{n-1+k}\,;q)_{n-k}}
  \frac{(q^{-n},q^{-x}\,;q)_k}{(q\,;q)_k}q^k\\
  &\phantom{\check{\tilde{P}}^{(1)}_n(x;\bm{\lambda})=}\quad
  \times\sum_{l=0}^{n-k-1}q^{-N+k+l}(q^{-N+k}\,;q)_l
  (q^{-N+k+l+1}\,;q)_{n-k-1-l},\\
  &\check{\tilde{P}}^{(2)}_m(x;\bm{\lambda})
  =\sum_{k=0}^m\frac{(aq^{N+1+k},q^{N+2+k}\,;q)_{m-k}}
  {(abq^{2N+1+m+k}\,;q)_{m-k}}
  \frac{(q^{-m},q^{-x+N+1}\,;q)_k}{(q\,;q)_k}q^k\\
  &\phantom{\check{\tilde{P}}^{(2)}_m(x;\bm{\lambda})=}\quad
  \times\biggl(\sum_{l=0}^{m-k-1}\Bigl(\frac{1}{1-aq^{N+1+k+l}}
  +\frac{1}{1-q^{N+2+k+l}}
  -\frac{2}{1-abq^{2N+1+m+k+l}}\Bigr)\\
  &\phantom{\check{\tilde{P}}^{(2)}_m(x;\bm{\lambda})=}\qquad\quad
  -\sum_{l=0}^{k-1}\frac{1}{1-q^{x-N-1-l}}\biggr).
\end{align*}

\subsubsection{$\bm{q}$-Krawtchouk ($\bm{q}$K)}
\label{app:qK}

Parameter range for the positivity \eqref{B,D>0}: $p>0$.
\begin{align*}
  &q^{\bm{\lambda}}=(p,q^N),\quad\bm{\delta}=(2,-1),\quad
  \bm{\bar{\delta}}=(0,-1),\quad\kappa=q^{-1},\quad\rho=q^{-1},\\
  &\mathcal{E}_n(\bm{\lambda})=(q^{-n}-1)(1+pq^n),\quad
  \eta(x;\bm{\lambda})=q^{-x}-1,\quad
  \varphi(x;\bm{\lambda})=q^{-x},\\
  &\check{P}_n(x;\bm{\lambda})={}_3\phi_2\Bigl(
  \genfrac{}{}{0pt}{}{q^{-n},\,q^{-x},\,-pq^n}
  {q^{-N},\,0}\Bigm|q\,;q\Bigr),\quad
  c_n(\bm{\lambda})=\frac{(-pq^n\,;q)_n}{(q^{-N}\,;q)_n},\\
  &\check{P}^{\text{monic}}_n(x;\bm{\lambda})
  =\sum_{k=0}^n\frac{(q^{-N+k}\,;q)_{n-k}}{(-pq^{n+k}\,;q)_{n-k}}
  \frac{(q^{-n},q^{-x}\,;q)_k}{(q\,;q)_k}q^k,\\
  &B(x;\bm{\lambda})=q^{x-N}-1,\quad
  D(x;\bm{\lambda})=p(1-q^x),\\
  &\phi_0(x;\bm{\lambda})^2=\qbinom{N}{x}p^{-x}q^{\frac12x(x-1)-xN},\\
  &d_n(\bm{\lambda})^2
  =\qbinom{N}{n}\frac{(-p\,;q)_n}{(-pq^{N+1}\,;q)_n\,p^nq^{\frac12n(n+1)}}\,
  \frac{1+pq^{2n}}{1+p}
  \times\frac{p^{N}q^{\frac12N(N+1)}}{(-pq\,;q)_N},\\
  &\check{\tilde{P}}^{(1)}_n(x;\bm{\lambda})
  =\sum_{k=0}^{n-1}\frac{1}{(-pq^{n+k}\,;q)_{n-k}}
  \frac{(q^{-n},q^{-x}\,;q)_k}{(q\,;q)_k}q^k\\
  &\phantom{\check{\tilde{P}}^{(1)}_n(x;\bm{\lambda})=}\quad
  \times\sum_{l=0}^{n-k-1}q^{-N+k+l}(q^{-N+k}\,;q)_l
  (q^{-N+k+l+1}\,;q)_{n-k-1-l},\\
  &\check{\tilde{P}}^{(2)}_m(x;\bm{\lambda})
  =\sum_{k=0}^m\frac{(q^{N+2+k}\,;q)_{m-k}}{(-pq^{2N+2+m+k}\,;q)_{m-k}}
  \frac{(q^{-m},q^{-x+N+1}\,;q)_k}{(q\,;q)_k}q^k\\
  &\phantom{\check{\tilde{P}}^{(2)}_m(x;\bm{\lambda})=}\quad
  \times\biggl(\sum_{l=0}^{m-k-1}\Bigl(\frac{1}{1-q^{N+2+k+l}}
  -\frac{1-pq^{2N+2+m+k+l}}{1+pq^{2N+2+m+k+l}}\Bigr)
  -\sum_{l=0}^{k-1}\frac{1}{1-q^{x-N-1-l}}\biggr).
\end{align*}

\subsubsection{quantum $\bm{q}$-Krawtchouk (q$\bm{q}$K)}
\label{app:qqK}

Parameter range for the positivity \eqref{B,D>0}: $p>q^{-N}$.
\begin{align*}
  &q^{\bm{\lambda}}=(p,q^N),\quad\bm{\delta}=(1,-1),\quad
  \bm{\bar{\delta}}=(0,-1),\quad\kappa=q,\quad\rho=q^{-1},\\
  &\mathcal{E}_n(\bm{\lambda})=1-q^n,\quad
  \eta(x;\bm{\lambda})=q^{-x}-1,\quad
  \varphi(x;\bm{\lambda})=q^{-x},\\
  &\check{P}_n(x;\bm{\lambda})={}_2\phi_1\Bigl(
  \genfrac{}{}{0pt}{}{q^{-n},\,q^{-x}}
  {q^{-N}}\Bigm|q\,;pq^{n+1}\Bigr),\quad
  c_n(\bm{\lambda})=\frac{p^nq^{n^2}}{(q^{-N}\,;q)_n},\\
  &\check{P}^{\text{monic}}_n(x;\bm{\lambda})
  =\sum_{k=0}^n(q^{-N+k}\,;q)_{n-k}
  \frac{(q^{-n},q^{-x}\,;q)_k}{(q\,;q)_k}p^{k-n}q^{(n+1)k-n^2},\\
  &B(x;\bm{\lambda})=p^{-1}q^x(q^{x-N}-1),\quad
  D(x;\bm{\lambda})=(1-q^x)(1-p^{-1}q^{x-N-1}),\\
  &\phi_0(x;\bm{\lambda})^2
  =\qbinom{N}{x}\frac{p^{-x}q^{x(x-1-N)}}{(p^{-1}q^{-N}\,;q)_x},\quad
  d_n(\bm{\lambda})^2
  =\qbinom{N}{n}\frac{p^{-n}q^{-Nn}}{(p^{-1}q^{-n}\,;q)_n}\,
  \times(p^{-1}q^{-N}\,;q)_N,\\
  &\check{\tilde{P}}^{(1)}_n(x;\bm{\lambda})
  =\sum_{k=0}^{n-1}
  \frac{(q^{-n},q^{-x}\,;q)_k}{(q\,;q)_k}p^{k-n}q^{(n+1)k-n^2}\\
  &\phantom{\check{\tilde{P}}^{(1)}_n(x;\bm{\lambda})=}\quad
  \times\sum_{l=0}^{n-k-1}q^{-N+k+l}(q^{-N+k}\,;q)_l
  (q^{-N+k+l+1}\,;q)_{n-k-1-l},\\
  &\check{\tilde{P}}^{(2)}_m(x;\bm{\lambda})
  =\sum_{k=0}^m(q^{N+2+k}\,;q)_{m-k}
  \frac{(q^{-m},q^{-x+N+1}\,;q)_k}{(q\,;q)_k}(pq^{N+1})^{k-m}q^{(m+1)k-m^2}\\
  &\phantom{\check{\tilde{P}}^{(2)}_m(x;\bm{\lambda})=}\quad
  \times\biggl(\sum_{l=0}^{m-k-1}\frac{1}{1-q^{N+2+k+l}}
  -\sum_{l=0}^{k-1}\frac{1}{1-q^{x-N-1-l}}\biggr).
\end{align*}

\subsubsection{affine $\bm{q}$-Krawtchouk (a$\bm{q}$K)}
\label{app:aqK}

Parameter range for the positivity \eqref{B,D>0}: $0<p<q^{-1}$.
\begin{align*}
  &q^{\bm{\lambda}}=(p,q^N),\quad\bm{\delta}=(1,-1),\quad
  \bm{\bar{\delta}}=(0,-1),\quad\kappa=q^{-1},\quad\rho=q^{-1},\\
  &\mathcal{E}_n(\bm{\lambda})=q^{-n}-1,\quad
  \eta(x;\bm{\lambda})=q^{-x}-1,\quad
  \varphi(x;\bm{\lambda})=q^{-x},\\
  &\check{P}_n(x;\bm{\lambda})={}_3\phi_2\Bigl(
  \genfrac{}{}{0pt}{}{q^{-n},\,q^{-x},\,0}
  {pq,\,q^{-N}}\Bigm|q\,;q\Bigr),\quad
  c_n(\bm{\lambda})=\frac{1}{(pq,q^{-N}\,;q)_n},\\
  &\check{P}^{\text{monic}}_n(x;\bm{\lambda})
  =\sum_{k=0}^n(pq^{1+k},q^{-N+k}\,;q)_{n-k}
  \frac{(q^{-n},q^{-x}\,;q)_k}{(q\,;q)_k}q^k,\\
  &B(x;\bm{\lambda})=(q^{x-N}-1)(1-pq^{x+1}),\quad
  D(x;\bm{\lambda})=pq^{x-N}(1-q^x),\\
  &\phi_0(x;\bm{\lambda})^2=\qbinom{N}{x}\frac{(pq\,;q)_x}{(pq)^x},\quad
  d_n(\bm{\lambda})^2
  =\qbinom{N}{n}\frac{(pq\,;q)_n}{(pq)^n}\times(pq)^N,\\
  &\check{\tilde{P}}^{(1)}_n(x;\bm{\lambda})
  =\sum_{k=0}^{n-1}(pq^{1+k}\,;q)_{n-k}
  \frac{(q^{-n},q^{-x}\,;q)_k}{(q\,;q)_k}q^k\\
  &\phantom{\check{\tilde{P}}^{(1)}_n(x;\bm{\lambda})=}\quad
  \times\sum_{l=0}^{n-k-1}q^{-N+k+l}(q^{-N+k}\,;q)_l
  (q^{-N+k+l+1}\,;q)_{n-k-1-l},\\
  &\check{\tilde{P}}^{(2)}_m(x;\bm{\lambda})
  =\sum_{k=0}^m(pq^{N+2+k},q^{N+2+k}\,;q)_{m-k}
  \frac{(q^{-m},q^{-x+N+1}\,;q)_k}{(q\,;q)_k}q^k\\
  &\phantom{\check{\tilde{P}}^{(2)}_m(x;\bm{\lambda})=}\quad
  \times\biggl(k+\sum_{l=0}^{m-k-1}\Bigl(\frac{1}{1-pq^{N+2+k+l}}
  +\frac{1}{1-q^{N+2+k+l}}\Bigr)
  +\sum_{l=0}^{k-1}\frac{1}{1-q^{-x+N+1+l}}\biggr).
\end{align*}

\subsubsection{$\bm{q}$-Racah ($\bm{q}$R)}
\label{app:qR}

We take $a=q^{-N}$ and define $\tilde{d}=abcd^{-1}q^{-1}$.\\
Parameter range for the positivity \eqref{B,D>0}:
$0<ab<d<1$, $qd<c<1$, ($d\neq q$).
\begin{align*}
  &q^{\bm{\lambda}}=(a,b,c,d),\quad\bm{\delta}=(1,1,1,1),\quad
  \bm{\bar{\delta}}=(1,0,0,1),\quad\kappa=q^{-1},\quad\rho=q^{-1},\\
  &\mathcal{E}_n(\bm{\lambda})=(q^{-n}-1)(1-\tilde{d}q^n),
  \ \eta(x;\bm{\lambda})=(q^{-x}-1)(1-dq^x),
  \ \varphi(x;\bm{\lambda})=q^{-x}\frac{1-dq^{2x+1}}{1-dq},\\
  &\check{P}_n(x;\bm{\lambda})={}_4\phi_3\Bigl(
  \genfrac{}{}{0pt}{}{q^{-n},\,\tilde{d}q^n,\,q^{-x},\,dq^x}
  {a,\,b,\,c}\Bigm|q\,;q\Bigr),\quad
  c_n(\bm{\lambda})=\frac{(\tilde{d}q^n\,;q)_n}{(a,b,c\,;q)_n},\\
  &\check{P}^{\text{monic}}_n(x;\bm{\lambda})
  =\sum_{k=0}^n\frac{(aq^k,bq^k,cq^k\,;q)_{n-k}}{(\tilde{d}q^{n+k}\,;q)_{n-k}}
  \frac{(q^{-n},q^{-x},dq^x\,;q)_k}{(q\,;q)_k}q^k,\\
  &B(x;\bm{\lambda})
  =-\frac{(1-aq^x)(1-bq^x)(1-cq^x)(1-dq^x)}{(1-dq^{2x})(1-dq^{2x+1})},\\
  &D(x;\bm{\lambda})
  =-\tilde{d}\,\frac{(1-a^{-1}dq^x)(1-b^{-1}dq^x)(1-c^{-1}dq^x)(1-q^x)}
  {(1-dq^{2x-1})(1-dq^{2x})},\\
  &\phi_0(x;\bm{\lambda})^2=\frac{(a,b,c,d\,;q)_x}
  {(a^{-1}dq,b^{-1}dq,c^{-1}dq,q\,;q)_x\,\tilde{d}^x}\,
  \frac{1-dq^{2x}}{1-d},\\
  &d_n(\bm{\lambda})^2
  =\frac{(a,b,c,\tilde{d}\,;q)_n}
  {(a^{-1}\tilde{d}q,b^{-1}\tilde{d}q,c^{-1}\tilde{d}q,q\,;q)_n\,d^n}\,
  \frac{1-\tilde{d}q^{2n}}{1-\tilde{d}}\\
  &\phantom{d_n(\bm{\lambda})^2=}
  \times
  \frac{(-1)^N(a^{-1}dq,b^{-1}dq,c^{-1}dq\,;q)_N\,\tilde{d}^Nq^{\frac12N(N+1)}}
  {(\tilde{d}q\,;q)_N(dq\,;q)_{2N}},\\
  &\check{\tilde{P}}^{(1)}_n(x;\bm{\lambda})
  =\sum_{k=0}^{n-1}\frac{(bq^k,cq^k\,;q)_{n-k}}{(\tilde{d}q^{n+k}\,;q)_{n-k}}
  \frac{(q^{-n},q^{-x},dq^x\,;q)_k}{(q\,;q)_k}q^k\\
  &\phantom{\check{\tilde{P}}^{(1)}_n(x;\bm{\lambda})=}\quad
  \times\sum_{l=0}^{n-k-1}\Bigl((q^{-N+k}\,;q)_l(q^{-N+k+l+1}\,;q)_{n-k-1-l}
  -\frac{(q^{-N+k}\,;q)_{n-k}}{1-\tilde{d}q^{n+k+l}}\Bigr),\\
  &\check{\tilde{P}}^{(2)}_m(x;\bm{\lambda})
  =\sum_{k=0}^m\frac{(q^{N+2+k},bq^{N+1+k},cq^{N+1+k}\,;q)_{m-k}}
  {(\tilde{d}q^{2N+2+m+k}\,;q)_{m-k}}
  \frac{(q^{-m},q^{-x+N+1},dq^{x+N+1}\,;q)_k}{(q\,;q)_k}q^k\\
  &\phantom{\check{\tilde{P}}^{(2)}_m(x;\bm{\lambda})=}\quad
  \times\biggl(\sum_{l=0}^{m-k-1}\Bigl(\frac{1}{1-q^{N+2+k+l}}
  +\frac{1}{1-bq^{N+1+k+l}}+\frac{1}{1-cq^{N+1+k+l}}\\
  &\phantom{\check{\tilde{P}}^{(2)}_m(x;\bm{\lambda})=}\qquad\qquad\qquad
  -\frac{1}{1-\tilde{d}q^{2N+2+m+k+l}}\Bigr)\\
  &\phantom{\check{\tilde{P}}^{(2)}_m(x;\bm{\lambda})=}\qquad\quad
  +\sum_{l=0}^{k-1}\Bigl(\frac{1}{1-q^{-x+N+1+l}}
  +\frac{1}{1-dq^{x+N+1+l}}\Bigr)\biggr).
\end{align*}

\subsubsection{dual $\bm{q}$-Hahn (d$\bm{q}$H)}
\label{app:dqH}

Parameter range for the positivity \eqref{B,D>0}:
$0<a,b<1$, ($ab\neq q,q^2$).
\begin{align*}
  &q^{\bm{\lambda}}=(a,b,q^N),\quad\bm{\delta}=(1,0,-1),\quad
  \bm{\bar{\delta}}=(0,1,-1),\quad\kappa=q^{-1},\quad\rho=q^{-1},\\
  &\mathcal{E}_n(\bm{\lambda})=q^{-n}-1,\quad
  \eta(x;\bm{\lambda})=(q^{-x}-1)(1-abq^{x-1}),\quad
  \varphi(x;\bm{\lambda})=q^{-x}\frac{1-abq^{2x}}{1-ab},\\
  &\check{P}_n(x;\bm{\lambda})={}_3\phi_2\Bigl(
  \genfrac{}{}{0pt}{}{q^{-n},\,abq^{x-1},\,q^{-x}}
  {a,\,q^{-N}}\Bigm|q\,;q\Bigr),\quad
  c_n(\bm{\lambda})=\frac{1}{(a,q^{-N}\,;q)_n},\\
  &\check{P}^{\text{monic}}_n(x;\bm{\lambda})
  =\sum_{k=0}^n(aq^k,q^{-N+k}\,;q)_{n-k}
  \frac{(q^{-n},abq^{x-1},q^{-x}\,;q)_k}{(q\,;q)_k}q^k,\\
  &B(x;\bm{\lambda})=
  \frac{(q^{x-N}-1)(1-aq^x)(1-abq^{x-1})}
  {(1-abq^{2x-1})(1-abq^{2x})},\\
  &D(x;\bm{\lambda})=aq^{x-N-1}
  \frac{(1-q^x)(1-abq^{x+N-1})(1-bq^{x-1})}
  {(1-abq^{2x-2})(1-abq^{2x-1})},\\
  &\phi_0(x;\bm{\lambda})^2
  =\qbinom{N}{x}\frac{(a,abq^{-1}\,;q)_x}{(abq^N,b\,;q)_x\,a^x}\,
  \frac{1-abq^{2x-1}}{1-abq^{-1}},\\
  &d_n(\bm{\lambda})^2
  =\qbinom{N}{n}\frac{(a\,;q)_n}{(bq^{N-n}\,;q)_n\,a^n}
  \times\frac{(b\,;q)_N\,a^N}{(ab\,;q)_N},\\
  &\check{\tilde{P}}^{(1)}_n(x;\bm{\lambda})
  =\sum_{k=0}^{n-1}(aq^k\,;q)_{n-k}
  \frac{(q^{-n},abq^{x-1},q^{-x}\,;q)_k}{(q\,;q)_k}q^k\\
  &\phantom{\check{\tilde{P}}^{(1)}_n(x;\bm{\lambda})=}\quad
  \times\sum_{l=0}^{n-k-1}q^{-N+k+l}(q^{-N+k}\,;q)_l
  (q^{-N+k+l+1}\,;q)_{n-k-1-l},\\
  &\check{\tilde{P}}^{(2)}_m(x;\bm{\lambda})
  =\sum_{k=0}^m(aq^{N+1+k},q^{N+2+k}\,;q)_{m-k}
  \frac{(q^{-m},abq^{x+N},q^{-x+N+1}\,;q)_k}{(q\,;q)_k}q^k\\
  &\phantom{\check{\tilde{P}}^{(2)}_m(x;\bm{\lambda})=}\quad
  \times\biggl(\sum_{l=0}^{m-k-1}\Bigl(\frac{1}{1-aq^{N+1+k+l}}
  +\frac{1}{1-q^{N+2+k+l}}\Bigr)\\
  &\phantom{\check{\tilde{P}}^{(2)}_m(x;\bm{\lambda})=}\qquad\quad
  +\sum_{l=0}^{k-1}\Bigl(\frac{1}{1-abq^{x+N+l}}+\frac{1}{1-q^{-x+N+1+l}}
  \Bigr)\biggr).
\end{align*}

\subsubsection{dual $\bm{q}$-Krawtchouk (d$\bm{q}$K)}
\label{app:dqK}

Parameter range for the positivity \eqref{B,D>0}: $p>0$.
\begin{align*}
  &q^{\bm{\lambda}}=(p,q^N),\quad\bm{\delta}=(1,-1),\quad
  \bm{\bar{\delta}}=(1,-1),\quad\kappa=q^{-1},\quad\rho=q^{-1},\\
  &\mathcal{E}_n(\bm{\lambda})=q^{-n}-1,\quad
  \eta(x;\bm{\lambda})=(q^{-x}-1)(1+pq^x),\quad
  \varphi(x;\bm{\lambda})=q^{-x}\frac{1+pq^{2x+1}}{1+pq},\\
  &\check{P}_n(x;\bm{\lambda})={}_3\phi_2\Bigl(
  \genfrac{}{}{0pt}{}{q^{-n},\,q^{-x},\,-pq^x}
  {q^{-N},\,0}\Bigm|q\,;q\Bigr),\quad
  c_n(\bm{\lambda})=\frac{1}{(q^{-N}\,;q)_n},\\
  &\check{P}^{\text{monic}}_n(x;\bm{\lambda})
  =\sum_{k=0}^n(q^{-N+k}\,;q)_{n-k}
  \frac{(q^{-n},q^{-x},-pq^x\,;q)_k}{(q\,;q)_k}q^k,\\
  &B(x;\bm{\lambda})=\frac{(q^{x-N}-1)(1+pq^x)}
  {(1+pq^{2x})(1+pq^{2x+1})},\quad
  D(x;\bm{\lambda})=pq^{2x-N-1}\frac{(1-q^x)(1+pq^{x+N})}
  {(1+pq^{2x-1})(1+pq^{2x})},\\
  &\phi_0(x;\bm{\lambda})^2
  =\qbinom{N}{x}
  \frac{(-p\,;q)_x\,p^{-x}q^{-\frac12x(x+1)}}{(-pq^{N+1}\,;q)_x}\,
  \frac{1+pq^{2x}}{1+p},\\
  &d_n(\bm{\lambda})^2
  =\qbinom{N}{n}p^{-n}q^{-Nn+\frac12n(n-1)}
  \times\frac{p^Nq^{\frac12N(N+1)}}{(-pq\,;q)_N},\\
  &\check{\tilde{P}}^{(1)}_n(x;\bm{\lambda})
  =\sum_{k=0}^{n-1}
  \frac{(q^{-n},q^{-x},-pq^x\,;q)_k}{(q\,;q)_k}q^k
  \sum_{l=0}^{n-k-1}q^{-N+k+l}(q^{-N+k}\,;q)_l
  (q^{-N+k+l+1}\,;q)_{n-k-1-l},\\
  &\check{\tilde{P}}^{(2)}_m(x;\bm{\lambda})
  =\sum_{k=0}^m(q^{N+2+k}\,;q)_{m-k}
  \frac{(q^{-m},q^{-x+N+1},-pq^{x+N+1}\,;q)_k}{(q\,;q)_k}q^k\\
  &\phantom{\check{\tilde{P}}^{(2)}_m(x;\bm{\lambda})=}\quad
  \times\biggl(\sum_{l=0}^{m-k-1}\frac{1}{1-q^{N+2+k+l}}
  +\sum_{l=0}^{k-1}\Bigl(-\frac{1}{1-q^{x-N-1-l}}+\frac{1}{1+pq^{x+N+1+l}}
  \Bigr)\biggr).
\end{align*}

\subsection{$\bm{\eta(x)}$}
\label{app:eta}

We consider five families of the sinusoidal coordinates \cite{os12}:
\begin{equation}
  \begin{array}{rll}
  \text{(\romannumeral1)}:&\eta(x)=x
  &:\text{H,\,K}\\[3pt]
  \text{(\romannumeral2)}:&\eta(x)=x(x+d)
  &:\text{R},\,\text{dH}(d=a+b-1)\\[3pt]
  \text{(\romannumeral3)}:&\eta(x)=1-q^x
  &:\text{dq$q$K}\\[3pt]
  \text{(\romannumeral4)}:&\eta(x)=q^{-x}-1
  &:\text{$q$H,\,$q$K,\,q$q$K,\,a$q$K}\\[3pt]
  \text{(\romannumeral5)}:&\eta(x)=(q^{-x}-1)(1-dq^x)
  &:\text{$q$R},\,\text{d$q$H}(d=abq^{-1}),\,\text{d$q$K}(d=-p).
  \end{array}
  \tag{\ref{etadef}}
\end{equation}
The polynomial in $x$ that is invariant under $x\to-x-d$ is a polynomial
in $x(x+d)$, and the Laurent polynomial in $q^x$ that is invariant under
$q^x\to q^{-x}d^{-1}$ is a polynomial in $(q^{-x}-1)(1-dq^x)$.
Thus we have
\begin{equation}
  \left.\begin{array}{rl}
  \text{(\romannumeral1)}:&(-x)_k\\[3pt]
  \text{(\romannumeral2)}:&(-x,x+d)_k\\[3pt]
  \text{(\romannumeral3)}:&(q^{-x}\,;q)_k\,q^{kx}\\[3pt]
  \text{(\romannumeral4)}:&(q^{-x}\,;q)_k\\[3pt]
  \text{(\romannumeral5)}:&(q^{-x},dq^x\,;q)_k
  \end{array}\right\}
  \begin{array}{l}
  =\text{a polynomial of degree $k$ in $\eta(x)$}\\
  =\eta(x)^k\times\left\{
  \begin{array}{ll}
  (-1)^k&\!\!\!:\text{(\romannumeral1)--(\romannumeral3)}\\
  (-1)^kq^{\binom{k}{2}}&\!\!\!:\text{(\romannumeral4)},\,\text{(\romannumeral5)}
  \end{array}\right.\!\!
  +(\text{lower degree terms}).
  \end{array}\!
\end{equation}

For an index set $\mathcal{D}=\{d_1,\ldots,d_M\}$ ($d_j\in\mathbb{Z}_{\geq 0}$
: mutually distinct), we have
\begin{align}
  &\quad\varphi_M(x;\bm{\lambda})^{-1}
  \text{W}_{\text{C}}[\eta^{d_1},\ldots,\eta^{d_M}](x;\bm{\lambda})\n
  &=\text{a polynomial of degree
  $\ell^{\text{KA}}_{\mathcal{D}}+M=\ell_{\mathcal{D}}$
  in $\eta\bigl(x;\bm{\lambda}+(M-1)\bm{\delta}\bigr)$}\n
  &=c^{\eta}_{\mathcal{D}}(\bm{\lambda})
  \eta\bigl(x;\bm{\lambda}+(M-1)\bm{\delta}\bigr)^{\ell_{\mathcal{D}}}
  +(\text{lower degree terms}).
\end{align}
Explicit forms of $c^{\eta}_{\mathcal{D}}(\bm{\lambda})$ are given by
\begin{align}
  \text{(\romannumeral1)}:&\ c^{\eta}_{\mathcal{D}}(\bm{\lambda})=
  \prod_{1\leq j<k\leq M}(d_k-d_j),\n
  \text{(\romannumeral2)}:&\ c^{\eta}_{\mathcal{D}}(\bm{\lambda})=
  \prod_{1\leq j<k\leq M}(d_k-d_j)\cdot\prod_{j=1}^{M-1}(d+1)_j,\n
  \text{(\romannumeral3)}:&\ c^{\eta}_{\mathcal{D}}(\bm{\lambda})=
  q^{-\binom{M}{3}}\prod_{1\leq j<k\leq M}(q^{d_j}-q^{d_k}),
  \label{cetaD}\\
  \text{(\romannumeral4)}:&\ c^{\eta}_{\mathcal{D}}(\bm{\lambda})=
  q^{\binom{M}{3}}\prod_{1\leq j<k\leq M}(q^{-d_k}-q^{-d_j}),\n
  \text{(\romannumeral5)}:&\ c^{\eta}_{\mathcal{D}}(\bm{\lambda})=
  q^{\binom{M}{3}}\prod_{1\leq j<k\leq M}(q^{-d_k}-q^{-d_j})\cdot
  \prod_{j=1}^{M-1}(dq\,;\,q)_j.
  \nonumber
\end{align}
Similarly, we define $c^{\eta}_{\mathcal{D},n}(\bm{\lambda})$
($n\in\mathbb{Z}_{\geq0}\backslash\mathcal{D}$) as
\begin{align}
  &\quad\varphi_{M+1}(x;\bm{\lambda})^{-1}
  \text{W}_{\text{C}}[\eta^{d_1},\ldots,\eta^{d_M},\eta^n](x;\bm{\lambda})\n
  &=\text{a polynomial of degree $\ell^{\text{KA}}_{\mathcal{D}}+n$
  in $\eta(x;\bm{\lambda}+M\bm{\delta})$}\n
  &=c^{\eta}_{\mathcal{D},n}(\bm{\lambda})
  \eta(x;\bm{\lambda}+M\bm{\delta})^{\ell^{\text{KA}}_{\mathcal{D}}+n}
  +(\text{lower degree terms}).
\end{align}
Since they are related as
\begin{equation}
  c^{\eta}_{\mathcal{D},n}(\bm{\lambda})=
  c^{\eta}_{\mathcal{D}'}(\bm{\lambda}),\quad
  \mathcal{D}'=\{d_1,\ldots,d_M,n\},
\end{equation}
explicit forms of $c^{\eta}_{\mathcal{D}}(\bm{\lambda})$ are given by
\begin{align}
  \text{(\romannumeral1)}:&\ c^{\eta}_{\mathcal{D},n}(\bm{\lambda})=
  \prod_{1\leq j<k\leq M}(d_k-d_j)\cdot\prod_{j=1}^M(n-d_j),\n
  \text{(\romannumeral2)}:&\ c^{\eta}_{\mathcal{D},n}(\bm{\lambda})=
  \prod_{1\leq j<k\leq M}(d_k-d_j)\cdot\prod_{j=1}^M(n-d_j)\cdot
  \prod_{j=1}^M(d+1)_j,\n
  \text{(\romannumeral3)}:&\ c^{\eta}_{\mathcal{D},n}(\bm{\lambda})=
  q^{-\binom{M+1}{3}}\prod_{1\leq j<k\leq M}(q^{d_j}-q^{d_k})\cdot
  \prod_{j=1}^M(q^{d_j}-q^n),
  \label{cetaDn}\\
  \text{(\romannumeral4)}:&\ c^{\eta}_{\mathcal{D},n}(\bm{\lambda})=
  q^{\binom{M+1}{3}}\prod_{1\leq j<k\leq M}(q^{-d_k}-q^{-d_j})\cdot
  \prod_{j=1}^M(q^{-n}-q^{-d_j}),\n
  \text{(\romannumeral5)}:&\ c^{\eta}_{\mathcal{D},n}(\bm{\lambda})=
  q^{\binom{M+1}{3}}\prod_{1\leq j<k\leq M}(q^{-d_k}-q^{-d_j})\cdot
  \prod_{j=1}^M(q^{-n}-q^{-d_j})\cdot
  \prod_{j=1}^M(dq\,;\,q)_j.
  \nonumber
\end{align}
For $n=d_i\in\mathcal{D}$, we define 
$c^{\prime\,\eta}_{\mathcal{D},n}(\bm{\lambda})$ as
\begin{equation}
  c^{\prime\,\eta}_{\mathcal{D},n}(\bm{\lambda})
  \eqdef c^{\eta}_{\mathcal{D},n}(\bm{\lambda})\times\left\{
  \begin{array}{ll}
  (n-d_i)^{-1}&:\text{(\romannumeral1)},\,\text{(\romannumeral2)}\\
  (q^{d_i}-q^n)^{-1}&:\text{(\romannumeral3)}\\
  (q^{-n}-q^{-d_i})^{-1}&:\text{(\romannumeral4)},\,\text{(\romannumeral5)}
  \end{array}\right.,
  \label{cetaDn'}
\end{equation}
namely in the products $\prod_{j=1}^M$ containing $n$ in \eqref{cetaDn},
the $j=i$ term is omitted.

\subsection{$\bm{\Lambda(x)}$}
\label{app:Lam}

For five families of $\eta(x)$ \eqref{etadef}, the function
$\Lambda(x;\bm{\lambda})$ is defined by
\begin{equation}
  \Lambda(x;\bm{\lambda})\eqdef\prod_{k=0}^N\bigl(
  \eta(x;\bm{\lambda})-\eta(k;\bm{\lambda})\bigr).
  \tag{\ref{Lambdadef}}
\end{equation}
We remark that the parameter $N$ is contained in $\bm{\delta}$ and the
components of $\bm{\delta}$ corresponding to the parameters $N$ and $d$ are
$-1$ and $1$, respectively.
For example, we have
\begin{equation}
  \Lambda(x;\bm{\lambda}+M\bm{\delta})=\prod_{k=0}^{N-M}\bigl(
  \eta(x;\bm{\lambda}+M\bm{\delta})-\eta(k;\bm{\lambda}+M\bm{\delta})\bigr)
  \quad(M\leq N).
\end{equation}
The functions $\Lambda(x;\bm{\lambda})$ are monic polynomials of degree $N+1$ in
$\eta(x;\bm{\lambda})$ and their explicit forms are given by
\begin{align}
  \text{(\romannumeral1)}:
  &\ \ \Lambda(x;\bm{\lambda})=\prod_{k=0}^N(x-k)
  =(-1)^{N+1}(-x)_{N+1},\n
  \text{(\romannumeral2)}:
  &\ \ \Lambda(x;\bm{\lambda})=\prod_{k=0}^N(x-k)(x+k+d)
  =(-1)^{N+1}(-x,x+d)_{N+1},\n
  \text{(\romannumeral3)}:
  &\ \ \Lambda(x;\bm{\lambda})=\prod_{k=0}^N(q^k-q^x)
  =(-1)^{N+1}(q^{-x}\,;q)_{N+1}\,q^{(N+1)x},
  \label{Lamx}\\
  \text{(\romannumeral4)}:
  &\ \ \Lambda(x;\bm{\lambda})=\prod_{k=0}^N(q^{-x}-q^{-k})
  =(-1)^{N+1}q^{-\frac12N(N+1)}(q^{-x}\,;q)_{N+1},\n
  \text{(\romannumeral5)}:
  &\ \ \Lambda(x;\bm{\lambda})=\prod_{k=0}^N(q^{-x}-q^{-k})(1-dq^{x+k})
  =(-1)^{N+1}q^{-\frac12N(N+1)}(q^{-x},dq^x\,;q)_{N+1}.
  \nonumber
\end{align}
We can show that
\begin{equation}
  \prod_{j=1}^{M+1}\Lambda(x+j-1;\bm{\lambda})
  :\text{a polynomial of degree $(M+1)(N+1)$ in
  $\eta(x;\bm{\lambda}+M\bm{\delta})$},
  \label{Lam_poly}
\end{equation}
by showing the invariance under (\romannumeral2) $x\to-x-(d+M)$ or
(\romannumeral5) $q^x\to q^{-x}(dq^M)^{-1}$.
We also have
\begin{align}
  &\quad\prod_{j=1}^M\Lambda(x+j-1;\bm{\lambda}+\bm{\delta})
  =\text{a polynomial of degree $MN$
  in $\eta(x;\bm{\lambda}+M\bm{\delta})$}\n
  &=\rho^{\binom{M}{2}N}\eta(x;\bm{\lambda}+M\bm{\delta})^{MN}
  +(\text{lower order terms}).
  \label{Lam_poly2}
\end{align}

In the rest of this subsection we assume $M\leq N$.

By using the explicit forms \eqref{Lamx}, we can show that
\begin{equation}
  \text{common factor of }\Lambda(x+j-1;\bm{\lambda})\ (j=1,2,\ldots,M+1)
  =\Lambda(x;\bm{\lambda}+M\bm{\delta}).
  \label{Lamcomfac}
\end{equation}
The components of $\bm{\bar{\delta}}$ corresponding to the parameters
$N$ and $d$ are $-1$ and $1$, respectively.
Explicit forms of $\Lambda(x+M;\bm{\lambda}-M\bm{\bar{\delta}})$ are given by
\begin{align}
  \text{(\romannumeral1)}:
  &\ \ \Lambda(x+M;\bm{\lambda}-M\bm{\bar{\delta}})
  =\prod_{k=-M}^N(x-k)=(-1)^{N+M+1}(-x-M)_{N+M+1},\n
  \text{(\romannumeral2)}:
  &\ \ \Lambda(x+M;\bm{\lambda}-M\bm{\bar{\delta}})
  =\prod_{k=-M}^N(x-k)\cdot\prod_{k=0}^{N+M}(x+d+k)\n
  &\phantom{\ \ \Lambda(x+M;\bm{\lambda}-M\bm{\bar{\delta}})}
  =(-1)^{N+M+1}(-x-M,x+d)_{N+M+1},\n
  \text{(\romannumeral3)}:
  &\ \ \Lambda(x+M;\bm{\lambda}-M\bm{\bar{\delta}})
  =q^{M(N+M+1)}\prod_{k=-M}^N(q^k-q^x)\n
  &\phantom{\ \ \Lambda(x+M;\bm{\lambda}-M\bm{\bar{\delta}})}
  =(-1)^{N+M+1}(q^{-x-M}\,;q)_{N+M+1}\,q^{(N+M+1)(x+M)},
  \label{Lam(x+M)}\\
  \text{(\romannumeral4)}:
  &\ \ \Lambda(x+M;\bm{\lambda}-M\bm{\bar{\delta}})
  =q^{-M(N+M+1)}\prod_{k=-M}^N(q^{-x}-q^{-k})\n
  &\phantom{\ \ \Lambda(x+M;\bm{\lambda}-M\bm{\bar{\delta}})}
  =(-1)^{N+M+1}q^{-\frac12(N+M)(N+M+1)}(q^{-x-M}\,;q)_{N+M+1},\n
  \text{(\romannumeral5)}:
  &\ \ \Lambda(x+M;\bm{\lambda}-M\bm{\bar{\delta}})
  =q^{-M(N+M+1)}\prod_{k=-M}^N(q^{-x}-q^{-k})\cdot
  \prod_{k=0}^{N+M}(1-dq^{x+k})\n
  &\phantom{\ \ \Lambda(x+M;\bm{\lambda}-M\bm{\bar{\delta}})}
  =(-1)^{N+M+1}q^{-\frac12(N+M)(N+M+1)}(q^{-x-M},dq^x\,;q)_{N+M+1},
  \nonumber
\end{align}
which are polynomials of degree $N+M+1$ in $\eta(x;\bm{\lambda}+M\bm{\delta})$
and vanish at $x\in\{-M,-M+1,\ldots,N\}$.
We can show that
\begin{equation}
  \frac{\prod_{j=1}^{M+1}\Lambda(x+j-1;\bm{\lambda})}
  {\prod_{j=1}^M\Lambda(x+j-1;\bm{\lambda}+\bm{\delta})}
  =\Lambda(x+M;\bm{\lambda}-M\bm{\bar{\delta}})\times\left\{
  \begin{array}{ll}
  1&:\text{(\romannumeral1)\,(\romannumeral2)}\\
  q^{-\frac12M(M+1)}&:\text{(\romannumeral3)}\\
  q^{\frac12M(M+1)}&:\text{(\romannumeral4),\,(\romannumeral5)}
  \end{array}\right.,
  \label{prodLam/Lam}
\end{equation}
by using explicit forms of $\Lambda(x)$ and the following formulas ($M\leq N$),
\begin{align}
  \prod_{j=1}^{M+1}\prod_{k=1-j}^{N+1-j}a_k
  &=\prod_{k=-M}^{-1}a_k^{M+1+k}\cdot\prod_{k=0}^{N-M}a_k^{M+1}\cdot
  \prod_{k=N-M+1}^Na_k^{N+1-k},\n
  \prod_{j=1}^{M+1}\prod_{k=j-1}^{N+j-1}a_k
  &=\prod_{k=0}^{M-1}a_k^{k+1}\cdot\prod_{k=M}^Na_k^{M+1}\cdot
  \prod_{k=N+1}^{N+M}a_k^{N+M+1-k}.
\end{align}
Let us define $\Lambda_M(x;\bm{\lambda})$ as follows,
\begin{equation}
  \Lambda_M(x;\bm{\lambda})\eqdef
  \biggl(\frac{\prod_{j=1}^M\Lambda(x+j-1;\bm{\lambda}+\bm{\delta})}
  {\prod_{j=1}^{M+1}\Lambda(x+j-1;\bm{\lambda})}\biggr)^2
  \Lambda(x;\bm{\lambda})\Lambda(x+M;\bm{\lambda}).
  \label{LambdaMdef}
\end{equation}
Then from \eqref{Lam(x+M)}--\eqref{prodLam/Lam} and \eqref{Lamx}, we have
\begin{align}
  \text{(\romannumeral1)}:
  &\ \ \Lambda_M(x;\bm{\lambda})=\frac{1}{(x+1,x-N)_M},\n
  \text{(\romannumeral2)}:
  &\ \ \Lambda_M(x;\bm{\lambda})=\frac{1}{(x+1,x-N,x+d,x+N+1+d)_M},\n
  \text{(\romannumeral3)}:
  &\ \ \Lambda_M(x;\bm{\lambda})=\frac{q^{-2MN}}{(q^{x+1},q^{x-N}\,;q)_M},
  \label{LambdaM}\\
  \text{(\romannumeral4)}:
  &\ \ \Lambda_M(x;\bm{\lambda})
  =\frac{q^{M(M+N)}q^{2Mx}}{(q^{x+1},q^{x-N}\,;q)_M},\n
  \text{(\romannumeral5)}:
  &\ \ \Lambda_M(x;\bm{\lambda})
  =\frac{q^{M(M+N)}q^{2Mx}}{(q^{x+1},q^{x-N},dq^x,dq^{x+N+1}\,;q)_M}.
  \nonumber
\end{align}

\subsection{$\bm{\varphi_M(x)}$}
\label{app:varphiM}

For five families of $\eta(x)$ \eqref{etadef},
auxiliary functions $\varphi(x;\bm{\lambda})$ are defined by \cite{os12}
\begin{equation}
  \varphi(x;\bm{\lambda})\eqdef
  \frac{\eta(x+1;\bm{\lambda})-\eta(x;\bm{\lambda})}{\eta(1;\bm{\lambda})},
  \label{varphidef}
\end{equation}
and their explicit forms are
\begin{align}
  \text{(\romannumeral1)}:
  &\ \ \varphi(x;\bm{\lambda})=1,\n
  \text{(\romannumeral2)}:
  &\ \ \varphi(x;\bm{\lambda})=\frac{2x+1+d}{1+d},\n
  \text{(\romannumeral3)}:
  &\ \ \varphi(x;\bm{\lambda})=q^x,
  \label{varphi}\\
  \text{(\romannumeral4)}:
  &\ \ \varphi(x;\bm{\lambda})=q^{-x},\n
  \text{(\romannumeral5)}:
  &\ \ \varphi(x;\bm{\lambda})=q^{-x}\frac{1-dq^{2x+1}}{1-dq}.
  \nonumber
\end{align}
They satisfy
\begin{equation}
  \frac{\eta(x+\alpha;\bm{\lambda})-\eta(x;\bm{\lambda})}
  {\eta(\alpha;\bm{\lambda})}
  =\varphi\bigl(x;\bm{\lambda}+(\alpha-1)\bm{\delta}\bigr)\quad
  (\alpha\in\mathbb{R}).
  \label{varphiprop1}
\end{equation}
Auxiliary functions $\varphi_M(x;\bm{\lambda})$ ($M\in\mathbb{Z}_{\geq 0}$)
are defined by \cite{os22}
\begin{align}
  \varphi_M(x;\bm{\lambda})&\eqdef\prod_{1\leq j<k\leq M}
  \frac{\eta(x+k-1;\bm{\lambda})-\eta(x+j-1;\bm{\lambda})}
  {\eta(k-j;\bm{\lambda})}
  \label{varphiMdef}\\
  &=\prod_{1\leq j<k\leq M}
  \varphi\bigl(x+j-1;\bm{\lambda}+(k-j-1)\bm{\delta}\bigr),
  \label{varphiMdef2}
\end{align}
($\varphi_0(x)=\varphi_1(x)=1$). From this definition we have the following
properties:
\begin{align}
  \frac{\varphi_{M+1}(x;\bm{\lambda})}{\varphi_M(x;\bm{\lambda})}
  &=\prod_{j=1}^M\varphi\bigl(x+j-1;\bm{\lambda}+(M-j)\bm{\delta}\bigr),\n
  \frac{\varphi_{M+1}(x;\bm{\lambda})}{\varphi_M(x+1;\bm{\lambda})}
  &=\prod_{k=1}^M\varphi\bigl(x;\bm{\lambda}+(k-1)\bm{\delta}\bigr),
  \label{varphiM+1M}\\
  \frac{\varphi_{M+1}(x;\bm{\lambda})}{\varphi_M(x;\bm{\lambda}+\bm{\delta})}
  &=\prod_{j=1}^M\varphi(x+j-1;\bm{\lambda}).
  \nonumber
\end{align}
Explicit forms of $\varphi_M(x;\bm{\lambda})$ are
\begin{align}
  \text{(\romannumeral1)}:
  &\ \ \varphi_M(x;\bm{\lambda})=1,\n
  \text{(\romannumeral2)}:
  &\ \ \varphi_M(x;\bm{\lambda})
  =\frac{\prod_{j=1}^{[\frac{M}{2}]}(2x+d+2j-1)_{2M-4j+1}}
  {\prod_{j=1}^{M-1}(d+1)_j},\n
  \text{(\romannumeral3)}:
  &\ \ \varphi_M(x;\bm{\lambda})=q^{\binom{M}{2}x+\binom{M}{3}},\\
  \text{(\romannumeral4)}:
  &\ \ \varphi_M(x;\bm{\lambda})=q^{-\binom{M}{2}x-\binom{M}{3}},\n
  \text{(\romannumeral5)}:
  &\ \ \varphi_M(x;\bm{\lambda})
  =\frac{\prod_{j=1}^{[\frac{M}{2}]}(dq^{2x+2j-1}\,;q)_{2M-4j+1}}
  {\prod_{j=1}^{M-1}(dq\,;q)_j}q^{-\binom{M}{2}x-\binom{M}{3}}.
  \nonumber
\end{align}
By using \eqref{varphi}, we can show that
\begin{equation}
  \varphi(x-N-1;\bm{\lambda}')
  =\frac{\varphi(x;\bm{\lambda})}{\varphi(N+1;\bm{\lambda})},\quad
  \bm{\lambda}'=\bm{\lambda}+(N+1)(\bm{\delta}+\bm{\bar{\delta}}).
  \label{varphi/varphi}
\end{equation}
From this and \eqref{varphiMdef2}, we obtain
\begin{equation}
  \frac{\varphi_M(x-N-1;\bm{\lambda}')}{\varphi_M(0;\bm{\lambda}')}
  =\frac{\varphi_M(x;\bm{\lambda})}{\varphi_M(N+1;\bm{\lambda})},\quad
  \bm{\lambda}'=\bm{\lambda}+(N+1)(\bm{\delta}+\bm{\bar{\delta}}).
  \label{varphiM/varphiM}
\end{equation}

\subsection{$\bm{B(x)}$ and $\bm{D(x)}$}
\label{app:BandD}

The potential functions $B(x;\bm{\lambda})$ and $D(x;\bm{\lambda})$ satisfy
\begin{equation}
  \frac{B(x+1;\bm{\lambda})}{B(x;\bm{\lambda}+\bm{\delta})}
  =\kappa\frac{\varphi(x;\bm{\lambda})}{\varphi(x+1;\bm{\lambda})},\quad
  \frac{D(x;\bm{\lambda})}{D(x;\bm{\lambda}+\bm{\delta})}
  =\kappa\frac{\varphi(x;\bm{\lambda})}{\varphi(x-1;\bm{\lambda})}.
  \label{BDprop}
\end{equation}
By using this and induction on $j$ ($j\in\mathbb{Z}_{\geq 0}$), we can show that
\begin{align}
  B(x+j;\bm{\lambda})&=B(x;\bm{\lambda}+j\bm{\delta})
  \prod_{l=1}^j\kappa
  \frac{\varphi(x+l-1;\bm{\lambda}+(j-l)\bm{\delta})}
  {\varphi(x+l;\bm{\lambda}+(j-l)\bm{\delta})},\n
  D(x;\bm{\lambda})&=D(x;\bm{\lambda}+j\bm{\delta})
  \prod_{l=1}^j\kappa
  \frac{\varphi(x;\bm{\lambda}+(l-1)\bm{\delta})}
  {\varphi(x-1;\bm{\lambda}+(l-1)\bm{\delta})}.
  \label{B(x+j),D(x)}
\end{align}
By using \eqref{Lamx}, we have
\begin{align}
  B(x+M;\bm{\lambda}-M\bm{\bar{\delta}})
  &=\kappa^MB(x;\bm{\lambda}+M\bm{\delta})
  \frac{\Lambda(x;\bm{\lambda})}{\Lambda(x+M;\bm{\lambda})}
  \frac{\Lambda(x+M;\bm{\lambda}+\bm{\delta})}
  {\Lambda(x;\bm{\lambda}+\bm{\delta})},\n
  D(x+M;\bm{\lambda}-M\bm{\bar{\delta}})
  &=\kappa^MD(x;\bm{\lambda}+M\bm{\delta})
  \frac{\Lambda(x+M;\bm{\lambda})}{\Lambda(x;\bm{\lambda})}
  \frac{\Lambda(x-1;\bm{\lambda}+\bm{\delta})}
  {\Lambda(x+M-1;\bm{\lambda}+\bm{\delta})}.
  \label{B(x+M),D(x+M)}
\end{align}

\subsection{$\bm{\phi_0(x)}$}
\label{app:phi0}

The ground state $\phi_0(x)$ is given by \eqref{phi0def}.
By using explicit forms of $B(x)$, $\phi_0(x)$ and $\varphi(x)$, we have
\cite{os12}
\begin{equation}
  \varphi(x;\bm{\lambda})=\sqrt{\frac{B(0;\bm{\lambda})}{B(x;\bm{\lambda})}}
  \frac{\phi_0(x;\bm{\lambda}+\bm{\delta})}{\phi_0(x;\bm{\lambda})}.
  \label{varphiprop2}
\end{equation}
This property and induction on $M\in\mathbb{Z}_{\geq 0}$ give the following,
\begin{equation}
  \sqrt{\prod_{k=1}^MB\bigl(x;\bm{\lambda}+(k-1)\bm{\delta}\bigr)}\,
  \phi_0(x;\bm{\lambda})
  =\prod_{k=1}^M\frac{\sqrt{B(0;\bm{\lambda}+(k-1)\bm{\delta})}}
  {\varphi\bigl(x;\bm{\lambda}+(k-1)\bm{\delta}\bigr)}\cdot
  \phi_0(x;\bm{\lambda}+M\bm{\delta}).
\end{equation}
For $M\leq N$, we can show that
\begin{equation}
  \phi_0(x;\bm{\lambda}+M\bm{\delta})^2\,\Lambda_M(x;\bm{\lambda})
  =\frac{\phi_0(x+M;\bm{\lambda}-M\bm{\bar{\delta}})^2}
  {\phi_0(M;\bm{\lambda}-M\bm{\bar{\delta}})^2}
  \Lambda_M(0;\bm{\lambda}),
  \label{phi0Lam}
\end{equation}
by using explicit forms of $\phi_0(x;\bm{\lambda})$ and $\Lambda_M(x)$
\eqref{LambdaM} and the following formulas,
\begin{equation}
  \frac{(\alpha+M)_x}{(\alpha-M)_{x+M}}
  =\frac{(\alpha+x)_M}{(\alpha-M)_{2M}},\quad
  \frac{(\alpha q^M\,;q)_x} {(\alpha q^{-M}\,;q)_{x+M}}
  =\frac{(\alpha q^x\,;q)_M} {(\alpha q^{-M}\,;q)_{2M}}.
\end{equation}
The equation \eqref{phi0Lam} is shown for $x\in\mathbb{Z}_{\geq0}$, but
the domain of $\phi_0(x)$ can be extended to $x\in\mathbb{R}$ ($x\in\mathbb{Z}$)
as mentioned in \S\,\ref{sec:sabuneq}.
After canceling the factors in
$\phi_0(x;\bm{\lambda}+M\bm{\delta})^2\,\Lambda_M(x;\bm{\lambda})$,
it gives $(\text{const})\times\phi_0(x+M;\bm{\lambda}-M\bm{\bar{\delta}})^2$,
which is well defined for $x\geq-M$.

\section{``\,$\bm{n\to d_i}$ Limit\,''}
\label{app:n->di}

In this appendix we discuss an appropriate ``$n\to d_i$ limit'' of
$\check{Q}^{\text{monic}}_{\mathcal{D}',n}(x;\bm{\lambda})$ \eqref{cQmonic}.
We try to achieve this limit by shifting $N$ to $N+\varepsilon$ and taking
$\varepsilon\to 0$ limit.
 
The Casoratian in \eqref{cQmonic} is rewritten as
\begin{align}
  &\quad\text{W}_{\text{C}}[\check{P}^{\text{monic}}_{d_1},\ldots,
  \check{P}^{\text{monic}}_{d_M},
  \check{P}^{\text{monic}}_n](x;\bm{\lambda})\n
  &=\left|
  \begin{array}{cccc}
  &\vdots&&\vdots\\
  \cdots&\rho^{(N+1)m_k}\Lambda(x_j;\bm{\lambda})
  \check{P}^{\text{monic}}_{m_k}(x_j-N-1;\bm{\lambda}')&\cdots
  &\check{P}^{\text{monic}}_n(x_j;\bm{\lambda})\\
  &\vdots&&\vdots\\
  \end{array}\right|\n
  &=\Lambda(x;\bm{\lambda}+M\bm{\delta})^M\left|
  \begin{array}{cccc}
  &\vdots&&\vdots\\
  \cdots&\rho^{(N+1)m_k}
  \frac{\Lambda(x_j;\bm{\lambda})}{\Lambda(x;\bm{\lambda}+M\bm{\delta})}
  \check{P}^{\text{monic}}_{m_k}(x_j-N-1;\bm{\lambda}')&\cdots
  &\check{P}^{\text{monic}}_n(x_j;\bm{\lambda})\\
  &\vdots&&\vdots\\
  \end{array}\right|\n
  &=\Lambda(x;\bm{\lambda}+M\bm{\delta})^M
  \prod_{j=1}^{M+1}\frac{\Lambda(x_j;\bm{\lambda})}
  {\Lambda(x;\bm{\lambda}+M\bm{\delta})}\n
  &\quad\times\left|
  \begin{array}{cccc}
  &\vdots&&\vdots\\
  \cdots&\rho^{(N+1)m_k}
  \check{P}^{\text{monic}}_{m_k}(x_j-N-1;\bm{\lambda}')&\cdots
  &\frac{\Lambda(x;\bm{\lambda}+M\bm{\delta})}{\Lambda(x_j;\bm{\lambda})}
  \check{P}^{\text{monic}}_n(x_j;\bm{\lambda})\\
  &\vdots&&\vdots\\
  \end{array}\right|,
  \label{WCd1dMn}
\end{align}
where $x_j=x+j-1$.
For $n=d_i=N+1+m_i$, the $(j,M+1)$-element is expressed as
\begin{equation}
  \frac{\Lambda(x;\bm{\lambda}+M\bm{\delta})}{\Lambda(x_j;\bm{\lambda})}
  \check{P}^{\text{monic}}_n(x_j;\bm{\lambda})
  =\frac{\Lambda(x;\bm{\lambda}+M\bm{\delta})}{\Lambda(x_j;\bm{\lambda})}
  \Lambda(x_j;\bm{\lambda})\rho^{(N+1)m_i}
  \check{P}^{\text{monic}}_{m_i}(x_j-N-1;\bm{\lambda}').
  \label{(j,M+1)}
\end{equation}
Let us consider the contributions from the following three factors,
\begin{equation}
  \text{(a): }
  \frac{\Lambda(x;\bm{\lambda}+M\bm{\delta})}{\Lambda(x_j;\bm{\lambda})},\quad
  \text{(b): }
  \check{P}^{\text{monic}}_n(x_j;\bm{\lambda}),\quad
  \text{(c): }
  \check{P}^{\text{monic}}_{m_i}(x_j-N-1;\bm{\lambda}').
  \label{3factors}
\end{equation}
In the next three subsections, we will see how these quantities change under
the shift $N\to N+\varepsilon$.

In this appendix we assume $M\leq N$ and take $\bm{\lambda}'$ \eqref{lambda'def}
and set $A$ as follows,
\begin{equation}
  A=\left\{\begin{array}{ll}
  \varepsilon&:\text{non $q$-polynomial}\\
  1-q^{\varepsilon}&:\text{$q$-polynomial}
  \end{array}\right..
\end{equation}
We will use the following formulas,
\begin{align}
  (\alpha+\varepsilon)_n&=\prod_{j=0}^{n-1}(\alpha+\varepsilon+j)
  =\prod_{j=0}^{n-1}(\alpha+j)\Bigl(1+\frac{\varepsilon}{\alpha+j}\Bigr)
  =(\alpha)_n\Bigl(1+\varepsilon\sum_{j=0}^{n-1}\frac{1}{\alpha+j}
  +O(\varepsilon^2)\Bigr),\n
  (\alpha q^{\varepsilon}\,;q)_n&=\prod_{j=0}^{n-1}(1-\alpha q^{\varepsilon}q^j)
  =\prod_{j=0}^{n-1}(1-\alpha q^j)
  \Bigl(1+\frac{(1-q^{\varepsilon})\alpha q^j}{1-\alpha q^j}\Bigr)\n
  &=(\alpha\,;q)_n\Bigl(1+(1-q^{\varepsilon})
  \sum_{j=0}^{n-1}\frac{\alpha q^j}{1-\alpha q^j}+O(\varepsilon^2)\Bigr),
  \label{qpoche}\\
  q^{\alpha\varepsilon}&=1-(1-q^{\varepsilon})\alpha+O(\varepsilon^2),\quad
  1-q^{-\varepsilon}=-(1-q^{\varepsilon})+O(\varepsilon^2).
  \nonumber
\end{align}

\subsection{(a) Contribution from $\bm{\Lambda(x)}$}
\label{app:(a)Lambda}

{}From \eqref{Lamx} and \eqref{Lamcomfac},
$\Lambda(x+j-1;\bm{\lambda})/\Lambda(x;\bm{\lambda}+M\bm{\delta})$
($j=1,2,\ldots,M+1$) are expressed as
\begin{align}
  \text{(\romannumeral1)}:
  &\ \ \frac{\Lambda(x+j-1;\bm{\lambda})}
  {\Lambda(x;\bm{\lambda}+M\bm{\delta})}=
  (-1)^M(-x-j+1)_{j-1}(-x+N-M+1)_{M+1-j},\n
  \text{(\romannumeral2)}:
  &\ \ \frac{\Lambda(x+j-1;\bm{\lambda})}
  {\Lambda(x;\bm{\lambda}+M\bm{\delta})}=
  (-1)^M(-x-j+1,x+N+1+d)_{j-1}\n
  &\phantom{\ \ \frac{\Lambda(x+j-1;\bm{\lambda})}
  {\Lambda(x;\bm{\lambda}+M\bm{\delta})}=}
  \times(-x+N-M+1,x+j-1+d)_{M+1-j},\n
  \text{(\romannumeral3)}:
  &\ \ \frac{\Lambda(x+j-1;\bm{\lambda})}
  {\Lambda(x;\bm{\lambda}+M\bm{\delta})}=
  (-1)^Mq^{(j-1)(N+1)}q^{Mx}(q^{-x-j+1}\,;q)_{j-1}(q^{-x+N-M+1}\,;q)_{M+1-j},
  \label{Lam/Lam}\\
  \text{(\romannumeral4)}:
  &\ \ \frac{\Lambda(x+j-1;\bm{\lambda})}
  {\Lambda(x;\bm{\lambda}+M\bm{\delta})}=
  (-1)^Mq^{\frac12M(M-1)-MN}(q^{-x-j+1}\,;q)_{j-1}(q^{-x+N-M+1}\,;q)_{M+1-j},\n
  \text{(\romannumeral5)}:
  &\ \ \frac{\Lambda(x+j-1;\bm{\lambda})}
  {\Lambda(x;\bm{\lambda}+M\bm{\delta})}=
  (-1)^Mq^{\frac12M(M-1)-MN}(q^{-x-j+1},dq^{x+N+1}\,;q)_{j-1}\n
  &\phantom{\ \ \frac{\Lambda(x+j-1;\bm{\lambda})}
  {\Lambda(x;\bm{\lambda}+M\bm{\delta})}=}
  \times(q^{-x+N-M+1},dq^{x+j-1}\,;q)_{M+1-j}.
  \nonumber
\end{align}
The parameter $N$ is a positive integer, but the expression of r.h.s.\ of
\eqref{Lam/Lam} allow us to treat $N$ as a continuous real parameter.
So, by shifting $N$ by a small amount $\varepsilon$, we define
$X_{M,j}(x;\bm{\lambda})$ ($j=1,2,\ldots,M+1$) as follows:
\begin{equation}
  \frac{\Lambda(x;\bm{\lambda}+M\bm{\delta})}
  {\Lambda(x+j-1;\bm{\lambda})}\bigg|_{N\to N+\varepsilon}
  \biggl(\frac{\Lambda(x;\bm{\lambda}+M\bm{\delta})}
  {\Lambda(x+j-1;\bm{\lambda})}\biggr)^{-1}
  =1-AX_{M,j}(x;\bm{\lambda})+O(\varepsilon^2).
  \label{XMjdef}
\end{equation}
By using \eqref{qpoche}, explicit forms of $X_{M,j}(x;\bm{\lambda})$ are
\begin{align}
  \text{(\romannumeral1)}:
  &\ \ X_{M,j}(x;\bm{\lambda})=\sum_{l=0}^{M-j}\frac{1}{-x+N-M+1+l},\n
  \text{(\romannumeral2)}:
  &\ \ X_{M,j}(x;\bm{\lambda})=\sum_{l=0}^{M-j}\frac{1}{-x+N-M+1+l}
  +\sum_{l=0}^{j-2}\frac{1}{x+N+1+d+l},\n
  \text{(\romannumeral3)}:
  &\ \ X_{M,j}(x;\bm{\lambda})=1-j-\sum_{l=0}^{M-j}\frac{1}{1-q^{x-N+M-1-l}},
  \label{XMj}\\
  \text{(\romannumeral4)}:
  &\ \ X_{M,j}(x;\bm{\lambda})=j-1+\sum_{l=0}^{M-j}\frac{1}{1-q^{-x+N-M+1+l}},\n
  \text{(\romannumeral5)}:
  &\ \ X_{M,j}(x;\bm{\lambda})=\sum_{l=0}^{M-j}\frac{1}{1-q^{-x+N-M+1+l}}
  +\sum_{l=0}^{j-2}\frac{1}{1-dq^{x+N+1+l}}.
  \nonumber
\end{align}
Moreover we define $Y_{M,j}(x;\bm{\lambda})$ ($j=1,2,\ldots,M+1$) as
\begin{equation}
  Y_{M,j}(x;\bm{\lambda})\eqdef
  \Lambda(x+j-1;\bm{\lambda})X_{M,j}(x;\bm{\lambda}).
  \label{YMjdef}
\end{equation}
Their explicit forms are
\begin{align}
  \text{(\romannumeral1)}:
  &\ \ Y_{M,j}(x;\bm{\lambda})=(-1)^{N+1}
  \sum_{l=0}^{M-j}(-x-j+1)_{N-M+j+l}(-x+N-M+l+2)_{M-j-l},\n
  \text{(\romannumeral2)}:
  &\ \ Y_{M,j}(x;\bm{\lambda})=(-1)^{N+1}\Bigl(
  (x+j-1+d)_{N+1}\sum_{l=0}^{M-j}(-x-j+1)_{N-M+j+l}\n
  &\hspace{81mm}\times(-x+N-M+l+2)_{M-j-l}\n
  &\phantom{\ \ Y_{M,j}(x;\bm{\lambda})=}
  +(-x-j+1)_{N+1}\sum_{l=0}^{j-2}(x+j-1+d)_{N-j+2+l}
  (x+N+l+2+d)_{j-2-l}\Bigr),\n
  \text{(\romannumeral3)}:
  &\ \ Y_{M,j}(x;\bm{\lambda})=(1-j)\Lambda(x+j-1;\bm{\lambda})\n
  &\phantom{\ \ Y_{M,j}(x;\bm{\lambda})=}
  -q^{\frac12N(N+1)}
  \sum_{l=0}^{M-j}(q^{x+j-1-N}\,;q)_{M-j-l}(q^{x-N+M-l}\,;q)_{N-M+j+l},
  \label{YMj}\\
  \text{(\romannumeral4)}:
  &\ \ Y_{M,j}(x;\bm{\lambda})=(j-1)\Lambda(x+j-1;\bm{\lambda})\n
  &\phantom{\ \ Y_{M,j}(x;\bm{\lambda})=}
  +(-1)^{N+1}q^{-\frac12N(N+1)}
  \sum_{l=0}^{M-j}(q^{-x-j+1}\,;q)_{N-M+j+l}(q^{-x+N-M+l+2}\,;q)_{M-j-l},\n
  \text{(\romannumeral5)}:
  &\ \ Y_{M,j}(x;\bm{\lambda})=(-1)^{N+1}q^{-\frac12N(N+1)}\Bigl(
  (dq^{x+j-1}\,;q)_{N+1}\sum_{l=0}^{M-j}(q^{-x-j+1}\,;q)_{N-M+j+l}\n
  &\hspace{93mm}\times(q^{-x+N-M+l+2}\,;q)_{M-j-l}\n
  &\phantom{\ \ Y_{M,j}(x;\bm{\lambda})=}
  +(q^{-x-j+1}\,;q)_{N+1}\sum_{l=0}^{j-2}(dq^{x+j-1}\,;q)_{N-j+2+l}
  (dq^{x+N+l+2}\,;q)_{j-2-l}\Bigr).
  \nonumber
\end{align}

\subsection{(b) Contribution from $\bm{\check{P}^{\text{monic}}_n(x)}$}
\label{app:(b)cPn}

In the expression of $\check{P}^{\text{monic}}_n(x;\bm{\lambda})$,
the parameter $N$ appears as $(-N+k+\alpha)_{n-k}$ or
$(\alpha q^{-N+k}\,;q)_{n-k}$ ($\alpha$: $N$-independent, e.g. $\alpha=0$ or
$\alpha=1$), which are defined for a continuous real parameter $N$.
So, by shifting $N$ by a small amount $\varepsilon$, we define
$\check{\tilde{P}}^{(1)}_n(x;\bm{\lambda})$ as follows:
\begin{equation}
  \check{P}^{\text{monic}}_n(x;\bm{\lambda})\big|_{N\to N+\varepsilon}
  =\check{P}^{\text{monic}}_n(x;\bm{\lambda})
  -A\check{\tilde{P}}^{(1)}_n(x;\bm{\lambda})+O(\varepsilon^2).
  \label{ctP1def}
\end{equation}
This $\check{\tilde{P}}^{(1)}_n(x;\bm{\lambda})\eqdef
\tilde{P}^{(1)}_n(\eta(x;\bm{\lambda});\bm{\lambda})$ is a polynomial
of degree $n-1$ in $\eta(x;\bm{\lambda})$ and
0 for $n=0$.
By using \eqref{qpoche}, explicit forms of
$\check{\tilde{P}}^{(1)}_n(x;\bm{\lambda})$ are given in \S\,\ref{app:poly}.

\subsection{(c) Contribution from $\bm{\check{P}^{\text{monic}}_m(x)}$}
\label{app:(c)cPm}

Similar to \eqref{ctP1def}, we define
$\check{\tilde{P}}^{(2)}_m(x;\bm{\lambda})$ as follows:
\begin{equation}
  \check{P}^{\text{monic}}_m(x-N-1;\bm{\lambda}')\big|_{N\to N+\varepsilon}
  =\check{P}^{\text{monic}}_m(x-N-1;\bm{\lambda}')(1+AB)
  +A\check{\tilde{P}}^{(2)}_m(x;\bm{\lambda})+O(\varepsilon^2).
  \label{ctP2def}
\end{equation}
Here $B$ is introduced to simplify later equations,
$B=-2m$ for $q$R, d$q$H, a$q$K, $B=-m$ for d$q$K and $B=0$ for otherwise.
Note that the sign of the factor in front of
$\check{\tilde{P}}^{(2)}_m(x;\bm{\lambda})$ is taken as $+A$ instead of $-A$
for \eqref{XMjdef} and \eqref{ctP1def}.
This $\check{\tilde{P}}^{(2)}_m(x;\bm{\lambda})\eqdef
\tilde{P}^{(2)}_m(\eta(x;\bm{\lambda});\bm{\lambda})$ is a polynomial
of degree $m$ in $\eta(x;\bm{\lambda})$.
By using \eqref{qpoche}, explicit forms of
$\check{\tilde{P}}^{(2)}_m(x;\bm{\lambda})$ are given in \S\,\ref{app:poly}.

\subsection{$\bm{\check{Q}^{\text{monic}}_{\mathcal{D}',n}(x)}$ with
$\bm{n=d_i}$}
\label{app:cQDpn}

We have considered contributions from the $N$ shift coming from three factors
\eqref{3factors}.
These three contributions are not independent, but overlapping.
Trial-and-error calculations lead us to the following prescription;
add the contributions from (a) and (b) and subtract that of (c).
We replace the $(j,M+1)$-element \eqref{(j,M+1)} with
\begin{align}
  &\qquad\frac{\Lambda(x;\bm{\lambda}+M\bm{\delta})}{\Lambda(x_j;\bm{\lambda})}
  \bigl(1-AX_{M,j}(x;\bm{\lambda})+O(\varepsilon^2)\bigr)
  \check{P}^{\text{monic}}_n(x_j;\bm{\lambda})\n
  &\quad+\frac{\Lambda(x;\bm{\lambda}+M\bm{\delta})}{\Lambda(x_j;\bm{\lambda})}
  \bigl(\check{P}^{\text{monic}}_n(x_j;\bm{\lambda})
  -A\check{\tilde{P}}^{(1)}_n(x_j;\bm{\lambda})+O(\varepsilon^2)
  \bigr)\n
  &\quad-\frac{\Lambda(x;\bm{\lambda}+M\bm{\delta})}{\Lambda(x_j;\bm{\lambda})}
  \Lambda(x_j;\bm{\lambda})\rho^{(N+1)m_i}
  \bigl(\check{P}^{\text{monic}}_{m_i}(x_j-N-1;\bm{\lambda}')(1+AB)
  +A\check{\tilde{P}}^{(2)}_{m_i}(x_j;\bm{\lambda})+O(\varepsilon^2)\bigr)\n
  &=\frac{\Lambda(x;\bm{\lambda}+M\bm{\delta})}{\Lambda(x_j;\bm{\lambda})}
  \bigl(\check{P}^{\text{monic}}_n(x_j;\bm{\lambda})(1-AB)
  -AR_j(x)+O(\varepsilon^2)\bigr),
\end{align}
where $R_j(x)$ is given by
\begin{align}
  &\quad R_j(x)=R_{M,j,n}(x;\bm{\lambda})
  \tag{\ref{Rjdef}}\\
  &\eqdef\rho^{(N+1)m_i}Y_{M,j}(x;\bm{\lambda})
  \check{P}^{\text{monic}}_{m_i}(x_j-N-1;\bm{\lambda}')
  +\check{\tilde{P}}^{(1)}_n(x_j;\bm{\lambda})
  +\rho^{(N+1)m_i}\Lambda(x_j;\bm{\lambda})
  \check{\tilde{P}}^{(2)}_{m_i}(x_j;\bm{\lambda}).
  \nonumber
\end{align}
Then \eqref{WCd1dMn} becomes
\begin{align}
  &\quad\Lambda(x;\bm{\lambda}+M\bm{\delta})^M
  \prod_{j=1}^{M+1}\frac{\Lambda(x_j;\bm{\lambda})}
  {\Lambda(x;\bm{\lambda}+M\bm{\delta})}\n
  &\quad\times\left|
  \begin{array}{cccc}
  &\vdots&&\vdots\\
  \!\cdots\!\!\!&\rho^{(N+1)m_k}
  \check{P}^{\text{monic}}_{m_k}(x_j-N-1;\bm{\lambda}')&\!\!\!\cdots\!\!\!
  &\frac{\Lambda(x;\bm{\lambda}+M\bm{\delta})}{\Lambda(x_j;\bm{\lambda})}
  \bigl(\check{P}^{\text{monic}}_n(x_j;\bm{\lambda})(1-AB)
  -AR_j(x)\bigr)\!\!\\
  &\vdots&&\vdots\\
  \end{array}\right|\n
  &\quad+O(\varepsilon^2)\n
  &=\left|
  \begin{array}{cccc}
  &\vdots&&\vdots\\
  \cdots&\check{P}^{\text{monic}}_{d_k}(x_j;\bm{\lambda})&\cdots
  &\check{P}^{\text{monic}}_n(x_j;\bm{\lambda})(1-AB)-AR_j(x)\\
  &\vdots&&\vdots\\
  \end{array}\right|+O(\varepsilon^2)\n
  &=-A\left|
  \begin{array}{cccc}
  &\vdots&&\vdots\\
  \cdots&\check{P}^{\text{monic}}_{d_k}(x_j;\bm{\lambda})&\cdots
  &R_j(x)\\
  &\vdots&&\vdots\\
  \end{array}\right|+O(\varepsilon^2).
  \label{WCd1dMn2}
\end{align}

The normalization constant $c^{\eta}_{\mathcal{D},n}(\bm{\lambda})$ in
\eqref{cQmonic} for $n=d_i$ is rewritten as \eqref{cetaDn'},
\begin{equation}
  c^{\eta}_{\mathcal{D},n}(\bm{\lambda})
  =A'c^{\prime\,\eta}_{\mathcal{D},n}(\bm{\lambda}),\quad
  A'=\left\{
  \begin{array}{ll}
  n-d_i&:\text{(\romannumeral1)},\,\text{(\romannumeral2)}\\
  q^{d_i}-q^n&:\text{(\romannumeral3)}\\
  q^{-n}-q^{-d_i}&:\text{(\romannumeral4)},\,\text{(\romannumeral5)}
  \end{array}\right..
\end{equation}
Under the shift $N\to N+\varepsilon$ in $d_i=N+1+m_i$ while keeping $n=d_i$,
namely $(n,d_i)\to(d_i,d_i+\varepsilon)$, $A'$ behaves as
\begin{equation}
  A'\big|_{N\to N+\varepsilon}=-A\rho^{d_i}+O(\varepsilon^2).
\end{equation}
So $c^{\eta}_{\mathcal{D},n}(\bm{\lambda})$ with $n=d_i$ behaves as
\begin{equation}
  c^{\eta}_{\mathcal{D},n}(\bm{\lambda})\big|_{N\to N+\varepsilon}
  =-A\rho^{d_i}c^{\prime\,\eta}_{\mathcal{D},n}(\bm{\lambda})
  +O(\varepsilon^2),
  \label{cetaDn=-A}
\end{equation}
and this factor $A$ cancels the factor $A$ in \eqref{WCd1dMn2}.
Thus we obtain \eqref{cQmonic2} from \eqref{cQmonic} by taking
$\varepsilon\to 0$ limit.

\subsection{$\bm{\mathcal{E}_n-\mathcal{E}_{d_i}}$ and
$\bm{d^{\,\prime\,\text{monic}}_n(\bm{\lambda})^2}$}
\label{app:En-Edi}

For five families of $\mathcal{E}_n$ \eqref{Endef}, we have
\begin{align}
  \mathcal{E}_n(\bm{\lambda})-\mathcal{E}_{d_i}(\bm{\lambda})
  &=A''\times\left\{
  \begin{array}{ll}
  1&:\text{(\romannumeral1)}',\,\text{(\romannumeral3)}',\,
  \text{(\romannumeral4)}'\\
  n+d_i+\tilde{d}&:\text{(\romannumeral2)}'\\
  1-\tilde{d}q^{n+d_i}&:\text{(\romannumeral5)}'
  \end{array}\right.,\\
  A''&=\left\{
  \begin{array}{ll}
  n-d_i&:\text{(\romannumeral1)}',\,\text{(\romannumeral2)}'\\
  q^{d_i}-q^n&:\text{(\romannumeral3)}'\\
  q^{-n}-q^{-d_i}&:\text{(\romannumeral4)}',\,\text{(\romannumeral5)}'
  \end{array}\right..
\end{align}
For $n=d_i>N$, under the shift $N\to N+\varepsilon$ in $d_i=N+1+m_i$ while
keeping $n=d_i$, namely $(n,d_i)\to(d_i,d_i+\varepsilon)$, $A''$ behaves as
\begin{equation}
  A''\big|_{N\to N+\varepsilon}=-A\kappa^{d_i}+O(\varepsilon^2).
\end{equation}
Thus $\mathcal{E}_n-\mathcal{E}_{d_i}$ with $n=d_i>N$ behaves as
\begin{equation}
  \bigl(\mathcal{E}_n(\bm{\lambda})-\mathcal{E}_{d_i}(\bm{\lambda})\bigr)
  \bigm|_{N\to N+\varepsilon}
  =-A\kappa^{d_i}\times\left\{
  \begin{array}{ll}
  1&:\text{(\romannumeral1)}',\,\text{(\romannumeral3)}',
  \,\text{(\romannumeral4)}'\\
  2d_i+\tilde{d}&:\text{(\romannumeral2)}'\\
  1-\tilde{d}q^{2d_i}&:\text{(\romannumeral5)}'
  \end{array}\right.
  +O(\varepsilon^2).
  \label{En-Edi=-A}
\end{equation}

There is a factor $1/(-N)_n$ or $1/(q^{-N}\,;q)_n$ in the expression
$d^{\text{monic}}_n(\bm{\lambda})^2=c_n(\bm{\lambda})^2d_n(\bm{\lambda})^2$.
So $1/d^{\text{monic}}_n(\bm{\lambda})^2$ vanishes for $n>N$.
For $n>N$ ($n=N+1+m$), by shifting $N$ to $N+\varepsilon$, we have
\begin{align}
  (-N)_n\big|_{N\to N+\varepsilon}
  &=-\varepsilon(-N)_N(1)_m+O(\varepsilon^2),\n
  (q^{-N}\,;q)_n\big|_{N\to N+\varepsilon}
  &=-(1-q^{\varepsilon})(q^{-N}\,;q)_N(q\,;q)_m+O(\varepsilon^2).
\end{align}
Let us define $d^{\,\prime\,\text{monic}}_n(\bm{\lambda})^2$ for $n>N$
($n=N+1+m$) as follows,
\begin{align}
  &d^{\,\prime\,\text{monic}}_n(\bm{\lambda})^2
  \eqdef d^{\text{monic}}_n(\bm{\lambda})^2\big|_{\text{replacement}},\n
  &\qquad\text{replacement}:\left\{
  \begin{array}{ll}
  (-N)_n\to(-N)_N(1)_m&:\text{non $q$-polynomial}\\[2pt]
  (q^{-N}\,;q)_n\to(q^{-N}\,;q)_N(q\,;q)_m&:\text{$q$-polynomial}
  \end{array}\right..
  \label{d'nmonic}
\end{align}
Then $d^{\text{monic}}_n(\bm{\lambda})^2$ for $n>N$ behaves as
\begin{equation}
  d^{\text{monic}}_n(\bm{\lambda})^2\big|_{N\to N+\varepsilon}
  =\frac{1}{-A}d^{\,\prime\,\text{monic}}_n(\bm{\lambda})^2
  \times\bigl(1+O(\varepsilon)\bigr).
  \label{dnmonic=-A}
\end{equation}



\begin{thebibliography}{99}

\bibitem{gkm08}
D.\,G\'{o}mez-Ullate, N.\,Kamran and R.\,Milson,
%
``An extended class of orthogonal polynomials defined by a
Sturm-Liouville problem,''
J. Math. Anal. Appl. {\bf 359} (2009) 352-367,
{\tt arXiv:0807.3939[math-\hspace{0pt}ph]}.

\bibitem{q08}
C.\,Quesne,
``Exceptional orthogonal polynomials, exactly solvable potentials
and supersymmetry,''
J. Phys. {\bf A41} (2008) 392001 (6pp),
{\tt arXiv:0807.4087[quant-ph]}.

\bibitem{os16}
S.\,Odake and R.\,Sasaki,
``Infinitely many shape invariant potentials and new orthogonal polynomials,''
Phys. Lett. {\bf B679} (2009) 414-417,
{\tt arXiv:0906.0142[math-ph]}.

\bibitem{os19}
S.\,Odake and R.\,Sasaki,
``Another set of infinitely many exceptional ($X_{\ell}$) Laguerre
polynomials,''
Phys. Lett. {\bf B684} (2010) 173-176,
{\tt arXiv:0911.3442[math-ph]}.


\bibitem{gkm11_2} 
D.\,G\'{o}mez-Ullate, N.\,Kamran and R.\,Milson,
``Two-step Darboux transformations and exceptional Laguerre polynomials,''
J. Math. Anal. Appl. {\bf 387} (2012) 410-418,
{\tt arXiv:\hspace{0pt}1103.5724[math-ph]}.

\bibitem{os25}
S.\,Odake and R.\,Sasaki,
``Exactly solvable quantum mechanics and infinite families of
multi-indexed orthogonal polynomials,''
Phys. Lett. {\bf B702} (2011) 164-170,
{\tt arXiv:1105.\hspace{0pt}0508[math-ph]}.

\bibitem{os17}
S.\,Odake and R.\,Sasaki,
``Infinitely many shape invariant discrete quantum mechanical systems
and new exceptional orthogonal polynomials related to the Wilson and
Askey-Wilson polynomials,''
Phys. Lett. {\bf B682} (2009) 130-136,
{\tt arXiv:0909.3668[math-ph]}.

\bibitem{os27}
S.\,Odake and R.\,Sasaki,
``Multi-indexed Wilson and Askey-Wilson polynomials,''
J. Phys. {\bf A46} (2013) 045204 (22pp),
{\tt arXiv:1207.5584[math-ph]}.

\bibitem{os23}
S.\,Odake and R.\,Sasaki,
``Exceptional ($X_{\ell}$) ($q$)-Racah polynomials,''
Prog. Theor. Phys. {\bf 125} (2011) 851-870,
{\tt arXiv:1102.0812[math-ph]}.

\bibitem{os26}
S.\,Odake and R.\,Sasaki,
``Multi-indexed ($q$-)Racah polynomials,''
J. Phys. {\bf A 45} (2012) 385201 (21pp),
{\tt arXiv:1203.5868[math-ph]}.

\bibitem{ggm13}
D.\,G\'{o}mez-Ullate, Y.\,Grandati and R.\,Milson,
``Rational extensions of the quantum harmonic oscillator and exceptional
Hermite polynomials,''
J. Phys. {\bf A47} (2014) 015203 (27pp),
{\tt arXiv:1306.5143[math-ph]}.

\bibitem{d14}
A.\,J.\,Dur\'{a}n,
``Exceptional Meixner and Laguerre orthogonal polynomials,''
J. Approx. Theory {\bf 184} (2014) 176-208,
{\tt arXiv:1310.4658[math.CA]}.

\bibitem{d17}
A.\,J.\,Dur\'{a}n,
``Exceptional Hahn and Jacobi orthogonal polynomials,''
J. Approx. Theory {\bf 214} (2017) 9-48,
{\tt arXiv:1510.02579[math.CA]}.

\bibitem{os35}
S.\,Odake and R.\,Sasaki,
``Multi-indexed Meixner and Little $q$-Jacobi (Laguerre) Polynomials,''
J. Phys. {\bf A50} (2017) 165204 (23pp),
{\tt arXiv:1610.09854[math.CA]}.

\bibitem{idQMcH}
S.\,Odake,
``Exactly Solvable Discrete Quantum Mechanical Systems and Multi-indexed
Orthogonal Polynomials of the Continuous Hahn and Meixner-Pollaczek Types,''
Prog. Theor. Exp. Phy. {\bf 2019} (2019) 123A01 (20pp),
{\tt arXiv:1907.12218[math-ph]}.


\bibitem{ismail}
M.\,E.\,H.\,Ismail,
{\it Classical and Quantum Orthogonal Polynomials in One Variable\/},
vol. 98 of Encyclopedia of mathematics and its applications,
Cambridge Univ. Press, Cambridge (2005).

\bibitem{kls}
R.\,Koekoek, P.\,A.\,Lesky and R.\,F.\,Swarttouw,
{\it Hypergeometric orthogonal polynomials and their $q$-analogues,\/}
Springer-Verlag Berlin-Heidelberg (2010).

\bibitem{os24}
S.\,Odake and R.\,Sasaki,
``Discrete quantum mechanics,'' (Topical Review)
J. Phys. {\bf A44} (2011) 353001 (47pp),
{\tt arXiv:1104.0473[math-ph]}.

\bibitem{os29}
S.\,Odake and R.\,Sasaki,
``Krein-Adler transformations for shape-invariant potentials and pseudo
virtual states,''
J. Phys. {\bf A46} (2013) 245201 (24pp),
{\tt arXiv:1212.6595[math-\hspace{0pt}ph]}.

\bibitem{os30}
S.\,Odake and R.\,Sasaki,
``Casoratian Identities for the Wilson and Askey-Wilson Polynomials,''
J. Approx. Theory {\bf 193} (2015) 184-209,
{\tt arXiv:1308.4240[math-ph]}.

\bibitem{casoidrdqm}
S.\,Odake,
``Casoratian Identities for the Discrete Orthogonal Polynomials
in Discrete Quantum Mechanics with Real Shifts,''
Prog. Theor. Exp. Phy. {\bf 2017(12)} (2017) 123A02 (30pp),
{\tt arXiv:1708.01830[math-ph]}.

\bibitem{os28}
S.\,Odake and R.\,Sasaki,
``Extensions of solvable potentials with finitely many discrete eigenstates,''
J. Phys. {\bf A46} (2013) 235205 (15pp),
{\tt arXiv:1301.3980[math-ph]}.

\bibitem{mtv22}
H.\,Miki,\ S.\,Tsujimoto and L.\,Vinet,
``The single-indexed exceptional Krawtchouk polynomials,''
J. Difference Equ. Appl. {\bf 29} (2023) 344–365,
{\tt arXiv:2201.12359[math.CA]}.

\bibitem{os40}
S.\,Odake and R.\,Sasaki,
````Diophantine'' and Factorisation Properties of Finite Orthogonal Polynomials
in the Askey Scheme,''
{\tt arXiv:2207.14479[math.CA]}.

\bibitem{os22}
S.\,Odake and R.\,Sasaki,
``Dual Christoffel transformations,''
Prog. Theor. Phys. {\bf 126} (2011) 1-34,
{\tt arXiv:1101.5468[math-ph]}.

\bibitem{os12}
S.\,Odake and R.\,Sasaki,
``Orthogonal Polynomials from Hermitian Matrices,''
J. Math. Phys. {\bf 49} (2008) 053503 (43pp),
{\tt arXiv:0712.4106[math.CA]}.


\bibitem{os7}
S.\,Odake and R.\,Sasaki,
``Unified theory of annihilation-creation operators for solvable
(`discrete') quantum mechanics,''
J. Math. Phys. {\bf 47} (2006) 102102 (33pp),
{\tt arXiv:\hspace{0pt}quant-ph/0605215}.
%

\bibitem{wcid}
S.\,Odake,
``Wronskian/Casoratian Identities and their Application to Quantum Mechanical
Systems,''
J. Phys. {\bf A53} (2020) 365202 (21pp),
{\tt arXiv:2003.00219[math-ph]}.

\bibitem{rrmiop5}
S.\,Odake,
``Recurrence Relations of the Multi-Indexed Orthogonal Polynomials $\V$ :
Racah and $q$-Racah types,''
J. Math. Phys. {\bf 60} (2019) 023508 (30pp),
{\tt arXiv:1804.10352\hspace{0mm}[math-ph]}.

\end{thebibliography}
\end{document}